\RequirePackage{fix-cm}
\documentclass[twocolumn,epjc3]{svjour3}  
\smartqed  
\RequirePackage{graphicx}
\RequirePackage{color}
\RequirePackage{mathtools}
\RequirePackage{snapshot}
\RequirePackage[english]{babel}
\RequirePackage[utf8]{inputenc}
\RequirePackage{cuted}
\RequirePackage{booktabs}
\RequirePackage{multirow}
\RequirePackage{url}
\RequirePackage[numbers,sort&compress]{natbib}
\RequirePackage[normalem]{ulem}
\newcommand{\PrePrintNumbersname}{{\bfseries Preprint numbers}}
\journalname{Eur. Phys. J. C}
\begin{document}

\title{The impact of top-quark modelling on the exclusion limits in $\boldsymbol{t\bar{t}}+\text{DM}$ searches at the LHC
}


\author{J. Hermann\thanksref{e1,addr1}
        \and
        M. Worek \thanksref{e2,addr1} 
}

\thankstext{e1}{e-mail: jonathan.hermann@rwth-aachen.de}
\thankstext{e2}{e-mail: worek@physik.rwth-aachen.de}


\institute{Institute for Theoretical Particle Physics and Cosmology, RWTH Aachen University, D-52056 Aachen, Germany \label{addr1}
}

\date{Received: date / Accepted: date}

\maketitle

\begin{abstract}
New Physics searches at the LHC rely very heavily on the precision and accuracy of Standard Model background predictions. 
Applying the spin-0 $s$-channel mediator model, we assess the importance of properly modelling such backgrounds in $t\bar{t}$ associated Dark Matter production. Specifically, we discuss higher-order corrections and off-shell effects for the two dominant background processes $t\bar{t}$ and $t\bar{t}Z$  in the presence of extremely exclusive cuts.
Exclusion limits are calculated for state-of-the-art NLO full off-shell $t\bar{t}$ and $t\bar{t}Z$ predictions and compared to those computed with backgrounds in the NWA and / or at LO. 
We perform the same comparison for several new-physics sensitive observables and evaluate which of them are affected by the top-quark modelling. Additionally, we make suggestions as to which observables should be used to obtain the most stringent limits assuming integrated luminosities of $300$ fb$^{-1}$ and $3000$ fb$^{-1}$.
\noindent \PrePrintNumbersname \enspace TTK-21-29 $\,\cdot\,$ P3H-21-053
\end{abstract}

\section{Introduction} \label{intro}
Even though most of our knowledge of Dark Matter (DM) stems from astrophysical observations, DM  sear-ches at the Large Hadron Collider (LHC) \citep{LHC} play a key role in finding DM particles, or failing that, constraining their properties. 
Both \textsc{CMS} \citep{CMS} and \textsc{ATLAS} \citep{ATLAS} are well suited for detecting the expected missing transverse energy signatures and many analyses with various visible final states have been undertaken by both collaborations \citep{CMS_EFT_1, ATLAS_EFT_1,CMS_simp_DM_1,CMS_simp_DM_2,CMS_simp_DM_3,CMS_simp_DM_4,ATLAS_simp_DM_1,ATLAS_simp_DM_2,ATLAS_simp_DM_3,ATLAS_excl_limits,CMS_excl_limits}. 
So far, the aim of detecting DM has proven to be an elusive goal but even the fact that it has not been detected yet can give us constraints on the properties of potential DM particles.
Naturally, such limits depend heavily on the considered DM model of which there are plenty to choose from. 
The most general approach is to use Effective Field Theories (EFTs), see e.g. Refs. \citep{CMS_EFT_1, ATLAS_EFT_1}, but over the last few years it has become increasingly popular to use so-called simplified models \citep{DM_simp_1,DM_simp_2,DM_simp_3,DM_simp_4} to interpret the data, see e.g. Refs. \citep{CMS_simp_DM_1,CMS_simp_DM_2,CMS_simp_DM_3,CMS_simp_DM_4,ATLAS_simp_DM_1,ATLAS_simp_DM_2,ATLAS_simp_DM_3,ATLAS_excl_limits,CMS_excl_limits}. 
\begin{figure*}
	\centering
	\includegraphics[width=\linewidth]{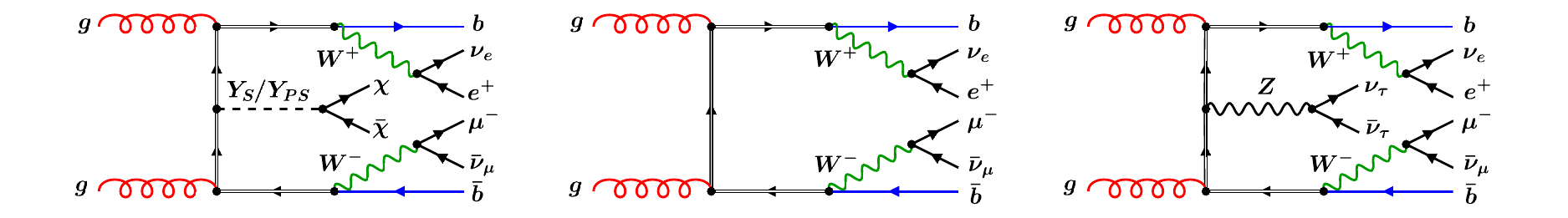}
	\caption{\textit{Leading order Feynman diagrams for the signal process (left) as well as for the two dominant background processes, $t\bar{t}$ (center) and $t\bar{t}Z$ (right).}}
	\label{fig:All_diagrams}
\end{figure*}

In this analysis, we employ  the latter in form of the simplified spin-0 $s$-channel mediator model.
This model extends the Standard Model (SM) by a fermionic DM particle $\chi$ and a scalar or pseudoscalar mediator $Y$ that couples the SM to the dark sector. 
Apart from the masses $m_\chi$ and $m_Y$ of the two new particles, the model is only characterised by the mediator-DM and mediator-quark couplings. 
In principle, the latter could adhere to any hierarchy but flavour measurements suggest that the only sources of flavour symmetry breaking are the quark masses, just like in the SM Yukawa couplings. 
In order to fulfil these requirements, one typically postulates Minimal Flavour Violation (MFV) \citep{MFV} which implies that the mediator-quark couplings should be proportional to the quark masses. 
As a result, we are left with four independent parameters to describe our model, the two masses, the SM-mediator coupling $g_\chi$ and a flavour universal mediator-quark coupling $g_q$. 

Since the mediator-quark couplings have the same hierarchy as the quark masses, the mediator will primarily couple to top quarks. 
So, just like for the SM Higgs, the main production modes are top-loop induced gluon fusion or top-quark pair associated production.
As the former leads to rather complicated $\text{jets} \,+ $ missing transverse momentum signatures due to overwhelming QCD backgrounds, we concentrate on $t\bar{t}+\text{DM}$ signals and focus on leptonic decay modes of the top quarks. 
In addition to the DM particle pair, we find two $b$-jets, two oppositely charged leptons as well as their corresponding (anti-)neutrinos in the final state. Hence, we are considering signal processes of the form $p p \to b \bar{b} l^- l'^+ + p_{T,\text{miss}}$ where $l$ and $l'$ are either electrons or muons since $\tau$ leptons decay further and are thus studied separately. The missing transverse momentum $p_{T,\text{miss}}$ encompasses the momenta of the invisible particles, i.e. the neutrinos and DM particles.

The leptonic channel is not only experimentally clean but it also gives us access to leptonic variables such as $\cos ( \theta^*_{ll}) = \tanh(|\eta_{l_1} - \eta_{l_2}|/2)$ \citep{cos_intro} and $\Delta \phi_{l,\text{miss}}$.
As the flight directions of top quarks and leptons are heavily correlated, these observables provide us with indirect information on the corresponding top-quark distributions.
The former of the two observables has also been shown to be a promising observable in $t\bar{t}+\text{DM}$ searches, both in separating the signal from the background and 
in differentiating scalar and pseudoscalar mediator models \citep{cos_intro,Haisch_analysis}.
Apart from these two angular observables, there are many more that are sensitive to new physics (NP) and in particular DM signatures. 
The most obvious one is the missing transverse momentum $p_{T,\text{miss}}$ as this is the primary observable to which DM particles would contribute directly. 
Other prominent variables include the stransverse masses $M_{T2,t}$ and $M_{T2,W}$ \citep{MT2_1,MT2_2,MT2_3}. These are generalizations of the transverse mass of either the top quark or the $W$-boson if these occur in pairs. 

\sloppy By making use of the differences in distribution shapes between the DM signal and the SM background we can separate the two through event selection cuts. 
However, the effectiveness of such cuts varies considerably depending on the background process. 
For our $p p \to b \bar{b} l^- l'^+ + p_{T,\text{miss}}$ signature, we can classify the SM background into three categories, see e.g. Ref. \citep{Haisch_analysis}: top-quark ($t\bar{t}$, $tW$), reducible ($WW$, $ZZ$, $WZ$, $Z+\text{jets}$) and irreducible backgrounds ($t\bar{t}Z$, $t\bar{t}W$). 
As the name suggests, the reducible backgrounds can be eliminated rather easily and simply requiring two $b$-jets, two leptons and a large $p_{T,\text{miss}}$ is enough to do so. 
The same is true for the $t\bar{t}W$ process as there will be too many light-jets or leptons. Furthermore, at leading order (LO) in QCD, $t\bar{t}W$ can only occur via $q\bar{q}^\prime$ annihilation. This is different to $t\bar{t}Z$ where gluon-gluon fusion production is accessible already at LO. Consequently, the contribution of $t\bar{t}W$ to the background process is suppressed with respect to  $t\bar{t}Z$. The only LO processes with exactly the same final state are $t\bar{t}$ and $t\bar{t}Z$ with $Z$ decaying into a neutrino pair.
Exemplary Feynman diagrams of the two background processes as well as for the signal process are depicted in Figure \ref{fig:All_diagrams}. All diagrams have been created with the help of \textsc{FeynGame} \citep{FeynGame}.

At next-to-leading order (NLO) in QCD, $tW$ has the same final state as $t\bar{t}$ at LO but since $t\bar{t}$ and $tW$ have interference effects at NLO, the latter is automatically included in the $t\bar{t}$ predictions if we consider full off-shell effects.
On the other hand, $t\bar{t}Z$ is classified as an irreducible background as it mimics the signal's structure quite closely which makes it rather hard to suppress through selection cuts. 
In contrast, the a priori dominant top-quark backgrounds can be reduced significantly due to the generally smaller $p_{T,\text{miss}}$ as well as a kinematic edge in $M_{T2,W}$. 

In both cases, precise predictions and a proper modelling of unstable particles play a vital role as the shapes of the above mentioned distributions are very sensitive to higher-order corrections as well as the modelling of top quarks and vector bosons. 
The most complete way of treating unstable particles is the full off-shell description, i.e. describing their propagators through Breit-Wigner distributions and considering all Feynman  diagrams of the same perturbative order with the same final state, irrespective of the number of top$-$, $W-$, and $Z$-resonances. NLO QCD corrections in the full off-shell treatment have already been calculated several years ago for $t\bar{t}$ \citep{tt_NLO_1,tt_NLO_2,tt_NLO_3,tt_NLO_new,tt_NLO_4,tt_NLO_5} but for the more complicated $t\bar{t}Z$ process they have been computed for the first time rather recently in Ref. \citep{ttZ}.
These full off-shell calculations can become very involved for processes with many final state particles, especially at higher-orders in perturbation theory. 
A common simplification known as the narrow-width approximation (NWA) can be used instead. The latter not only sets resonant particles on-shell but also discards all singly- and non-resonant Feynman diagrams which simplifies the calculation considerably. 
In the case of $t\bar{t}$, this has even enabled the calculation of next-to-next-to leading order (NNLO) corrections\footnote{If not stated otherwise, NLO and NNLO always refer to higher-order corrections in QCD.} \citep{tt_NNLO_1,tt_NNLO_2,tt_NNLO_3}. 

Both the order in perturbation theory and the treatment of unstable particles that one uses for the calculation can have profound implications for the size and shape of the background. 
Thus, one of the primary goals of this paper is to quantify off-shell effects and higher-order corrections in NP-sensitive observables. 
As these build the foundation for any Beyond the Standard Model (BSM) analysis, we want to further evaluate the impact of these changes in light of a typical search for a $t\bar{t}+\text{DM}$ signature. 
To this end, we compare the different distributions after applying very exclusive selection cuts that are designed to disentangle signal and background. 
These cuts are based on the analysis presented in Ref. \citep{Haisch_analysis}. We then use the resulting distributions to calculate exclusion limits for the signal strength depending on the mediator mass. 
For this, we employ both dimensionless and dimensionful observables and assess which of these yields the most stringent exclusion limits. But here, too, our focus will be on the background modelling and the ramifications of using an inadequate description.

Let us mention that Ref. \citep{Haisch_analysis} presents an analysis that is admittedly closer to experiment as it also incorporates the above described reducible and $t\bar{t}W$ backgrounds, parton shower as well as very roughly estimated detector effects. However, the dominant $t\bar{t}Z$ background is only modelled at LO (with NLO normalisation\footnote{The normalisation is computed with \textsc{MadGraph5\_aMC@NLO} \citep{MadGraph} which only computes the $t\bar{t}Z$ production at NLO while the decays are modeled at LO.}) and scale uncertainties are only taken into account as flat percentages in combination with the detector uncertainties. In our analysis, we try to mitigate the last two points. Having said this, the main goal of this paper is not to give accurate limits but rather to assess how changes in modelling the background can effect these exclusion limits and to sensitise the reader to these effects. For this, we essentially assume perfect detector performance.

The most stringent experimental limits on the considered DM model are currently provided by the \textsc{CMS} collaboration \citep{CMS_excl_limits}. In their analysis, $t+\text{DM}$ signatures are also considered in addition to the $t\bar{t}+\text{DM}$ signal we are analysing here. To calculate the exclusion limits, the $p_{T,\text{miss}}$ distribution is used. For a signal strength of $\mu = 1$, scalar (pseudoscalar) mediator masses up to $290$ ($300$) GeV can be excluded with a confidence level of $95 \%$ assuming a DM mass of $m_\chi = 1$ GeV and couplings of $g_q = g_\chi = 1$. The limits provided by the \textsc{ATLAS} collaboration are comparable at $250$ ($300$) GeV \citep{ATLAS_excl_limits}.

This paper is structured as follows. In Section \ref{sec:Signal} we discuss the applied DM model in more detail and describe the typical behavior of NP-sensitive observables depending on the mass of the mediator. We then compare these to the behavior of the SM background processes $t\bar{t}$ and $t\bar{t}Z$ in Section \ref{sec:Background}. Both higher-order and off-shell corrections to the background will be assessed with special emphasis on the phase-space regions where DM signatures might appear. In Section \ref{sec:Cut_effects} we outline the selection cuts and discuss the effects that these have on signal and background cross sections and distributions. We also study whether the cuts have any effect on the size of the corrections. These results are then used in Section \ref{sec:Limits} to compute signal strength exclusion limits depending on the mediator mass. We discuss the effects of different background modelling approaches, central scale choices and integrated luminosities and make suggestions as to which observables should be used to obtain the most stringent limits. Finally, in Section \ref{sec:Summary} we recapitulate our main results.

\section{The Dark Matter signal} \label{sec:Signal}

\subsection{The Dark Matter Model} \label{subsec:Model}
As we already mentioned before, we use the simplified spin-0 $s$-channel mediator model which consists of a fermionic DM particle $\chi$ and a mediator $Y$. As a spin-0 particle, the mediator can either be a scalar (S) or a pseudoscalar (PS) particle which we will denote as $Y_S$ and $Y_{PS}$ in the following.
As suggested in Ref. \citep{MFV}, MFV implies that the mediator-quark couplings are proportional to the SM Yukawa couplings $y_q = \sqrt{2} m_q / v$ where $v$ is the vacuum expectation value of the Higgs boson. With this the interaction Lagrangian of the mediator takes the form
\begin{equation} \label{eq:L_S}
\mathcal{L}_{\text{S}} \supset - g_{\chi} Y_{\text{S}} \bar{\chi} \chi - \frac{1}{\sqrt{2}} Y_{\text{S}} \sum_{q=u,d,c,s,t,b} g_q y_q \bar{q} q,
\end{equation}
and
\begin{equation} \label{eq:L_PS}
\mathcal{L}_{\text{PS}} \supset - i g_{\chi} Y_{\text{PS}} \bar{\chi} \gamma_5 \chi -  \frac{i}{\sqrt{2}} Y_{\text{PS}} \sum_{q=u,d,c,s,t,b} g_q y_q \bar{q} \gamma_5 q
\end{equation}
for scalar and pseudoscalar mediators, respectively. Following the recommendations of Ref. \citep{DM_simp_1}, we take $g_q = g_\chi = 1$ for the couplings. The mass of the fermionic DM particle $\chi$ is fixed at $m_\chi = 1$ GeV while the mediator mass $m_Y$ is varied between $10$ GeV and $1$ TeV.

\subsection{Signal and background processes} \label{subsec:Processes}
\begin{figure*}
		\centering
		\includegraphics[width=\linewidth]{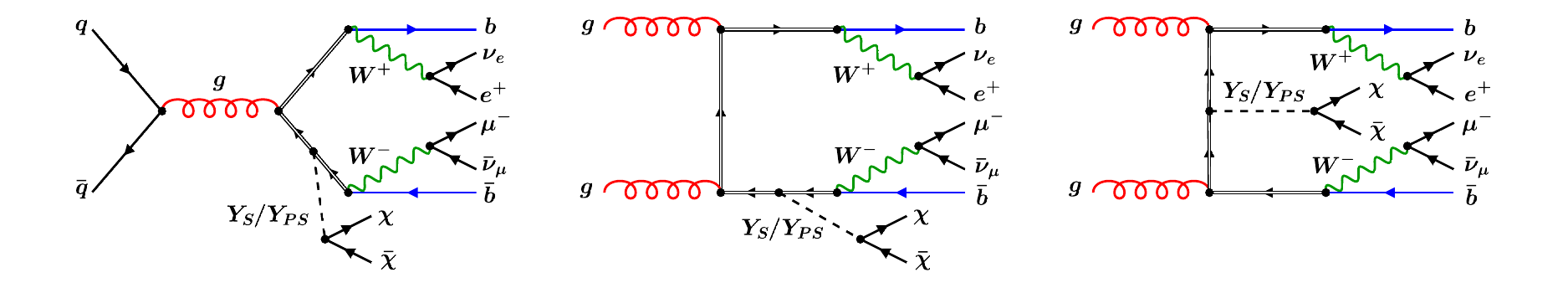}
	\caption{\textit{Leading order Feynman diagrams for the DM particle pair production via a scalar $Y_{\text{S}}$ or a pseudoscalar $Y_{\text{PS}}$ mediator in association with a top-quark pair.}}
	\label{fig:DM_diagrams}
\end{figure*}

Due to the Yukawa couplings appearing in Eqs. (\ref{eq:L_S}) and (\ref{eq:L_PS}), the mediator is primarily produced in association with top quarks. 
In this analysis, we specifically look at $t\bar{t} + Y$ production. 
Exemplary Feynman diagrams are shown in Figure \ref{fig:DM_diagrams}. 
The mediator is either radiated off one of the top quarks (left and central diagrams) or produced via top-quark fusion (right) and subsequently decays into the $\chi \bar{\chi}$ pair. 
For the top quarks we consider their leptonic decay modes so that we have $p p \to b \bar{b} l^+ l'^- \nu_l \bar{\nu}_{l'} \chi \bar{\chi}$ as our signal process where $l$ and $l'$ are either electrons or muons.
As the DM particles only appear in the form of additional missing transverse momentum $p_{T,\text{miss}}$, we have several SM processes with the same visible final state, most notably $t\bar{t}$ and $t\bar{t}Z$ production with $Z$ decaying into a $\nu\bar{\nu}$ pair.
More specifically, we consider $p p \to b \bar{b} e^+ \mu^- \nu_e \bar{\nu}_{\mu}$
and $p p \to b \bar{b} e^+ \mu^- \nu_e \bar{\nu}_{\mu} \nu_{\tau} \bar{\nu}_{\tau}$ production. As interference effects from $\gamma,Z \to l^+ l^-$ splitting are at the per-mille level \citep{ttZ}, we can get the full contributions by multiplying the results by $4$ for $t\bar{t}$ and the DM signal and by $12$ for $t\bar{t}Z$ .
If not stated otherwise, all results apart from the exclusion limits are presented without these lepton flavour factors.

\subsection{Basic setup} \label{subsec:Setup}
Before we can present any predictions for either the signal or the background, we must first discuss the setup.
As we use the NLO off-shell $t\bar{t}Z$ samples generated for Ref. \citep{ttZ}, we assume the same basic setup as presented there. 
Hence, we show all results for the LHC Run II center of mass energy of $\sqrt{s} = 13$ TeV. For the parameters describing the gauge bosons we use the $G_\mu$ scheme and fix the Fermi-constant to 
\begin{equation}
G_{\mu} = 1.166378 \times 10^{-5} \,\text{GeV}^{-2}
\end{equation}
and the masses of the massive gauge bosons to
\begin{equation}
m_W = 80.385 \,\text{GeV}, \;\;\;\;\;  m_Z = 91.1876 \,\text{GeV}.
\end{equation}
These then determine the the electroweak coupling $\alpha$ and mixing angle $\theta_W$:
\begin{equation} \label{eq:EW_coupling_and_sin_EW}
\alpha = \frac{\sqrt{2}}{\pi} G_{\mu} m_W^2 \left( 1-\frac{m_W^2}{m_Z^2}\right), \;\;\;\;\; \sin^2 (\theta_W) = 1-\frac{m_W^2}{m_Z^2}.
\end{equation}
For the gauge boson widths we take their NLO QCD values 
\begin{equation}
\Gamma_W = 2.0988\, \text{GeV}, \;\;\;\;\; \Gamma_Z = 2.50782\, \text{GeV}.
\end{equation}
The only other massive SM particle is the top quark for which we use $m_t = 173.2$ GeV\footnote{Note that we use lowercase $m$ for the input mass parameter and uppercase $M$ for the invariant mass to differentiate between the two.}. The top-quark width can then be calculated from the above parameters (see Ref. \citep{Denner_Dittmaier}) which results in 
\begin{equation} \label{eq:Gamma_t_Offshell}
\begin{split}
\Gamma_{t,\text{ Off-shell}}^{\text{LO}} &= 1.47848\, \text{GeV}, \\ \Gamma_{t,\text{ Off-shell}}^{\text{NLO}} &= 1.35159\, \text{GeV}
\end{split}
\end{equation}
in the full off-shell case and 
\begin{equation} \label{eq:Gamma_t_NWA}
\begin{split}
\Gamma_{t,\text{ NWA}}^{\text{LO}} &= 1.50176\, \text{GeV}, \\ \Gamma_{t,\text{ NWA}}^{\text{NLO}} &= 1.37279\, \text{GeV}
\end{split}
\end{equation}
in the NWA. 
All leptons as well as the remaining quarks are treated as massless particles. As this includes the $b$-quark, no Higgs boson diagrams contribute at LO. Due to their negligable contribution, we do not  take into account any loop diagrams that involve the Higgs boson for the higher-order calculations.
Additionally, setting $m_b$ to zero also entails that we must employ the $N_F = 5$ flavour scheme. The running of $\alpha_s$ at NLO (LO)  is provided with two loop (one loop) accuracy  by the \textsc{LHAPDF} interface \citep{LHAPDF}.

However, as the $b \bar{b}$ and $\bar{b} b$ initial state contributions are at the per-mill level and thus well within theory uncertainties, they are neglected throughout this analysis.
Let us also mention that we keep the Cabibbo-Kobayashi-Maskawa (CKM)-matrix diagonal so that at LO we only consider the subprocesses
\begin{align}
\begin{split}
g g &\rightarrow b \bar{b} \mu^- \bar{\nu}_{\mu} e^+ \nu_e \,(+\chi \bar{\chi} / +\nu_\tau \bar{\nu}_\tau) \\
q \bar{q} / \bar{q} q &\rightarrow b \bar{b} \mu^- \bar{\nu}_{\mu} e^+ \nu_e \,(+\chi \bar{\chi} / +\nu_\tau \bar{\nu}_\tau) \\
\end{split}
\end{align}
where $q = u,d,c,s$ and $+\chi \bar{\chi}$ and $+\nu_\tau \bar{\nu}_\tau$ simply indicate the additional final state particles occurring in $t\bar{t}Y$ and $t\bar{t}Z$ production. We should emphasise that in the full off-shell case we take into account any Feynman diagram of the order $\mathcal{O}(\alpha_s^2 \alpha^4)$ for $t\bar{t}$ and  $\mathcal{O}(\alpha_s^2 \alpha^6)$ for $t\bar{t}Z$ at LO.

At NLO we must also take into account the real radiation processes
\begin{align}
\begin{split}
g g &\rightarrow b \bar{b} \mu^- \bar{\nu}_{\mu} e^+ \nu_e g\,(+\chi \bar{\chi} / +\nu_\tau \bar{\nu}_\tau) \\
q \bar{q} / \bar{q} q &\rightarrow b \bar{b} \mu^- \bar{\nu}_{\mu} e^+ \nu_e g\,(+\chi \bar{\chi} / +\nu_\tau \bar{\nu}_\tau) \\
g q / q g &\rightarrow b \bar{b} \mu^- \bar{\nu}_{\mu} e^+ \nu_e q\,(+\chi \bar{\chi} / +\nu_\tau \bar{\nu}_\tau) \\
g \bar{q} / \bar{q} g &\rightarrow b \bar{b} \mu^- \bar{\nu}_{\mu} e^+ \nu_e \bar{q}\,(+\chi \bar{\chi} / +\nu_\tau \bar{\nu}_\tau)
\end{split}
\end{align}
in addition to those listed above.
In order to reduce the calculation time, we use PDF summation for the up-type ($u+c$) and down-type ($d+s$) quarks for the background processes. For the PDFs themselves we use the LO and NLO \textsc{CT14} \citep{CT14_PDF} PDF sets. They are obtained with $\alpha_s(m_Z) = 0.130$ at LO and  $\alpha_s(m_Z) = 0.118$ at NLO, respectively. The PDF uncertainties are calculated using the prescription outlined by the \textsc{CTEQ} group and are provided at $68 \%$ confidence level (CL). In practice, this means that we must re-scale the uncertainties by $1/1.645$ as they are originally given at $90 \%$ CL.
 
Like in any other fixed-order calculation, our results depend on the choice of the factorisation scale $\mu_F$ and the renormalisation scale $\mu_R$. In this analysis, we use three different types of scales: fixed scales, dynamical scales depending on the final state particles, and dynamical scales depending on the intermediate $t \bar{t} (Z / Y)$ particles. 
All of them are summarised in Table \ref{table:Central_scale} where we define 
\begin{equation} \label{eq:scale_def}
\begin{split}
E_T &= \sum_{i=t,\bar{t}(,Z,Y)} \sqrt{M_i^2 + p_{T,i}^2} \;\;\;\text{ and} \\
H_T &= p_{T,b} + p_{T,\bar{b}} + p_{T,\mu^-} + p_{T,e^+} + p_{T,\text{miss}}.
\end{split}
\end{equation}

For $E_T$ we use Monte-Carlo (MC) truth to reconstruct the four momenta of the intermediate particles. For example, we define the top-quark momentum as follows: $p_t = p_b + p_{e^+} + p_{\nu_e}$. In the off-shell case, we use the same procedure irrespective of whether the resonances actually occur. 
Note that we cannot define $H_T$ for the signal process as the calculation is split into the production and the decay in \textsc{MadGraph5\_aMC@NLO} \citep{MadGraph}, which we use to generate the signal. Thus, only the four-momenta of the top quarks and the mediator are known at the production stage. 
\begin{table}
	\caption{\textit{Summary of central scale settings for the three considered processes.}}
	\label{table:Central_scale}       
	\centering
	\begin{tabular}{l@{\hskip 10mm}lll}
		\hline\noalign{\smallskip}
		 Scale Setting & $\mu^{\text{DM}}_0$ & $\mu^{t\bar{t}}_0$ & $\mu^{t\bar{t}Z}_0$  \\
		\noalign{\smallskip}\midrule[0.5mm]\noalign{\smallskip}
		fixed & $m_t + m_Y/2$ & $m_t$ & $m_t + m_Z/2$ \\
		$E_T$ & $E_T/3$ & $E_T/4$ & $E_T/3$ \\
		$H_T$ & - & $H_T/4$ & $H_T/3$ \\
		\noalign{\smallskip}\hline\noalign{\smallskip}
		Default & $E_T/3$ & $H_T/4$ & $H_T/3$ \\
		\noalign{\smallskip}\hline
	\end{tabular}
\end{table}
If not stated otherwise, we use the $H_T$ scales for the background and $E_T$ for the signal as our central scale for both $\mu_R$ and $\mu_F$. The theoretical uncertainties associated with neglected higher-order terms in the perturbative expansion are estimated by varying the renormalisation
and factorisation scales in $\alpha_s$ and the PDFs by a factor of $2$ around $\mu_0$. Even though we set $\mu_0=\mu_R=\mu_F$,  we vary the two scales independently in the off-shell case. Specifically, 
we use the seven-point scale variation where we recalculate the cross sections for the following scale settings
\begin{equation} \label{eq:7point_scale}
\begin{split}
\left(\frac{\mu_R}{\mu_0}, \frac{\mu_F}{\mu_0} \right) = &\left(0.5, 0.5\right), \,\left(1, 0.5\right), \,\left(0.5,1\right), \,\left(1,1\right), \\ &\left(2,1\right), \,\left(1,2\right), \,\left(2,2\right)
\end{split}
\end{equation}
and take the envelope of the obtained results. For histograms this is done on a bin-by-bin basis.
In the NWA, \textsc{Helac-NLO} \citep{HELAC_NLO, HELAC_NWA}, which we employ to generate $t\bar{t}Z$ and $t\bar{t}$, requires $\mu_F$ and $\mu_R$ to be varied simultaneously. Therefore, we use the three-point scale variation 
\begin{equation} \label{eq:3point_scale}
\left(\frac{\mu_R}{\mu_0}, \frac{\mu_F}{\mu_0} \right) = \left(0.5, 0.5\right), \,\left(1,1\right), \,\left(2,2\right)
\end{equation}
in this case. We want to add here that the scale variation is driven by the changes in $\mu_R$, see Ref. \citep{ttZ}. Hence, the uncertainties will not change between the three- and seven-point scale variations.
To finalise the setup section, let us mention the cuts on the final state particles. For the two charged leptons we require 
\begin{equation} \label{eq:leptonic_cuts}
p_{T,l} > 30 \,\text{GeV}, \;\;\;\;\; |\eta_l| < 2.5 \;\;\;\;\; \text{and} \;\;\; \Delta R_{ll} > 0.4
\end{equation}
whilst the two $b$-jets should fulfil 
\begin{equation} \label{eq:bjet_cuts}
\begin{split}
p_{T,b} > 40 \,&\text{GeV}, \;\;\;\;\; |\eta_b| < 2.5  \;\;\;\;\; \Delta R_{b\bar{b}} > 0.4   \;\;\; \\ & \text{and} \;\;\; \Delta R_{lb} > 0.4. 
\end{split}
\end{equation}
These $b$-jets are reconstructed using the anti-$k_T$ jet algorithm \citep{anti-kT}  with a resolution parameter of $R = 0.4$ for partons with pseudorapidity $|\eta| < 5$. Since the computations are performed in the five flavour scheme and $b$-quarks are treated as massless, we define the $b$-jet flavour according to the following recombination rules:
\begin{equation}
 b g \to b, \;\;\;\;\; \bar{b} g \to \bar{b} \;\;\;\text{and}\;\;\; b \bar{b} \to g.
\end{equation} 
This ensures that the jet flavour definition is infrared safe at NLO. In addition to the cuts on leptons and $b$-jets, we ask for a missing transverse momentum of at least $p_{T,\text{miss}} > 50$ GeV. No cuts are placed on the potential extra light jet.

\subsection{Signal generation} \label{subsec:DM_gen}
To generate our signal samples, we use \textsc{MadGraph5\_aMC@NLO} \citep{MadGraph} together with the \textsc{DMsimp} \citep{DM_model_Backovic} implementation of the above described DM model for the $p p \to t \bar{t} \chi \bar{\chi}$ production. The $t \bar{t}$-pair is then decayed into the desired $b \bar{b} \mu^- \bar{\nu}_{\mu} e^+ \nu_e$ final state using \textsc{MadSpin} \citep{MadSpin}. This means that we only consider doubly resonant Feynman diagrams, just like in the NWA, but some finite width effects are recovered by introducing Breit-Wigner distributions (up to a cut-off) for the unstable particles. For the cut-off parameter in \textsc{MadGraph} we use $n_{BW} = 16$. Note that the decays can only be done at LO with \textsc{MadSpin} which means that whenever we refer to the NLO DM signal we actually mean NLO production with LO decays. The decay width of the mediator is calculated with \textsc{MadWidth} \citep{MadWidth}. For the top-quark decay width we use the LO off-shell value, as given in Eq. (\ref{eq:Gamma_t_Offshell}), contrary to the default \textsc{MadSpin} setup which uses the value in the NWA.
The clustering of the final state partons is then performed using \textsc{FastJet} \citep{FastJet_manual}. 

\subsection{DM production cross section} \label{subsec:DM_xSec}
\begin{figure}
	\includegraphics[width=1\linewidth]{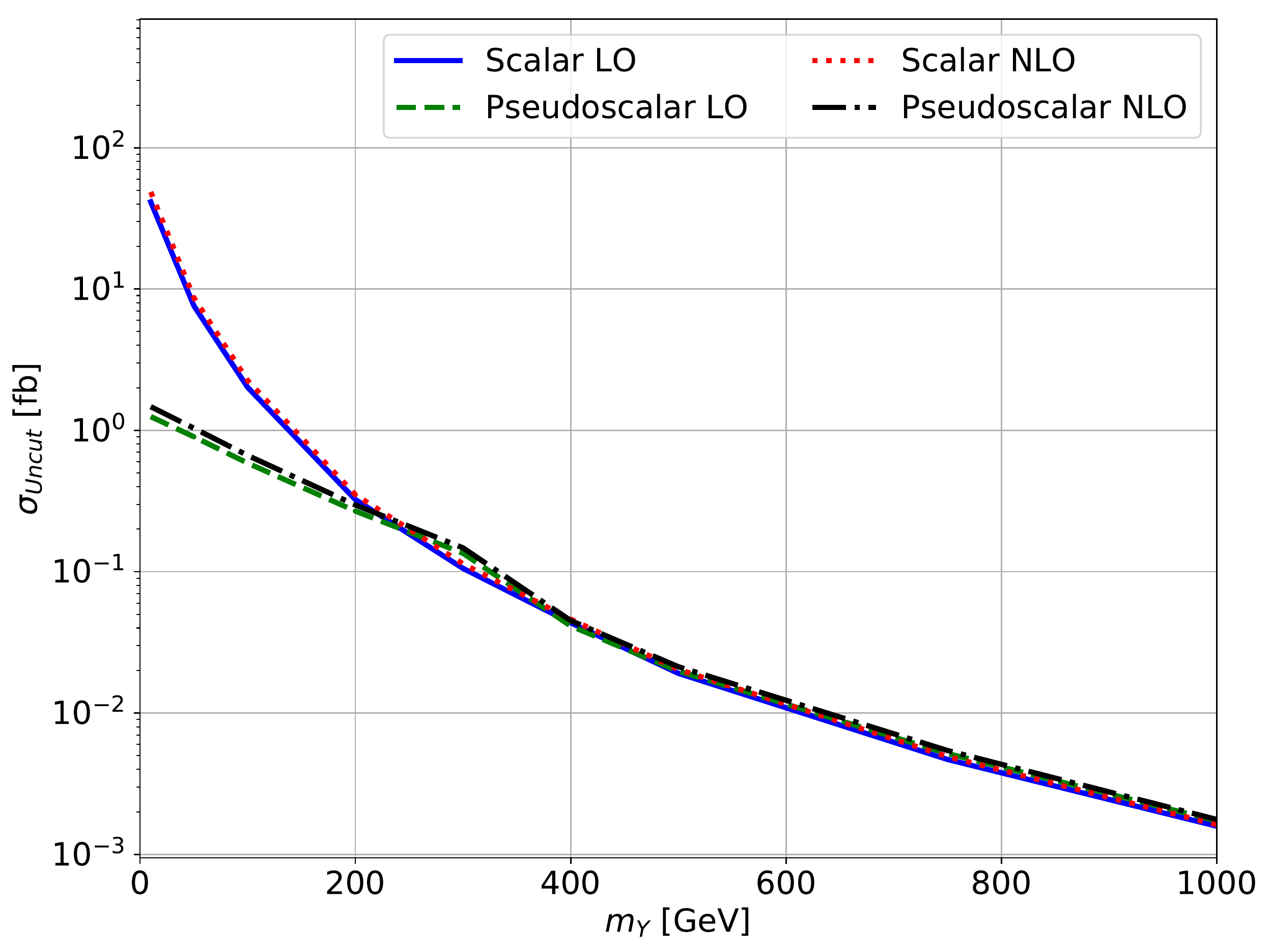}
	
	\caption{\textit{Production cross section for $p p \to b \bar{b} e^+ \mu^- \nu_e \bar{\nu}_\mu \chi \bar{\chi}$ for scalar and pseudoscalar mediators depending on the mass $m_Y$ of the mediator. The results have been generated using \textsc{MadGraph5\_aMC@NLO}  for the LO / NLO production and \textsc{MadSpin} for the (LO) decays with the respective LO and NLO \textsc{CT14} PDF sets and a central scale $\mu_0^{\text{DM}} = E_T/3$ for the LHC with center of mass energy $\sqrt{s} = 13$ TeV.}}
	\label{fig:DM_total_xSec}
\end{figure}

In Figure \ref{fig:DM_total_xSec} we present the integrated cross section for the scalar and pseudoscalar mediator scenarios at LO and NLO depending on the mediator mass $m_Y$. Irrespective of order and parity, the cross sections consistently decrease with increasing mediator mass. Both pseudoscalar curves exhibit the characteristic kink around $m_Y \sim 2 m_t$ \citep{DM_MatrixElement, Haisch_analysis} below which the cross sections in the scalar case far exceed those for the pseudoscalar one. For heavy mediators, on the other hand, the cross sections are largely the same. Higher-order corrections also depend heavily on the mediator mass with $K = \sigma_{NLO} / \sigma_{LO}$-factors ranging from $1.02$ for $m_Y = 1$ TeV to $1.18$  for $m_Y = 10$ GeV for both considered parities.

\subsection{Distribution shapes} \label{subsec:DM_shape}

\begin{figure*}
	\includegraphics[width=0.5\linewidth]{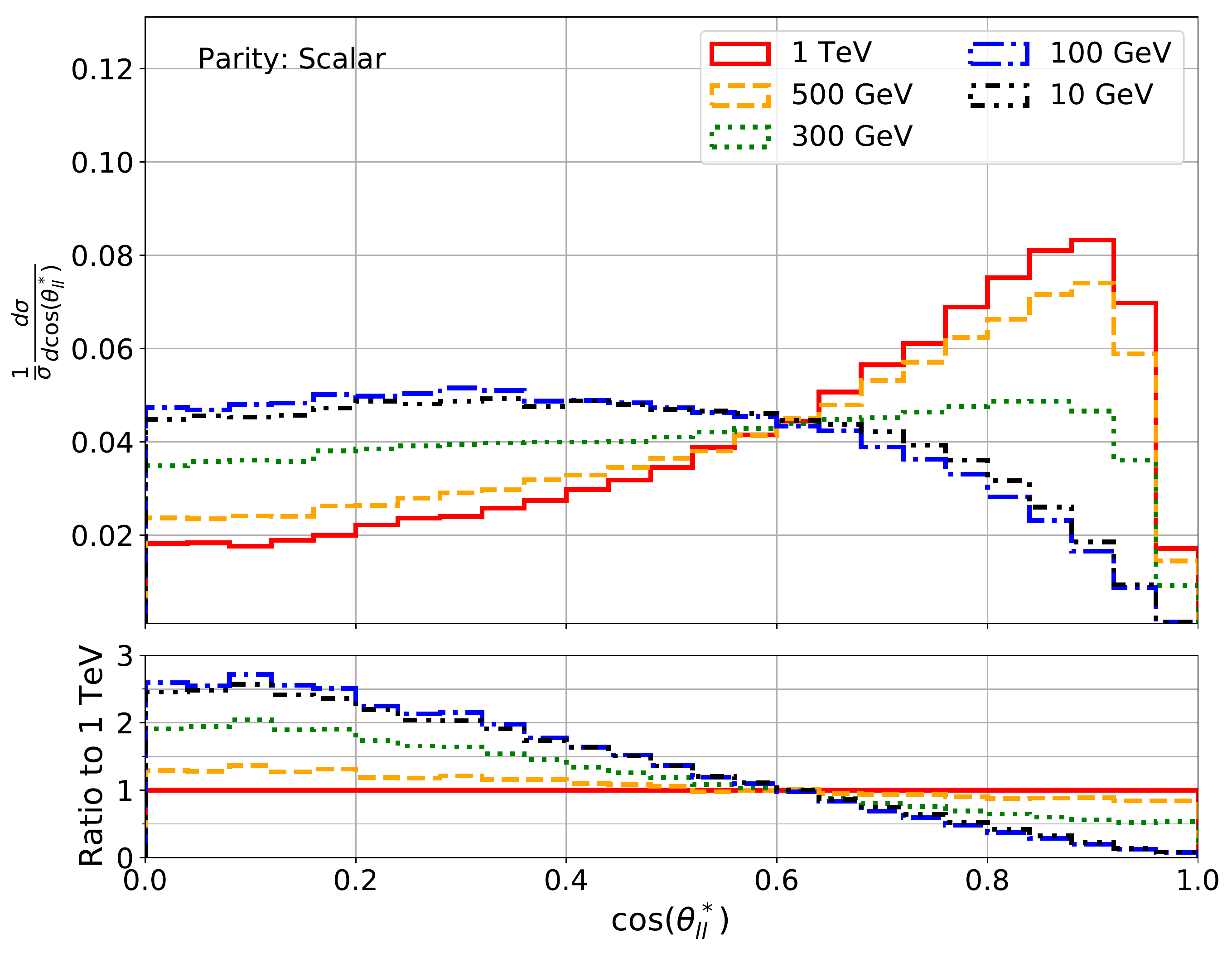}
	\includegraphics[width=0.5\linewidth]{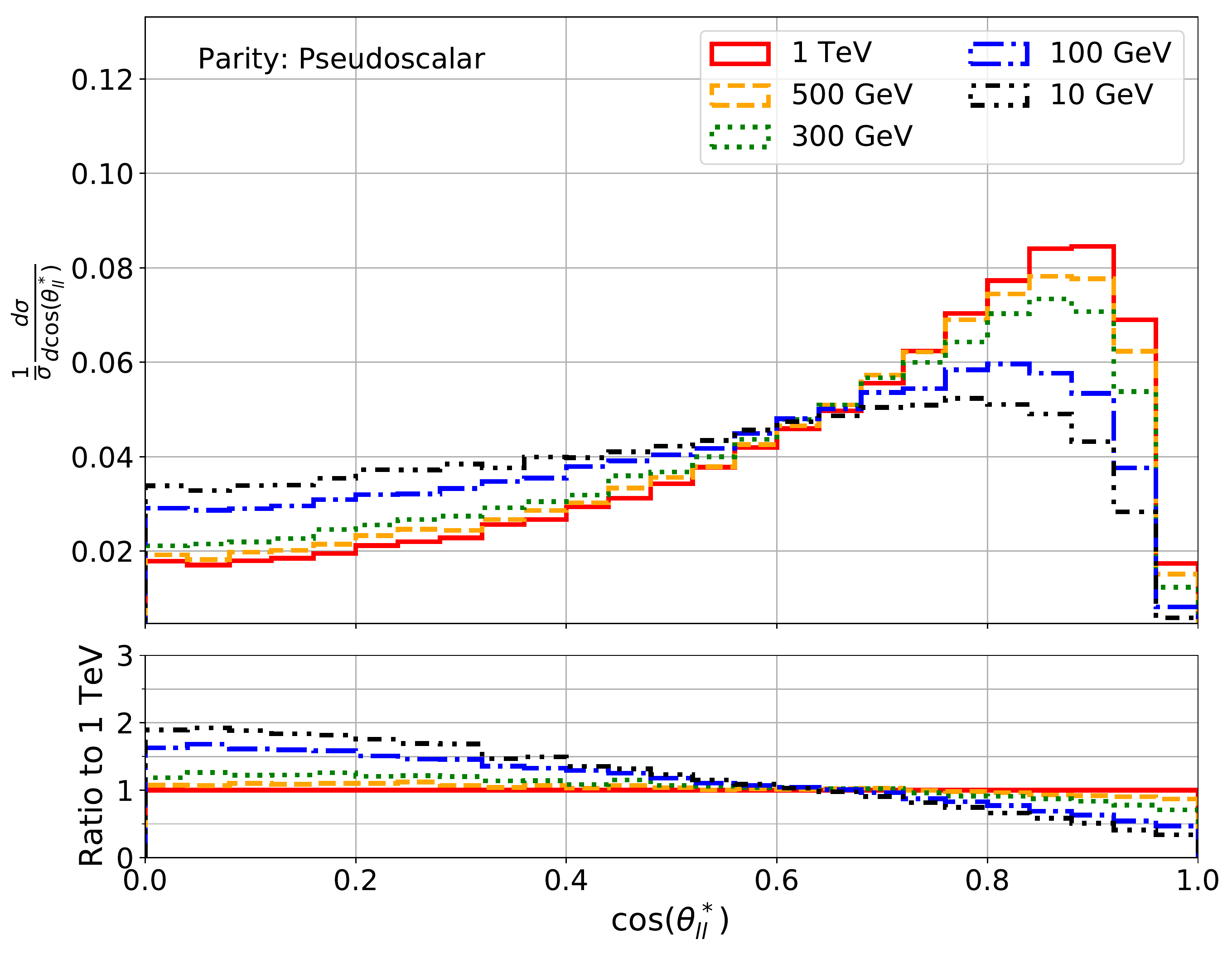}

	\caption{\textit{Comparison of normalised NLO differential $\cos ( \theta^*_{ll} )$ distributions for the scalar (left) and pseudoscalar (right) mediator scenarios for different masses $m_Y$. The samples have been  generated using a central scale of $\mu^{\text{DM}}_0 = E_T/3$ with the NLO \textsc{CT14} PDF set for the LHC with center of mass energy $\sqrt{s} = 13$ TeV. The lower panels depict the ratio to the distributions for $m_Y = 1$ TeV.}}
	\label{fig:DM_Uncut_norm_cosll_NLO}
\end{figure*}

\begin{figure*}
		\includegraphics[width=0.5\linewidth]{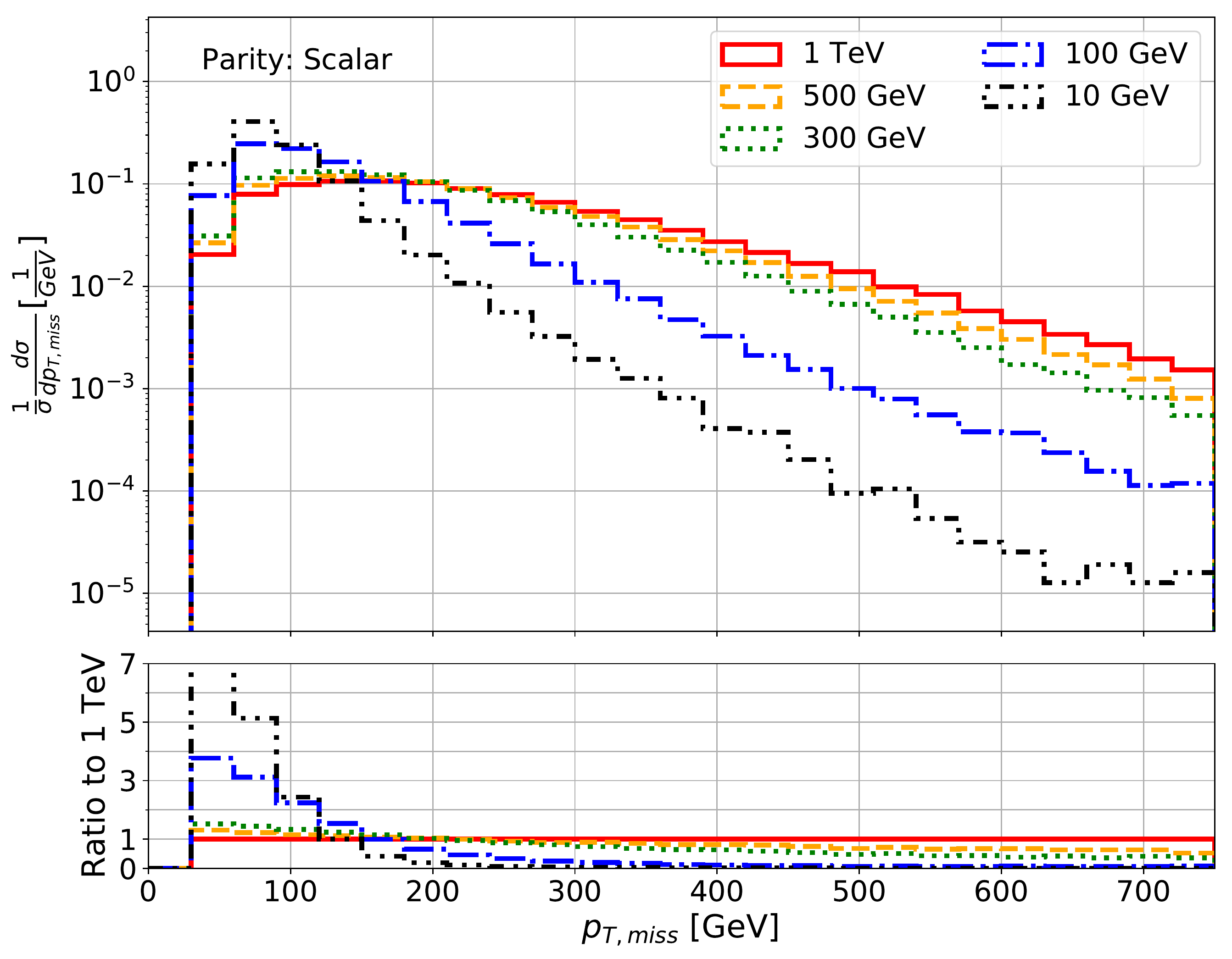}
		\includegraphics[width=0.5\linewidth]{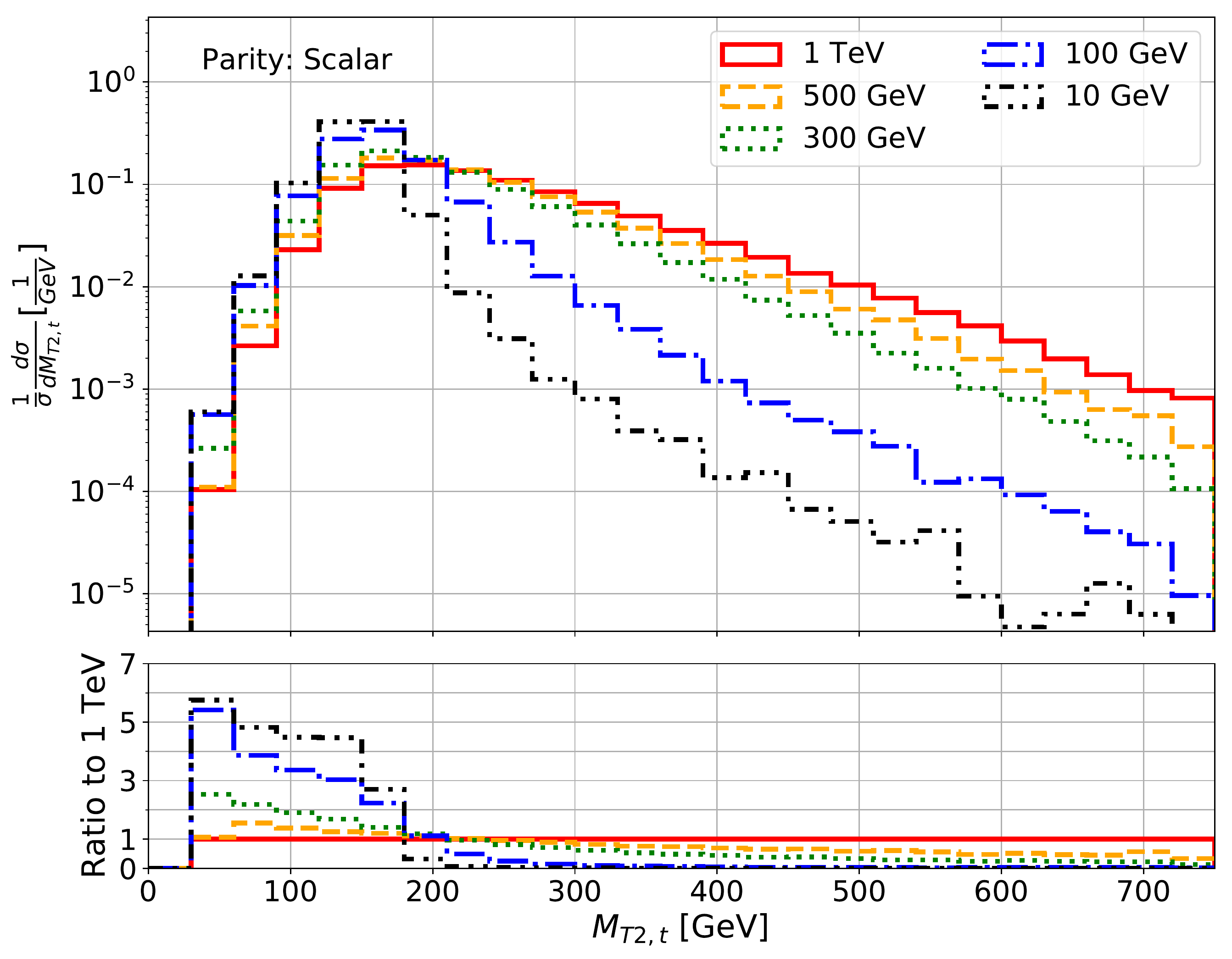}
		\includegraphics[width=0.5\linewidth]{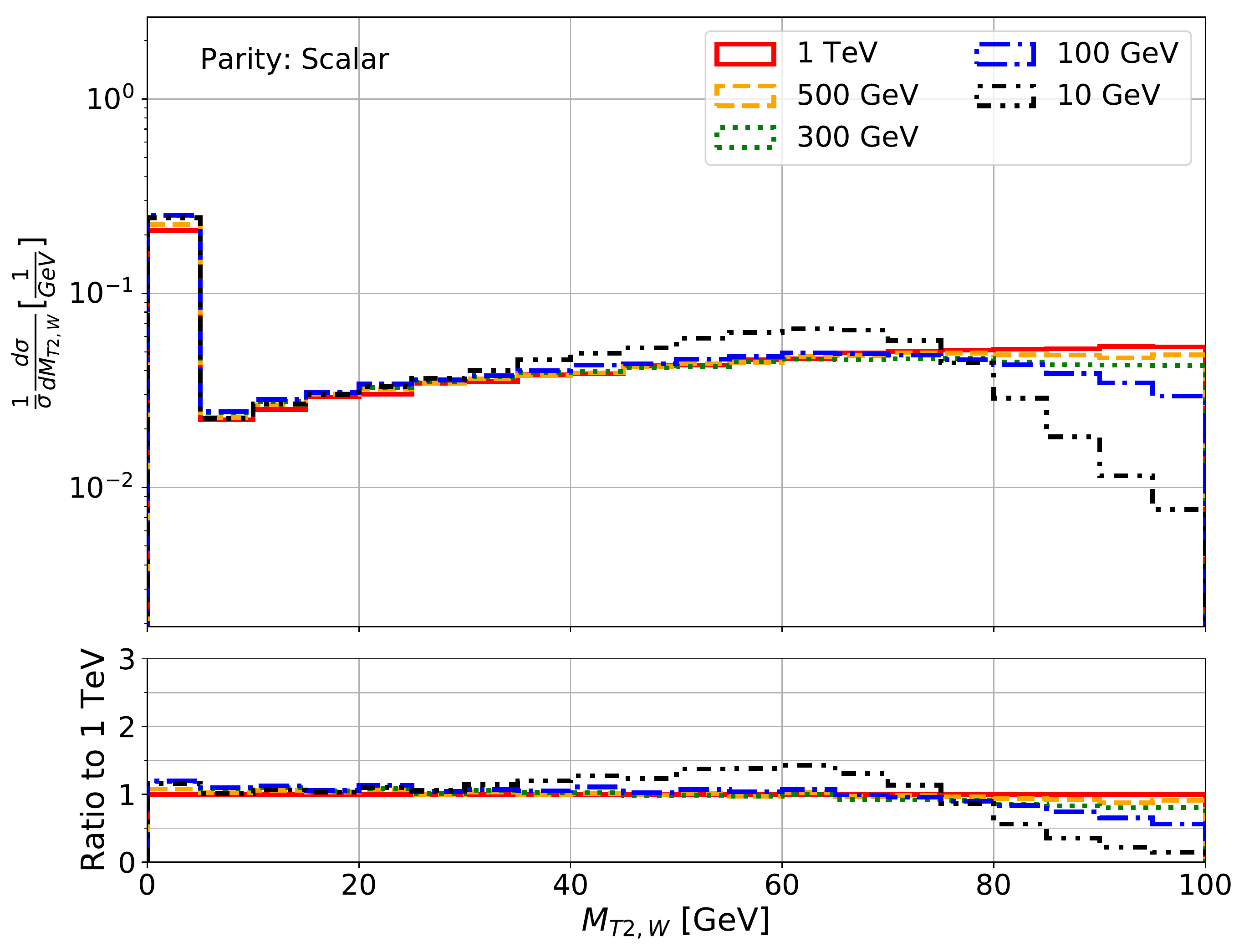}
		\includegraphics[width=0.5\linewidth]{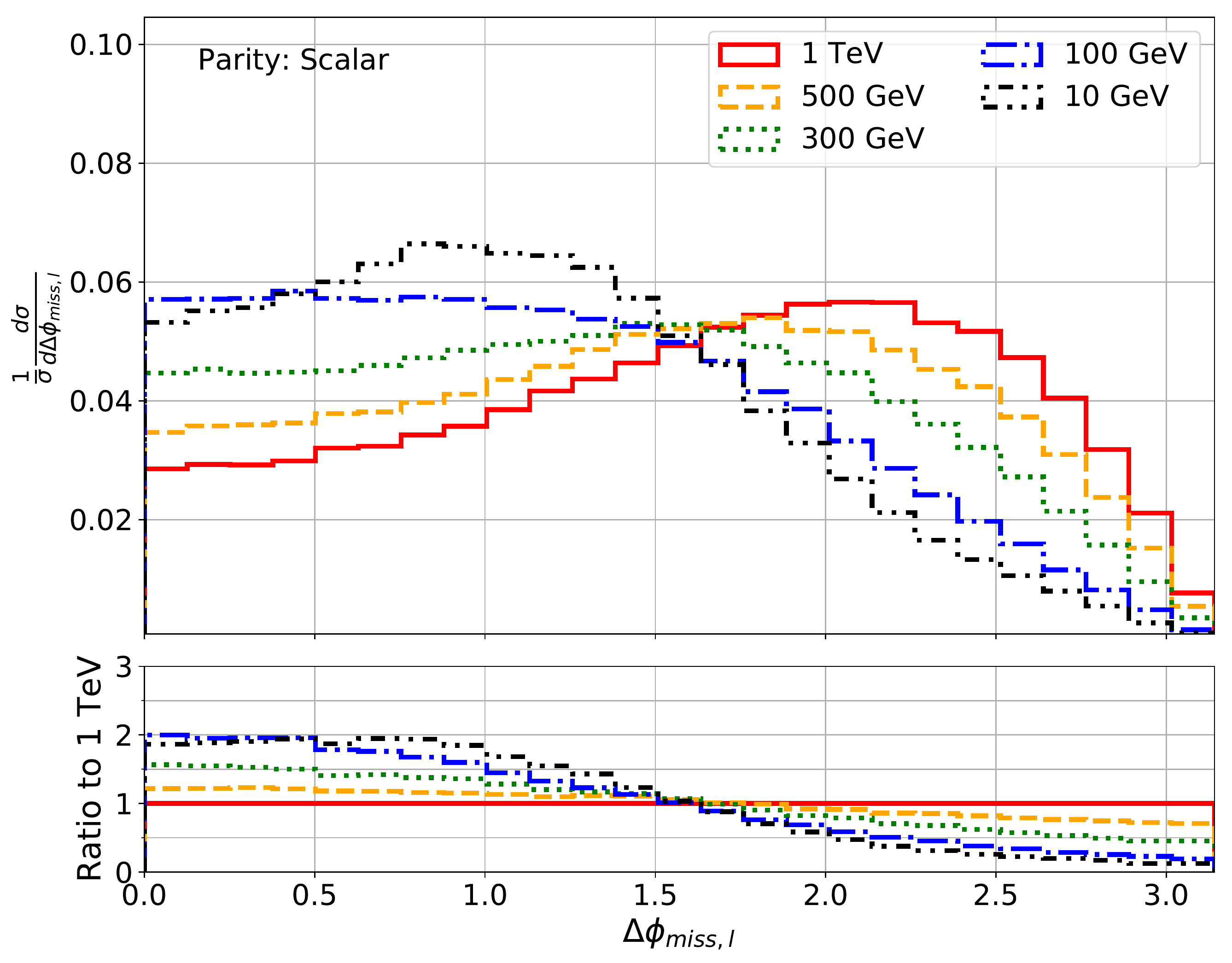}

	\caption{\textit{Comparison of normalised NLO differential distributions for the DM signal with a scalar mediator for different mediator masses. The samples have been generated using a central scale of $\mu^{\text{DM}}_0 = E_T/3$ with the NLO \textsc{CT14} PDF set for the LHC with center of mass energy $\sqrt{s} = 13$ TeV. The lower panels depict the ratio to the distributions for $m_Y = 1$ TeV.}}
	\label{fig:DM_Uncut_norm_mass_comp_S_NLO}
\end{figure*}

Even more interesting than the absolute size of the signal is the behavior of NP-sensitive observables. As the cross sections span many orders of magnitude, we normalise the differential NLO $\cos ( \theta^*_{ll})$ distributions which are depicted in Figure \ref{fig:DM_Uncut_norm_cosll_NLO} for scalar (left) and pseudoscalar (right) mediators of different masses. In the respective lower panels we show the comparison to the $m_Y = 1$ TeV case since it changes the least between the parities. For pseudoscalar mediators all distributions peak somewhere around $\sim0.9$ while this only happens for the heavier scalar mediators. Hence, this observable can be used as a CP discriminant if the mediators are not too heavy. Nevertheless, this observable can prove useful even for heavy mediators as its shape is also very different to the shape of the background processes as the latter do not exhibit the above described peak. One should also note that this peak is more pronounced the heavier the mediator is, irrespective of its parity. 

This is of course not the only relevant observable for DM analyses. Some additional ones are presented in Figure \ref{fig:DM_Uncut_norm_mass_comp_S_NLO} for the scalar mediator scenario. Ratios to $m_Y = 1$ TeV are again shown in the lower panels. The two `stransverse' masses $M_{T2,W}$ and $M_{T2,t}$ are defined as 

\begin{equation} \label{eq:MT2_W}
\begin{split}
M_{T2,W}^2 = \min_{\substack{\mathbf{p}^{\nu_1}_T + \mathbf{p}^{\nu_2}_T \\ = \mathbf{p}_{T,\text{\text{miss}}}}} [ \max \{ M_T^2 &\left( \mathbf{p}_T^{l_1}, \mathbf{p}^{\nu_1}_T \right),\\ &M_T^2 \left( \mathbf{p}_T^{l_2}, \mathbf{p}^{\nu_2}_T \right) \} ]
\end{split}
\end{equation}
and
\begin{equation} \label{eq:MT2_t}
\begin{split}
M_{T2,t}^2 = \min_{\substack{\mathbf{p}^{\nu_1}_T + \mathbf{p}^{\nu_2}_T \\ = \mathbf{p}_{T,\text{\text{miss}}}}} [ \max \{ M_T^2 &\left( \mathbf{p}_T^{(lb)_1}, \mathbf{p}^{\nu_1}_T \right),\\
 &M_T^2 \left( \mathbf{p}_T^{(lb)_2}, \mathbf{p}^{\nu_2}_T \right) \}  ]
\end{split}
\end{equation}
where
\begin{equation} \label{eq:MT_lb}
M_T^2 \left( \mathbf{p}_T^{(lb)_i}, \mathbf{p}^{\nu_i}_T \right) = M_{(lb)_i}^2 + 2 \left( E_T^{(lb)_i} E_T^{\nu_i} - \mathbf{p}_T^{(lb)_i} \cdot \mathbf{p}_T^{\nu_i} \right)
\end{equation}
is the transverse mass of the lepton$+$$b$-jet system in presence of a missing transverse momentum $p_{T}^{\nu_i}$, and similarly for $M_T^2 \left( \mathbf{p}_T^{(l)_i}, \mathbf{p}^{\nu_i}_T \right)$. Variables written in bold letters indicate three-vectors.
As we assume the charge of the $b$-jets to be untagged, we determine the appropriate combination of a $b$-jet and a lepton by minimising their invariant mass. 
More specifically, we take the smaller value of $M_{l1,b1} + M_{l2,b2}$ and $M_{l1,b2} + M_{l2,b1}$ in order to avoid one $b$-jet being associated with both leptons which might occur when just minimising $M_{l, b_i}$ for each of the leptons.
To calculate $M_{T2,t}$ and $M_{T2,W}$, we use the implementation presented in Ref. \citep{MT2_calculation}. 

We can also use $M_{T2,W}$ to define another useful observable\footnote{Note that there is a wrong sign in the definition given in Ref. \citep{Haisch_analysis}.} \citep{Haisch_analysis}, namely
\begin{equation}
C_{em,W} =  M_{T2,W} - 0.2\cdot (200 \text{ GeV} - p_{T,\text{miss}}). 
\end{equation}
Concerning the general behavior of the stransverse masses, let us mention that the peak in the first bin of $M_{T2,W}$ occurs because of the minimization procedure in Eq. (\ref{eq:MT2_W}). Indeed, $M_{T2,W}$ is only bounded from below by the lepton mass, which is zero in our case. 
This peak is absent in the $M_{T2,t}$ distributions since we have $M_{T2,t} \geq M_{lb}$. In this case, $M_{lb}$ is non-zero due to the cuts on $\Delta R_{lb}$, $p_{T,l}$ and $p_{T, b}$.
However, the most important feature of these two observables are the kinematic edges around $M_{T2,W} \sim m_W$ and $M_{T2,t} \sim m_t$ which can be used to determine the respective masses in $t\bar{t}$ production. However, if the mediators are light enough, the top-quark kinematics and the total $p_{T,\text{miss}}$ are only slightly changed by the addition of the mediator compared to $t\bar{t}$. As a result, one can still clearly observe the edges in $t\bar{t}Y$ production in such cases.


For both of the stransverse masses as well as for the missing transverse momentum we observe that the distribution tails are much more prominent for heavier mediators. Between the lightest and heaviest considered mediators the normalised distributions can differ by more than two orders of magnitude. 
This compensates some of the difference between the integrated cross sections but in absolute terms, the signal with $m_Y = 10$ GeV is still the largest one, even in the distribution tails. 

In all three dimensionful observables the shape differences are mostly down to the additional missing transverse momentum resulting from the mediator production which is harder the more massive the mediator is. 
Since $M_{T2,W}$ and $M_{T2,t}$ are both dependent on $p_{T,\text{miss}}$, they are also affected in a similar manner. 
This dependence on $p_{T,\text{miss}}$ also changes the appearance of the above mentioned kinematic edges in $M_{T2,W}$ and $M_{T2,t}$ which completely vanish for heavy mediators.

Just like for $\cos (\theta^*_{ll})$, the distributions for the pseudoscalar mediator scenario are almost identical to the ones shown in Figure \ref{fig:DM_Uncut_norm_mass_comp_S_NLO} for heavy mediators. Distributions for lighter mediators, however, tend to receive larger contributions from their tails than in the scalar case and are generally more akin to the heavy mediator distributions. The above mentioned kinematic edges are also only barely visible which is due to mediator radiation being harder in the pseudoscalar case, as discussed e.g. in Ref. \citep{Haisch_analysis}.

We also make use of one additional angular observable defined as
\begin{equation}
\Delta \phi_{\text{miss},l} = \min \{ \Delta \phi_{\text{miss},e^+}, \Delta \phi_{\text{miss},\mu^-} \}.
\end{equation}
Since the flight direction of the leptons and the top quarks are highly correlated, this gives us an idea of the azimuthal distance between the (anti-)top quark and the missing transverse momentum. As we can see from the bottom right plot in Figure \ref{fig:DM_Uncut_norm_mass_comp_S_NLO}, this distance tends to be larger for heavier mediators. The same behavior can be observed in the pseudoscalar case but the distributions for lighter mediators are slightly shifted towards larger angles.

\section{The Standard model background} \label{sec:Background}

After introducing the DM signal, we now turn our attention to the corresponding SM background. As stated above, we consider $t\bar{t}$ and $t\bar{t}Z$ production as these are the dominant background processes. 

Note that in principle, any process involving an additional, arbitrary number of $Z$-bosons could contribute to the background as well. However, even for just one more $Z$ boson, i.e. $t\bar{t}ZZ$ production, the cross section is three orders of magnitude smaller than the $t\bar{t}Z$ contribution.
As the latter is itself already four orders of magnitude smaller than the $t\bar{t}$ cross section (see Table \ref{table:total_xSec_FW_effects}), $t\bar{t}ZZ$ is not considered in this paper.

\subsection{Background generation and integrated cross sections} \label{subsec:Bkg_xSec_and_gen}

\begin{table*}
	\caption{\textit{Comparison of integrated background cross sections between the NWA and full off-shell predictions with their respective scale uncertainties at LO and NLO. All values are given for the LHC with a center of mass energy of $\sqrt{s} = 13\,\text{TeV}$. We employ the (N)LO \textsc{CT14} PDF set. For the $K$-factor in the NWA we give the values for the full NLO NWA result and the one with LO decays, the latter in parenthesis.}}
	\label{table:total_xSec_FW_effects}
	
	\centering
	\renewcommand{\arraystretch}{1.5}
	\begin{tabular}{ll@{\hskip 10mm}l@{\hskip 10mm}ll@{\hskip 10mm}l}
		\hline\noalign{\smallskip}
		Process & Scale &  & Off-shell & NWA & Off-shell effects  \\
		\noalign{\smallskip}\midrule[0.5mm]\noalign{\smallskip}
		\multirow[c]{3}{*}{$t\bar{t}$} & \multirow{3}{*}{$H_T/4$}    & $\sigma_{\text{LO}}$ [fb] & $ 1067^{+348(33\%)}_{-247(23\%)} $ & $ 1061^{+346(33\%)}_{-245(23\%)} $ & $ 0.6 \%$\\
		&             & $\sigma_{\text{NLO}}$ [fb] & $ 1101^{+19(2\%)}_{-57(5\%)} $ & $ 1097^{+0(0\%)}_{-56(5\%)} $ & $ 0.4 \%$\\
		&             & $\sigma_{\text{NLO}_{\text{LOdec}}}$ [fb] & - &  $1271^{+118(9\%)}_{-136(11\%)} $ & $ $\\
		\noalign{\smallskip}\hline\noalign{\smallskip}
		& & $K = \sigma_{\text{NLO}} / \sigma_{\text{LO}}$	& $1.03$ & $1.03 \;(\text{LOdec: } 1.20)$& \\

		\noalign{\smallskip}\midrule[0.5mm]\noalign{\smallskip}
		
		\multirow{3}{*}{$t\bar{t}Z$} & \multirow{3}{*}{$H_T/3$}    & $\sigma_{\text{LO}}$ [fb] & $ 0.1262^{+0.0439(35\%)}_{-0.0303(24\%)}$ & $ 0.1223^{+0.0422(35\%)}_{-0.0292(24\%)} $ & $ 3 \%$\\
		&             & $\sigma_{\text{NLO}}$ [fb] & $ 0.1269^{+0.0010(1\%)}_{-0.0085(7\%)} $ & $ 0.1226^{+0.0(0\%)}_{-0.0088(7\%)} $ & $ 4 \%$\\
		&             & $\sigma_{\text{NLO}_{\text{LOdec}}}$ [fb] & - &  $0.1364^{+0.0088(6\%)}_{-0.0136(10\%)}$ & $ $\\
		\noalign{\smallskip}\hline\noalign{\smallskip}
		& & $K = \sigma_{\text{NLO}} / \sigma_{\text{LO}}$	& $1.01$ & $1.0 \; (\text{LOdec: } 1.12)$& \\
		\noalign{\smallskip}\hline\noalign{\smallskip}
	\end{tabular}
\end{table*}

In Table \ref{table:total_xSec_FW_effects} we present integrated cross sections for the two background processes $t\bar{t}$ and $t\bar{t}Z$ at LO and NLO. 	
All of the results have been computed using \textsc{Helac-NLO} \citep{HELAC_NLO} which comprises \textsc{Helac-1Loop} \citep{HELAC_1loop} and \textsc{Helac-Dipoles} \citep{HELAC_Dipoles}. The former employs \textsc{CutTools} \citep{CutTools} and \textsc{OneLOop} \citep{OneLoop} to evaluate the virtual contributions. The \textsc{Helac-Dipoles} MC program, on the other hand, is used to calculate the real emission contributions. Two different subtractions schemes are applied here, Nagy-Soper \citep{NS_subtraction} for the off-shell results and Catani-Seymour \citep{CS_subtraction, CS_subtraction_massive} in the NWA. For further details concerning the calculation, we refer to Ref. \citep{ttZ}. 

In addition to the full off-shell results presented in Ref. \citep{ttZ}, we also show results using the NWA.
Theoretical predictions for $t\bar{t}Z$ production in the NWA are presented for the first time.
Note that if we use the NWA, we always put all of the resonant particles, i.e. $t$, $W$, and $Z$, on-shell.
We find that for the $t\bar{t}$ process, the off-shell effects are at the per-mille level for the integrated fiducial cross section. Specifically, they are of the order of $0.6 \%$ at LO and $0.4 \%$ at NLO. 
For the $t\bar{t}Z$ process they are slightly larger at $3\% - 4\%$. This is due to the additional effects coming from putting the $Z$-boson on-shell. For the latter, $\Gamma_Z / m_Z \sim 2.8 \%$ is rather large. 
In either case, the effects are well within the scale uncertainties, even at NLO. 
The higher-order corrections themselves are quite moderate at around $3 \%$ for $t\bar{t}$ and at $1 \%$ for $t\bar{t}Z$. This is mostly due to the judicious choice of dynamical scales, which are  designed to keep higher-order corrections small. For example, had we used $\mu_0=m_t + m_Z/2$ for $t\bar{t}Z$ instead, we would find significantly larger NLO corrections of $12 \%$ \citep{ttZ}.

In addition to the LO and NLO results in the NWA, we also compute NLO$_{\text{LOdec}}$ cross sections. These consist of NLO QCD corrections to the production while the top-quark decays are treated at LO. Spin correlations at LO are properly taken into account as well.
This is more in line with how the DM signal has been calculated with \textsc{MadGraph5\_aMC@NLO}   but for the latter some finite width effects are taken into account in \textsc{MadSpin}. 
We find that the NLO$_{\text{LOdec}}$ results are larger than the pure LO or NLO findings. 
The QCD corrections to just the production amount to $20 \%$ for $t\bar{t}$ and $12 \%$ for $t\bar{t}Z$.
Similar results have been found for $t\bar{t}\gamma$ and $t\bar{t}W^{\pm}$ in Refs. \citep{ttgamma, ttW}.

\sloppy Scale uncertainties also behave as expected. They decrease significantly from $(+33 \%,-23\%)$ at LO to $(+2\%,-5\%)$ at NLO for the top-quark background and similarly for $t\bar{t}Z$.
There is no significant difference for the scale uncertainties between the NWA and off-shell results with the exception that in the former case, the upper scale variation is zero for the two NWA predictions. 
To mitigate this, we adopt a conservative estimate of the uncertainties and take the maximal variation as our scale uncertainty.
The same is done for differential distributions on a bin-by-bin basis.
As expected, scale uncertainties for the NLO$_{\text{LOdec}}$ cross sections are larger than for the full NLO description at around $11 \%$.

As explained in the setup section, we calculate the internal PDF uncertainties of the \textsc{CT14} PDF sets for both $t\bar{t}$ and $t\bar{t}Z$ at NLO for the full off-shell case. 
They amount to $3\%$ for $t\bar{t}$ and $4 \%$ for $t\bar{t}Z$. 
We use these PDF uncertainties also for the NWA predictions since the modelling should not change the dependence on PDFs. 
Additionally, we also use them for the LO predictions as, firstly, there are no error-PDF sets provided for the LO \textsc{CT14} PDF set and secondly, the PDF uncertainties at LO are subdominant compared to the scale uncertainties.

\subsection{Distribution shapes} \label{subsec:norm_shape}

\begin{figure*}
	\includegraphics[width=0.48\linewidth]{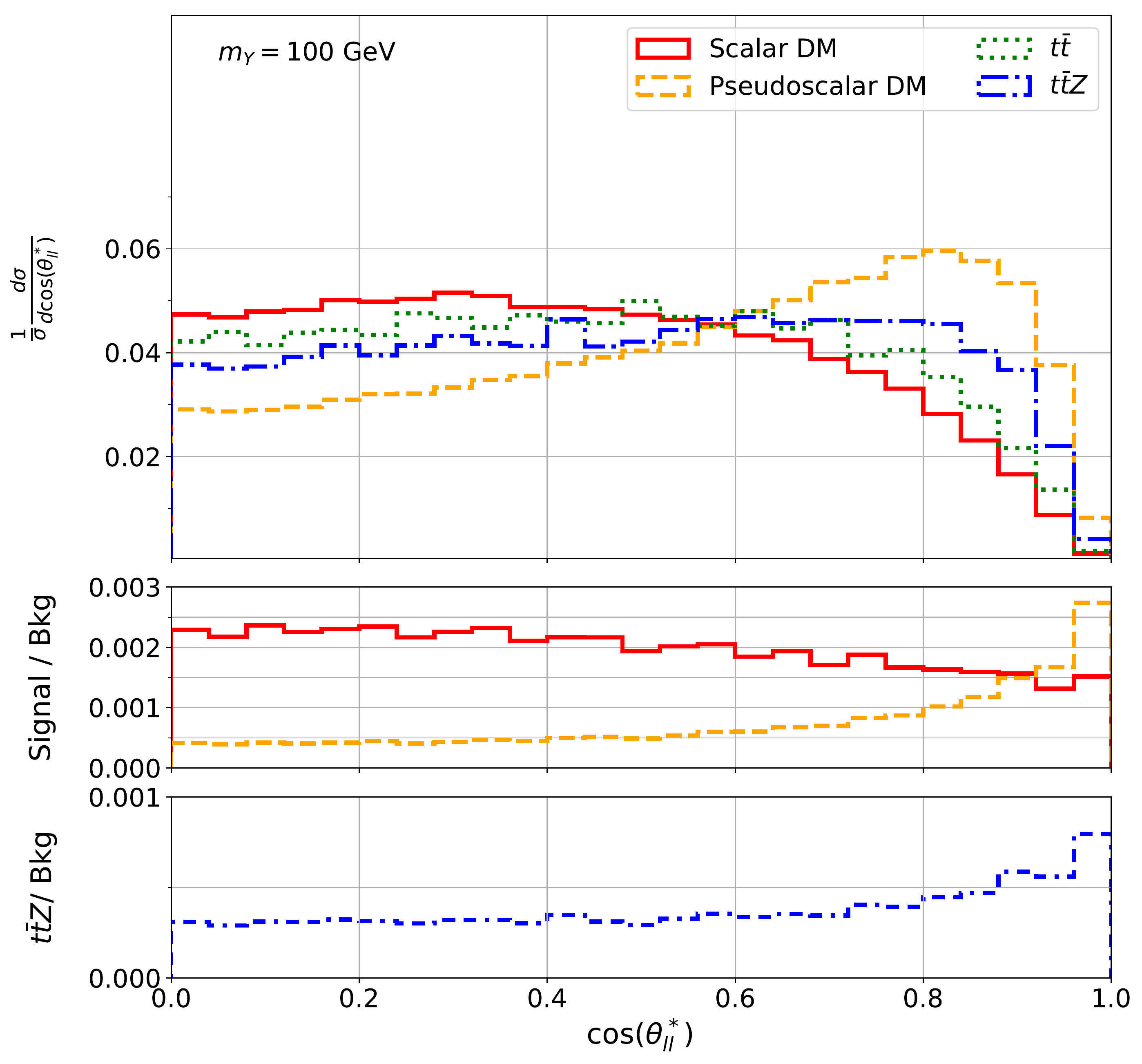}%
	\includegraphics[width=0.48\linewidth]{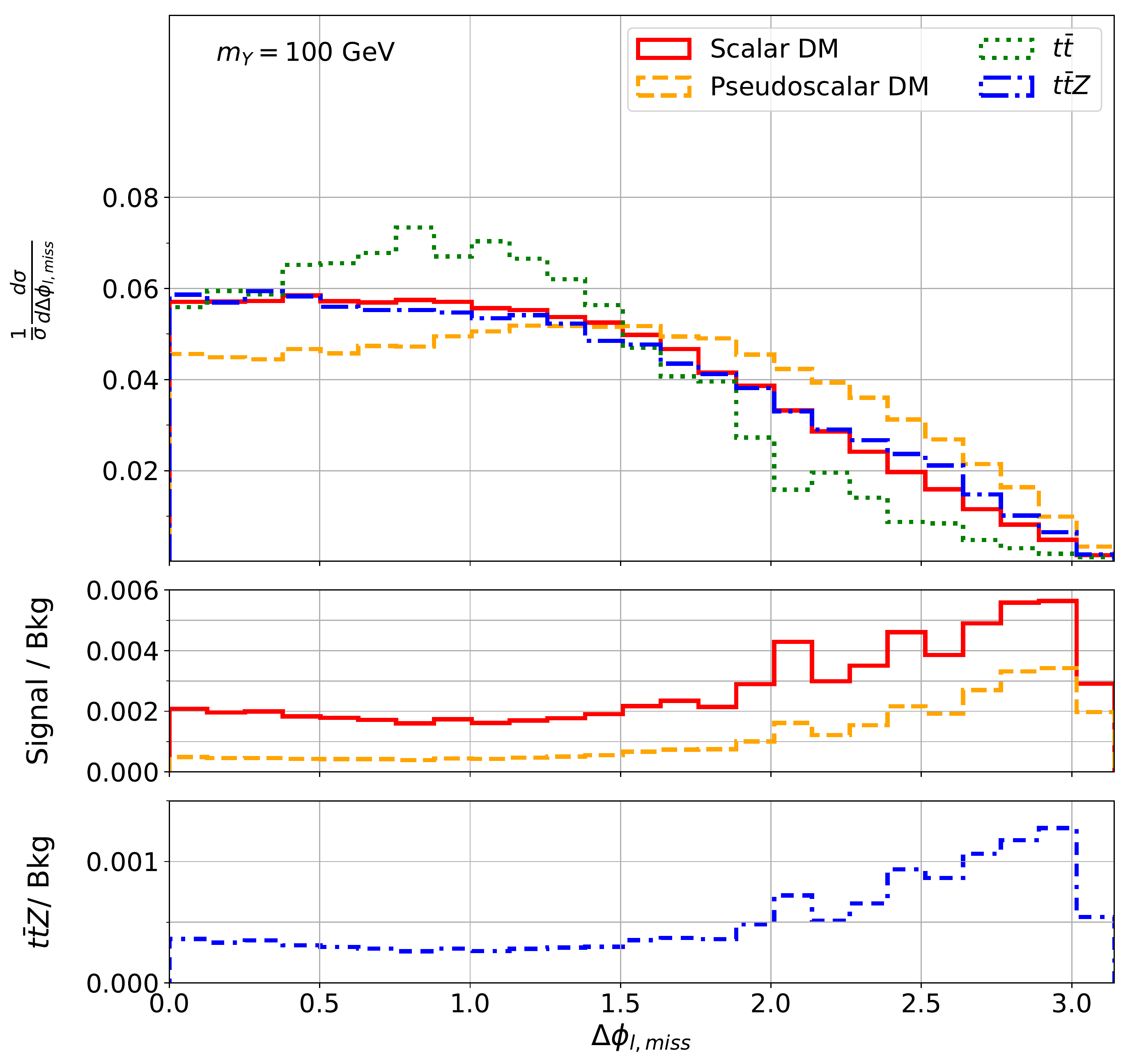}
	\includegraphics[width=0.48\linewidth]{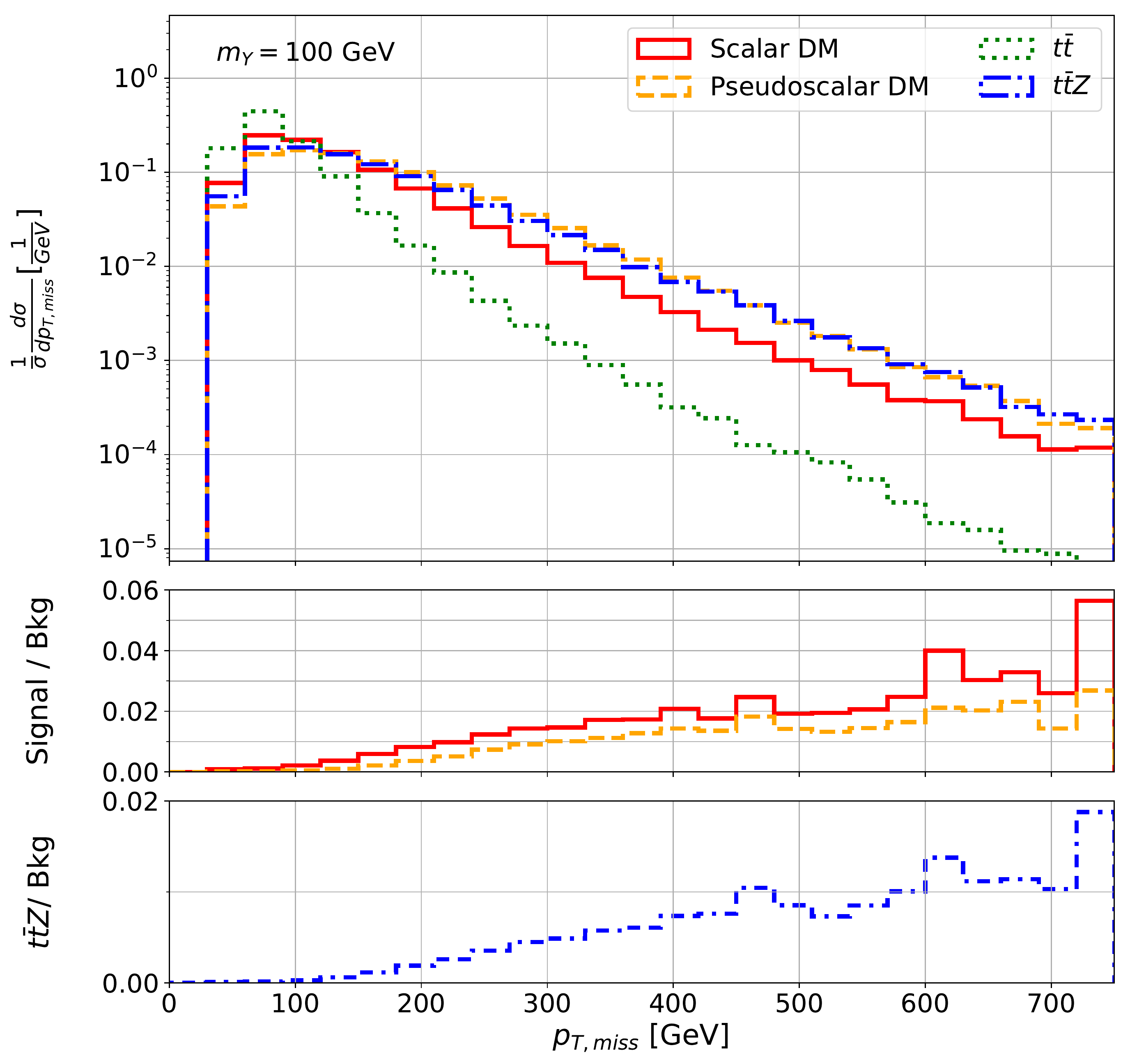}%
	\includegraphics[width=0.48\linewidth]{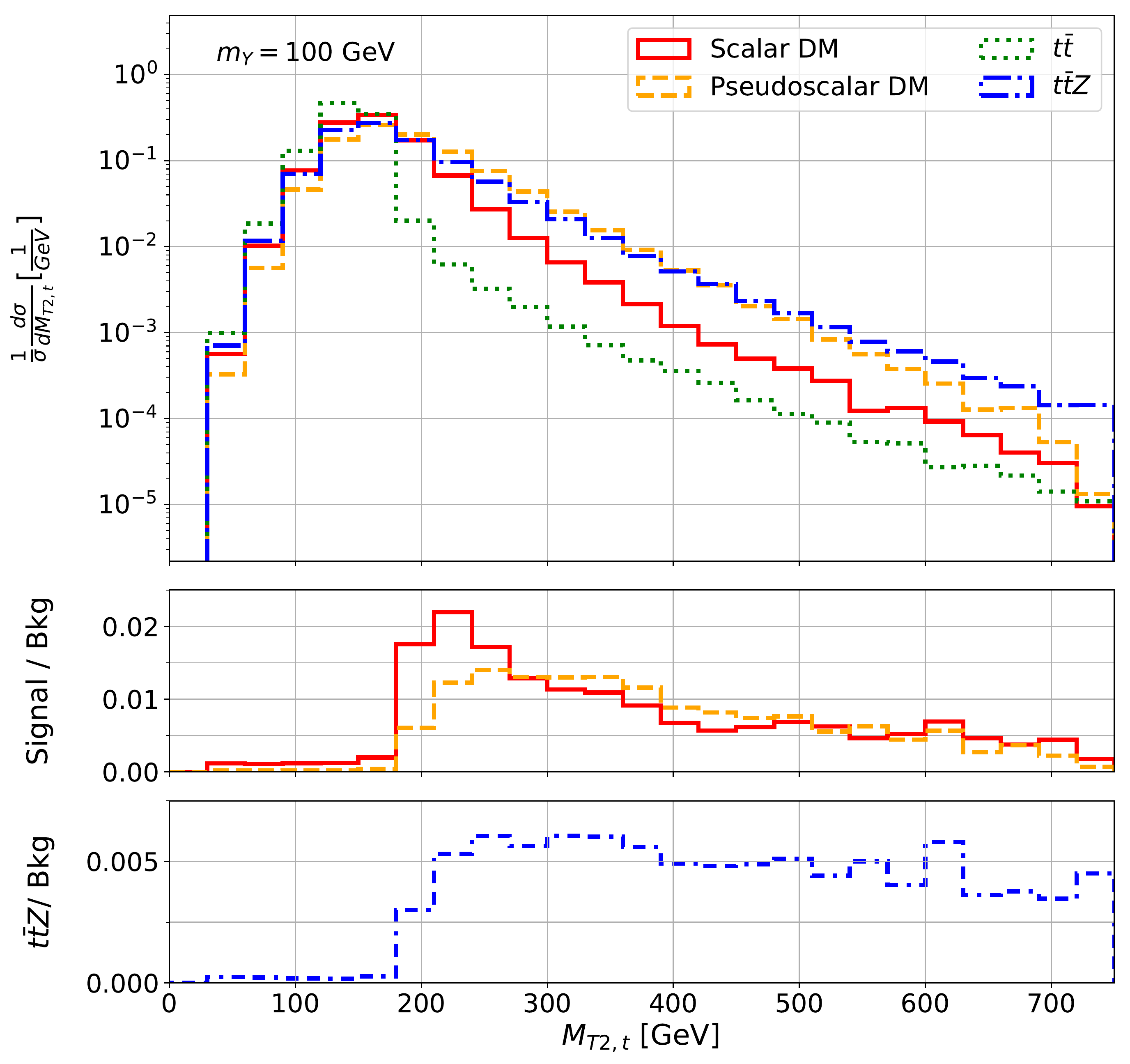}
	\includegraphics[width=0.48\linewidth]{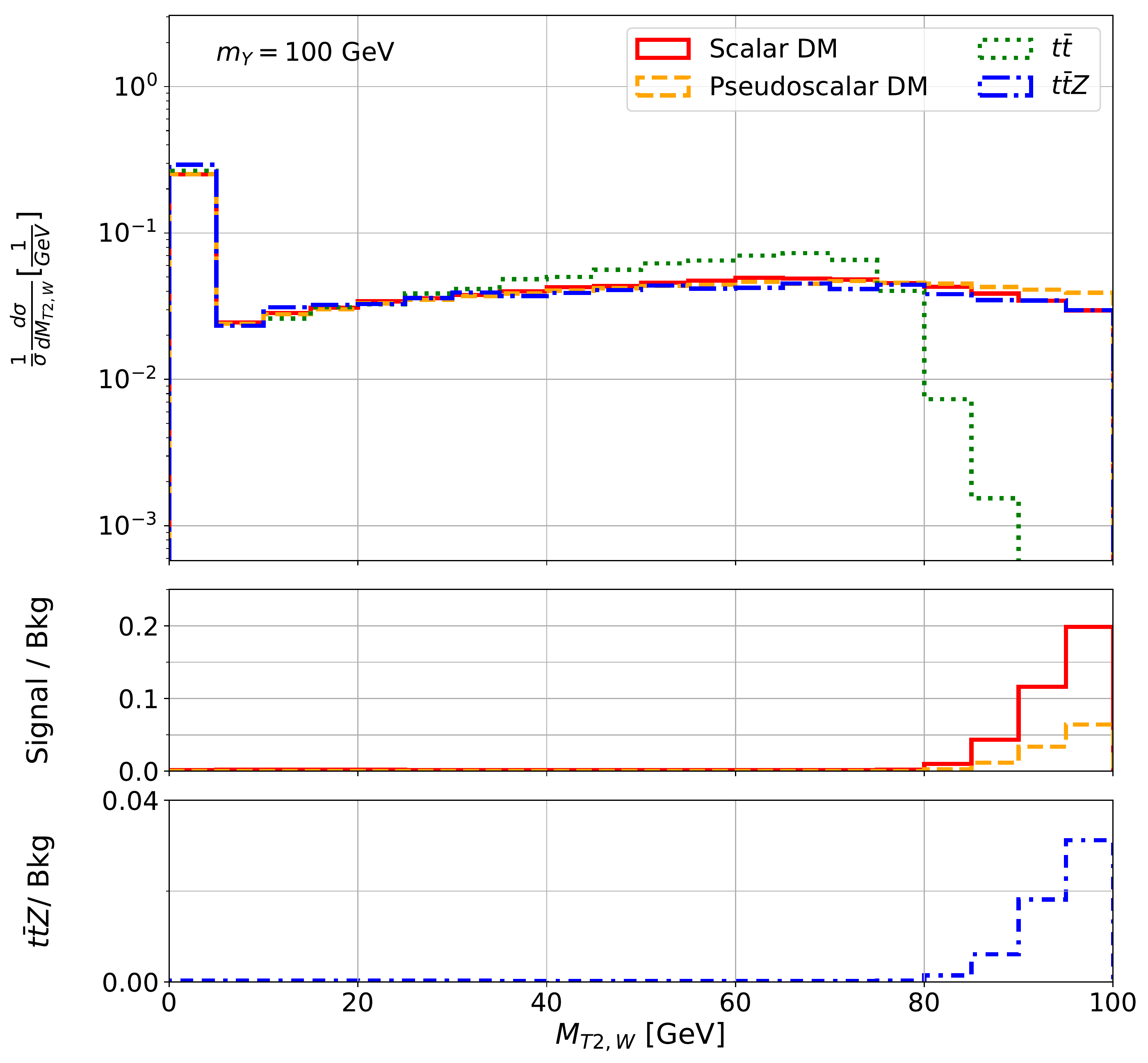}%
	\includegraphics[width=0.48\linewidth]{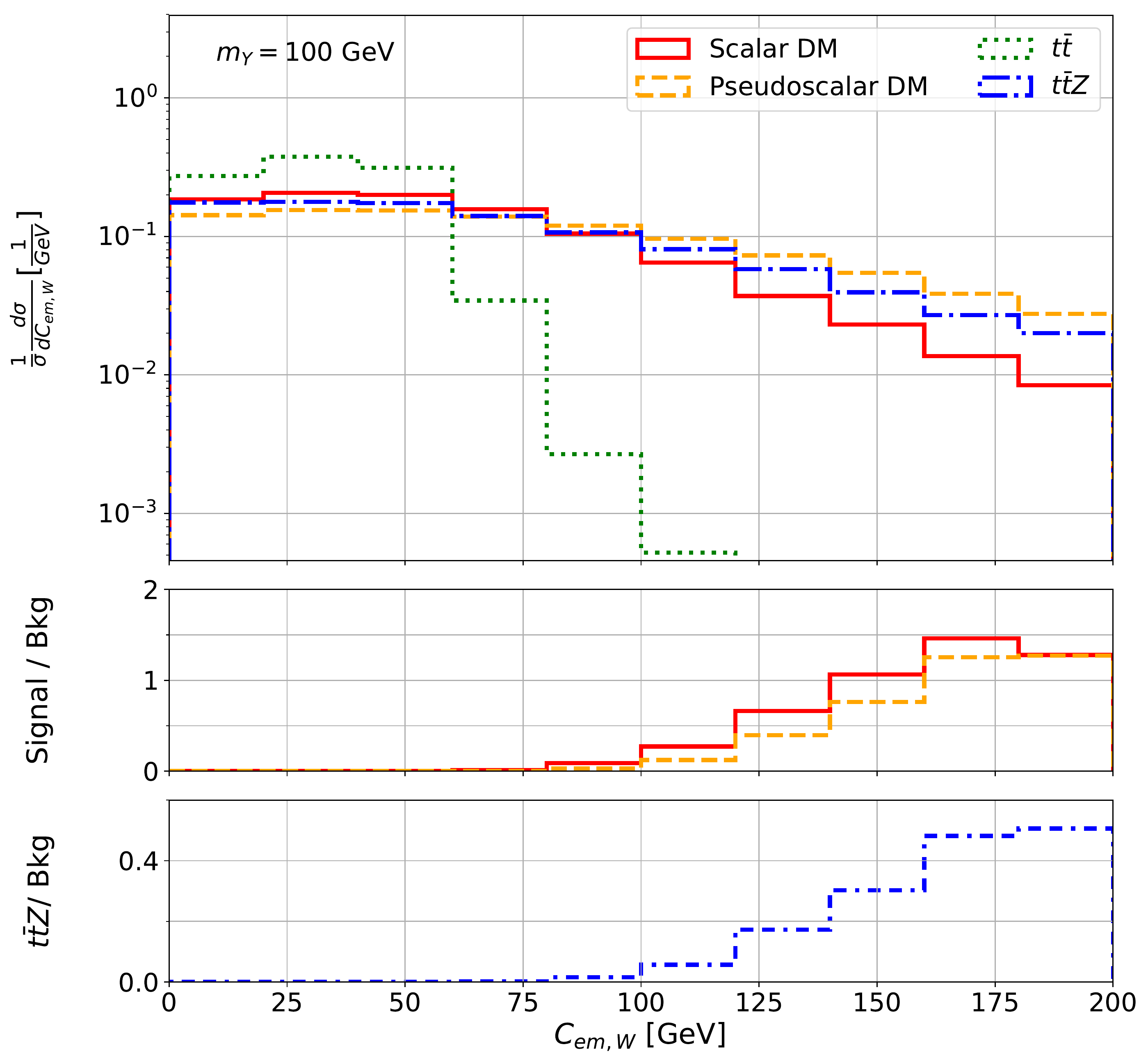}
	
	\caption{\textit{Comparison of normalised NLO differential distributions for the off-shell $t\bar{t}$ and $t\bar{t}Z$ background processes as well as scalar and pseudoscalar DM signals with $m_Y = 100$ GeV. The samples have been generated using the NLO \textsc{CT14} PDF set and our default scale choices for the LHC with center of mass energy $\sqrt{s} = 13$ TeV. In the central panels we show the signal-to-background ratio including the respective lepton flavour factors. The lower panels depict the fraction of the $t\bar{t}Z$ contribution to the total background.}}
	\label{fig:StBR_NLO_S_100_Uncut}
\end{figure*}

From Table \ref{table:total_xSec_FW_effects} we have seen that there is a clear hierarchy between the two background processes. 
However, they also differ substantially at the differential level in several key NP observables, as can be seen in Figure \ref{fig:StBR_NLO_S_100_Uncut}. Here, we present the normalised differential distributions for both background processes as well as a scalar and pseudoscalar DM signal with $m_Y= 100 \text{ GeV}$.
In each of the shown distributions, $t\bar{t}Z$ receives much larger contributions from the respective tails. While for angular observables the normalised distributions can already differ by around a factor $2$, the differences can far exceed an order of magnitude in $p_{T,\text{miss}}$, $M_{T2,W}$, $C_{em,W}$ and $M_{T2,t}$. This is not surprising as all of these are related to the missing transverse momentum which gets amplified substantially by the invisibly decaying $Z$ boson.

In $M_{T2,W}$, $C_{em,W}$ and $M_{T2,t}$ this is further enhanced by the kinematic edges we have already mentioned when discussing the signal.
For the $t\bar{t}$ process, we find sharp declines in the distributions around $M_{T2,W} \sim m_W$ and $M_{T2,t} \sim m_t$. However, these edges are completely absent in the case of $t\bar{t}Z$, just like for heavier mediators in Figure \ref{fig:DM_Uncut_norm_mass_comp_S_NLO}. A similar behavior can be observed in $C_{em,W}$ as it is connected to $M_{T2,W}$.

To emphasise the apparent similarities between the signal and the $t\bar{t}Z$ process, we also include distributions of DM models with $m_Y = 100$ GeV for both parities in Figure \ref{fig:StBR_NLO_S_100_Uncut}. It is immediately apparent that $t\bar{t}Z$ mimics the signal's behavior much more closely than $t\bar{t}$. Nevertheless, one can still observe some differences in the angular observables and in the tails of dimensionful ones.

For the calculation of exclusion limits, we will only compare the sum of both background processes to the signal. To already get an idea of the role that shape differences will play, we show the signal-to-background ratio
\begin{equation} \label{eq:StBR}
R = \frac{4 \cdot \sigma_{\text{DM}}}{4 \cdot \sigma_{t\bar{t}} + 12 \cdot \sigma_{t\bar{t}Z}}
\end{equation}
in the central panels of each plot. The factors $4$ and $12$ are the respective lepton flavour factors. This ratio underlines the above discussed shape differences between the signal and the SM background. They are clearly visible in all but one of the presented observables. The only exception is the scalar signal in $\cos (\theta^*_{ll})$ for which the ratio stays almost constant. 
In most phase-space regions, the denominator in Eq. (\ref{eq:StBR}) is dominated by the $t\bar{t}$ background which is why the ratio $R$ changes so much throughout the distributions. In the lower panels of each plot we additionally show what fraction of the background can be attributed to the $t\bar{t}Z$ process. These show that the $t\bar{t}Z$ process only really becomes relevant above the kinematic edges in $M_{T2,W}$ and $C_{em,W}$, and, to a lesser extend, in the high-$p_{T,\text{miss}}$ and -$M_{T2,t}$ regions. We have checked many more observables but found that $p_{T,\text{miss}}$, $M_{T2,t}$, $M_{T2,W}$, $C_{em,W}$, $\cos (\theta^*_{ll})$ and $\Delta \phi_{l,\text{miss}}$ exhibited the most significant shape differences  between the DM signal and the SM background. Hence, these observables are going to be analysed further.

\subsection{Modelling} \label{subsec:modeling}

\begin{figure*}
	\includegraphics[width=.5\linewidth]{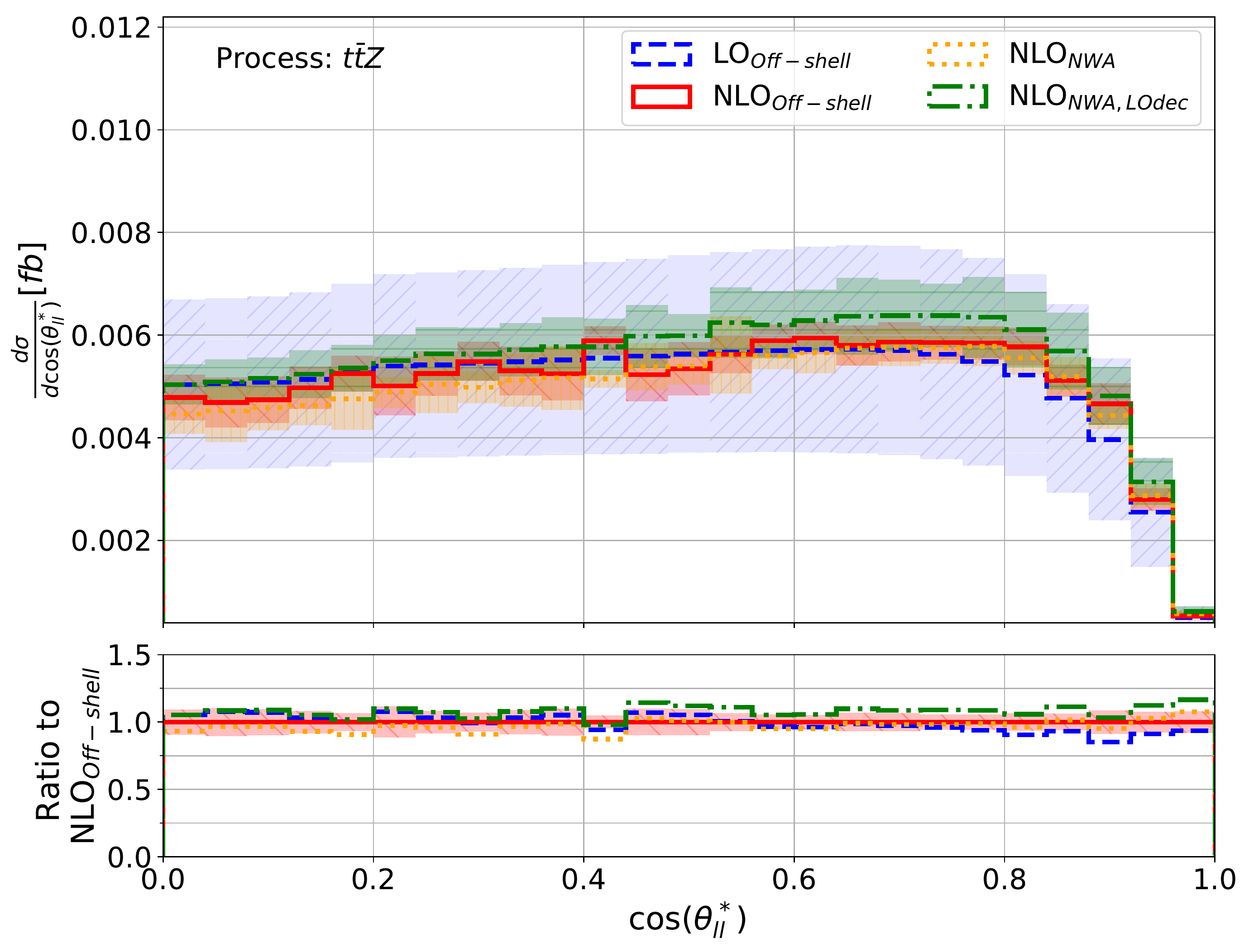}
	\includegraphics[width=.5\linewidth]{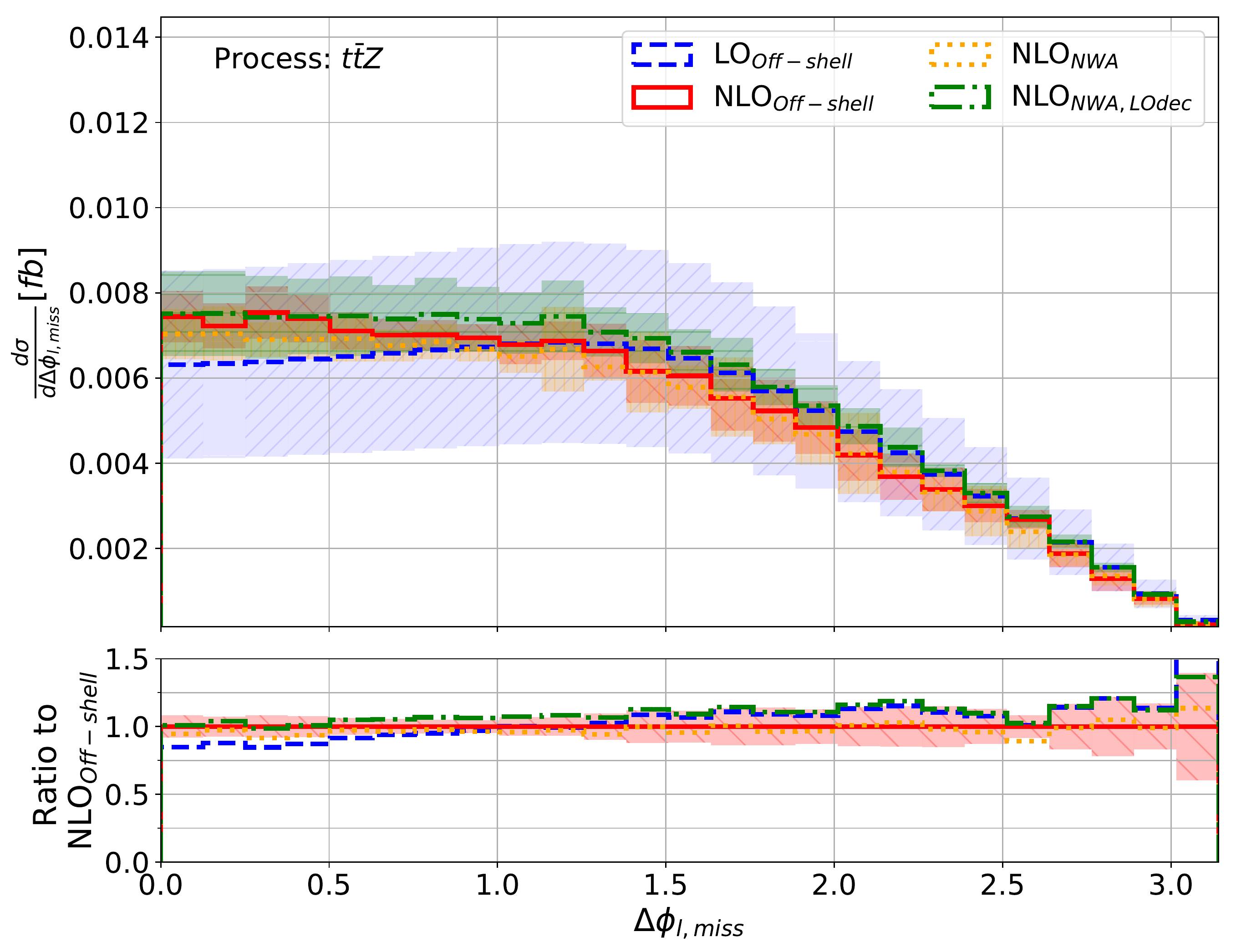}
	\includegraphics[width=.5\linewidth]{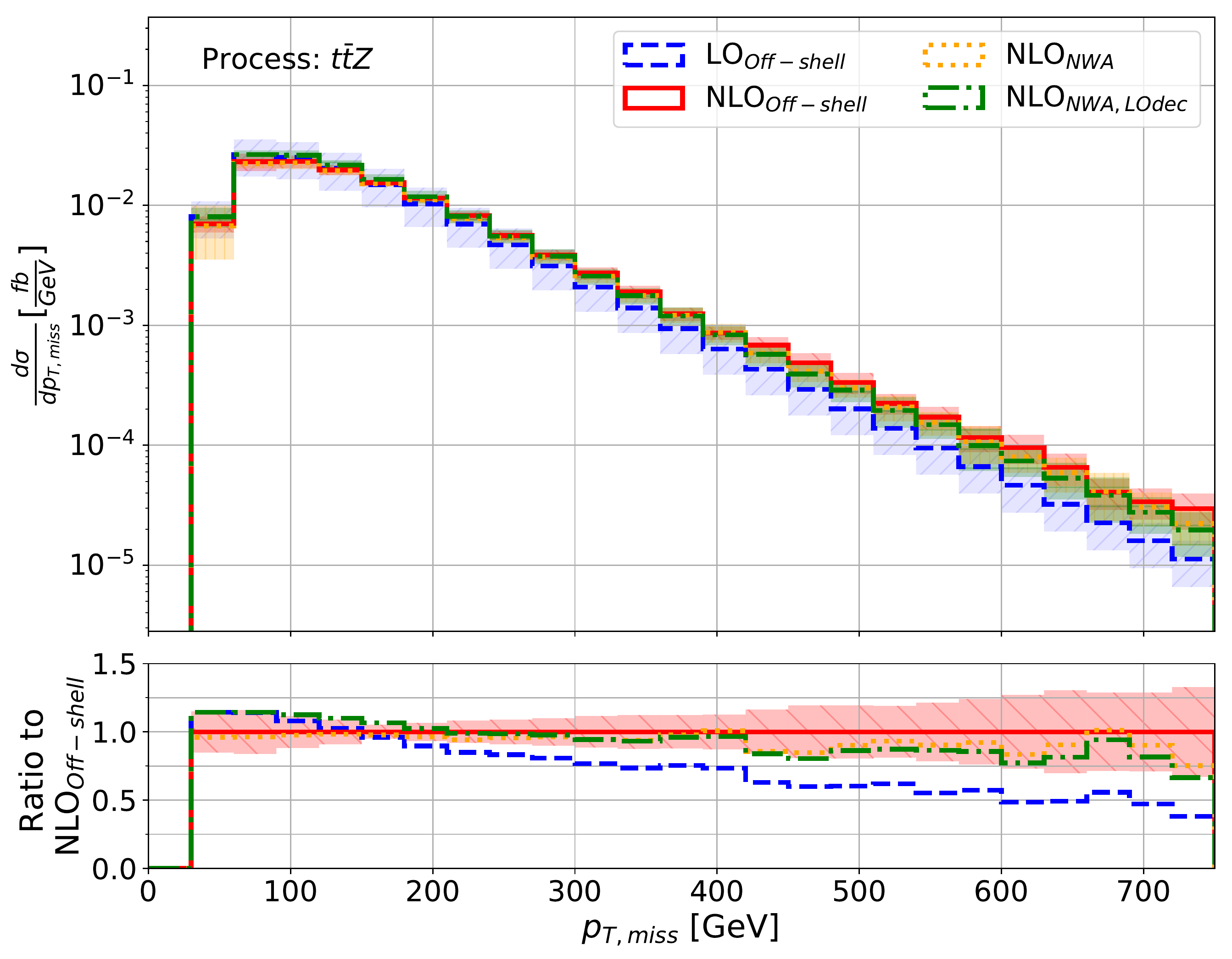}
	\includegraphics[width=.5\linewidth]{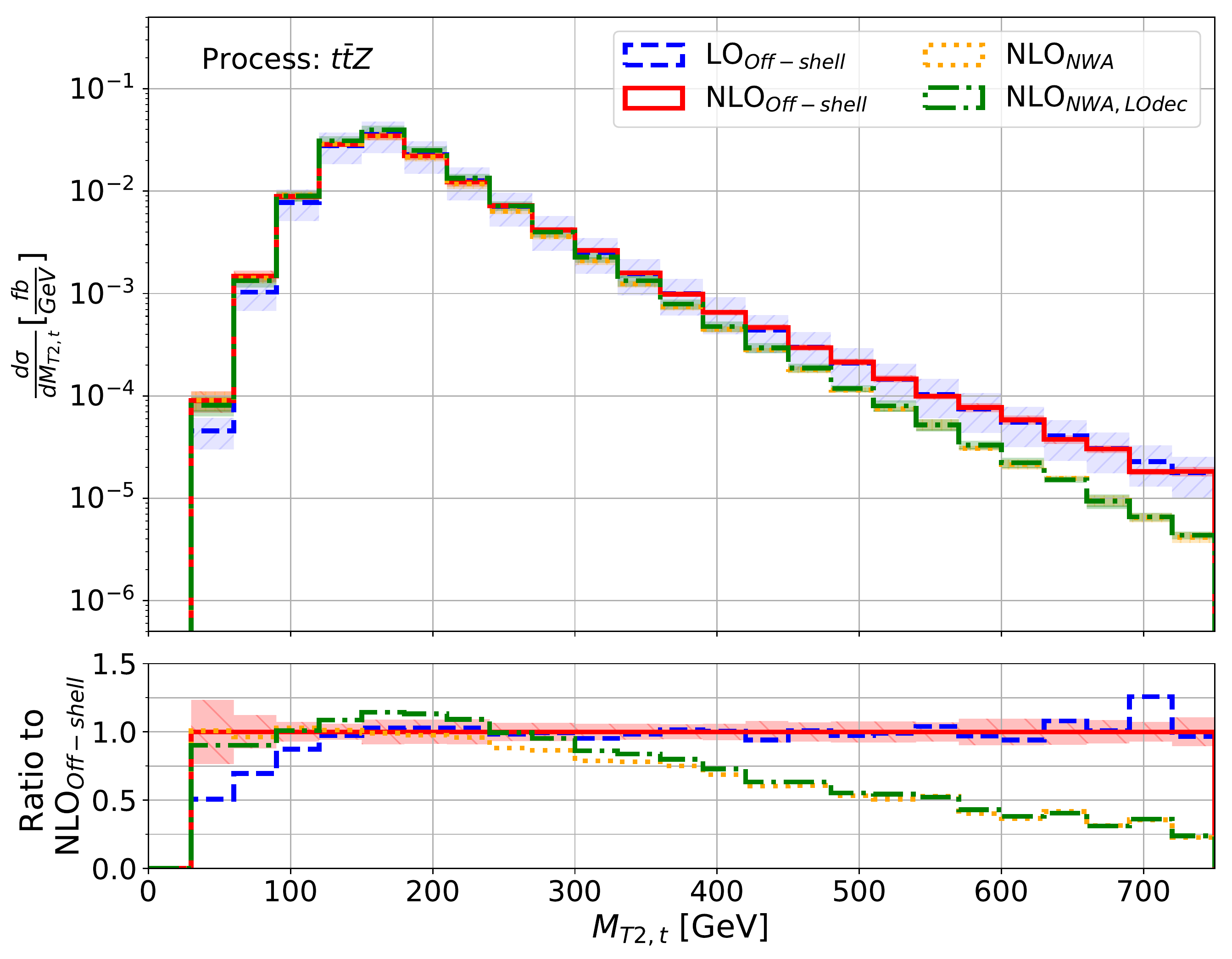}
	
	\caption{\textit{Comparison of differential distributions for the $t\bar{t}Z$ background process for different modelling approaches. The samples have been generated using a central scale of $\mu^{t\bar{t}Z}_0 = H_T/3$ with the NLO \textsc{CT14} PDF sets for the LHC with center of mass energy $\sqrt{s} = 13$ TeV. The error bands depict the respective scale uncertainties. In the lower panels we present the ratios to the NLO$_{\text{Off-shell}}$ results.}}
	\label{fig:Bkg_Uncut_Modeling_ttZ}
\end{figure*}
\begin{figure*}
	\includegraphics[width=.5\linewidth]{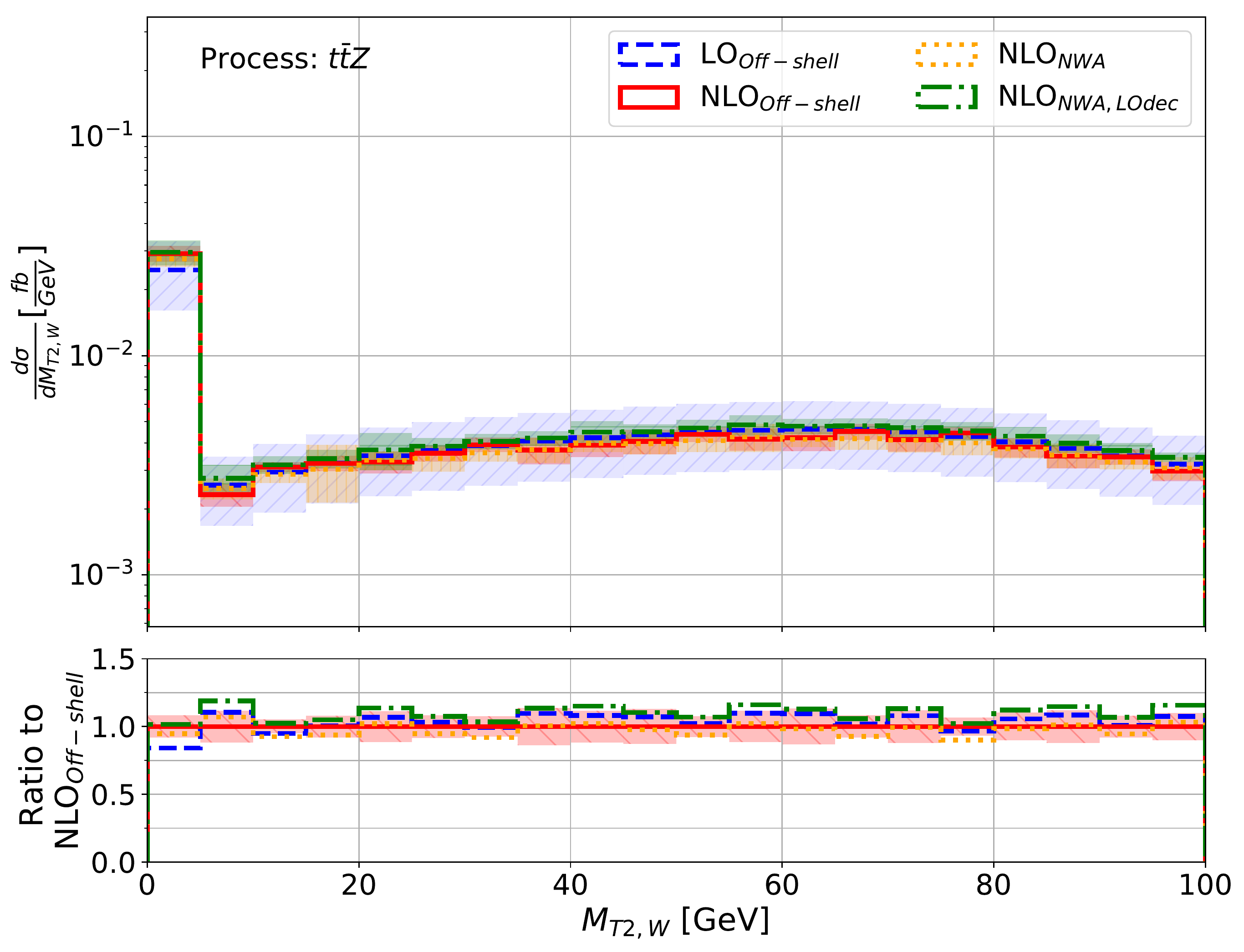}
	\includegraphics[width=.5\linewidth]{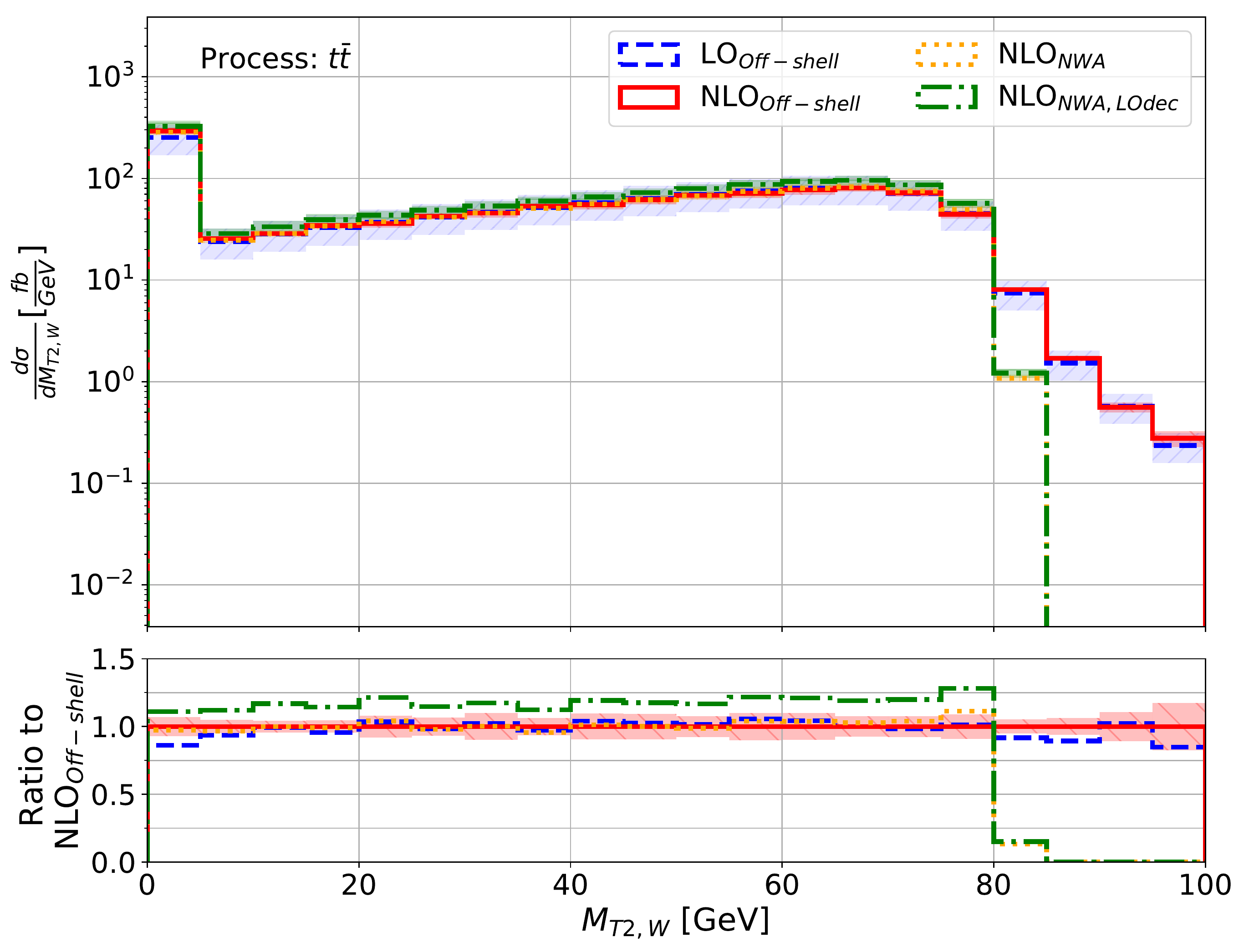}
	
	\caption{\textit{Comparison of differential distributions in $M_{T2,W}$ for the $t\bar{t}Z$ (left) and $t\bar{t}$ (right) background processes for different modelling approaches of the two processes. The results have been generated using central scales $\mu^{t\bar{t}Z}_0 = H_T/3$ and $\mu^{t\bar{t}}_0 = H_T/4$ with the NLO \textsc{CT14} PDF sets for the LHC with center of mass energy $\sqrt{s} = 13$ TeV. The error bands depict the respective scale uncertainties. In the lower panels we present the ratios to the NLO$_{\text{Off-shell}}$ results.}}
	\label{fig:Bkg_Uncut_Modeling_MTW}
\end{figure*}

As we have now established the most relevant observables, we turn our attention to off-shell and higher-order corrections at the differential level.
We focus here on the $t\bar{t}Z$ process as it is much more common for this one to be modelled at LO and / or without off-shell effects. Nevertheless, comments on the shape effects in $t\bar{t}$ are made where necessary.

In Figure \ref{fig:Bkg_Uncut_Modeling_ttZ} we compare the state-of-the-art NLO$_{\text{Off-shell}}$ $t\bar{t}Z$ predictions to LO$_{\text{Off-shell}}$, NLO$_{\text{NWA}}$ and NLO$_{\text{NWA,LOdec}}$. 
Their respective normalisations behave as outlined in the previous section and in Table \ref{table:total_xSec_FW_effects}. 
The same is true for the scale uncertainties which we indicate by the respective coloured and hatched bands. 
In the lower panels, we show the ratios to the NLO$_{\text{Off-shell}}$ prediction. This means that the LO$_{\text{Off-shell}}$ to NLO$_{\text{Off-shell}}$ ratio curve indicates the (inverse) $K$-factor and the respective NWA curves show the size of the off-shell effects.

For the most part, the higher-order corrections are well within the LO uncertainty bands. For $\cos (\theta^*_{ll})$, we find that the distribution is slightly shifted towards larger values while the opposite is true for $\Delta \phi_{l,\text{miss}}$. In both cases, the corrections stay within a few percent throughout most of the distribution but increase towards small and large $\Delta \phi_{l,\text{miss}}$ values. 

More significant changes can be observed in the missing transverse momentum distributions. Here, the NLO results are more than twice as large as at LO for large $p_{T,\text{miss}}$ and the $K$-factor increases consistently towards the tails. For $M_{T2,t}$, on the other hand, one can only really observe changes for low values while the tails are almost identical for LO and NLO off-shell predictions.

When we consider off-shell effects, the situation between $p_{T,\text{miss}}$ and $M_{T2,t}$ is essentially reversed. 
The tails of the latter are underestimated by up to $75 \%$ while for $p_{T,\text{miss}}$ the corrections only reach $25 \%$ in the depicted region. 
Just like the higher-order corrections, they increase consistently towards larger $p_{T,\text{miss}}$, but not to the same extend.
As one might expect, the effects on the angular observables are even smaller and only reach a few percent. More importantly, these corrections are rather stable and we observe no significant change in the overall shapes of the angular distributions.

In general, the $t\bar{t}$ distributions change similarly to those for the $t\bar{t}Z$ process. The only notable exception is $M_{T2,W}$ due to the kinematic edge at $m_W$ (see right side of Figure \ref{fig:Bkg_Uncut_Modeling_MTW}). In the case of $t\bar{t}$, $M_{T2,W}$ is bounded from above by $m_W$ in the NWA since we have $M_{T,W} \leq m_W$ for both of the two transverse $W$ masses considered in the definition of $M_{T2,W}$, which is given in
Eq. (\ref{eq:MT2_W}).  However, if we allow the $W$ to be off-shell, the transverse masses are instead limited by the invariant mass, i.e.  $M_{T,W} \leq M_W$ so that $M_{T2,W} \leq \max \{ M_{W^+}, M_{W^-}\}$. As a result, we still have events with $M_{T2,W} > m_W$ in the off-shell case. 
Let us mention that the same is in principle true for $M_{T2,t}$ and its edge around the top-quark mass. However, since we associate $b$-jets and leptons by minimising the invariant masses, the two might not actually originate from the same top quark and such events must not necessarily adhere to the limit $M_{T2,t} \leq m_t$. This, in turn, allows for events with $M_{T2,t} > m_t$ even in the NWA, albeit much fewer than in the off-shell case. For the latter, single and non-resonant diagrams also contribute above this edge which further amplifies the tails compared to the NWA.

From this discussion we can conclude that higher-order corrections and off-shell effects can both substantially alter the behavior of the NP observables. This is particularly true for the tails of dimensionful observables.  As this is also the region which is used to distinguish the signal from the SM background, we should expect the modelling to have a significant impact on the event selection and the calculation of exclusion limits.

\subsection{Different central scale choices} \label{subsec:scale_choice}

\begin{figure*}
	\includegraphics[width=.5\linewidth]{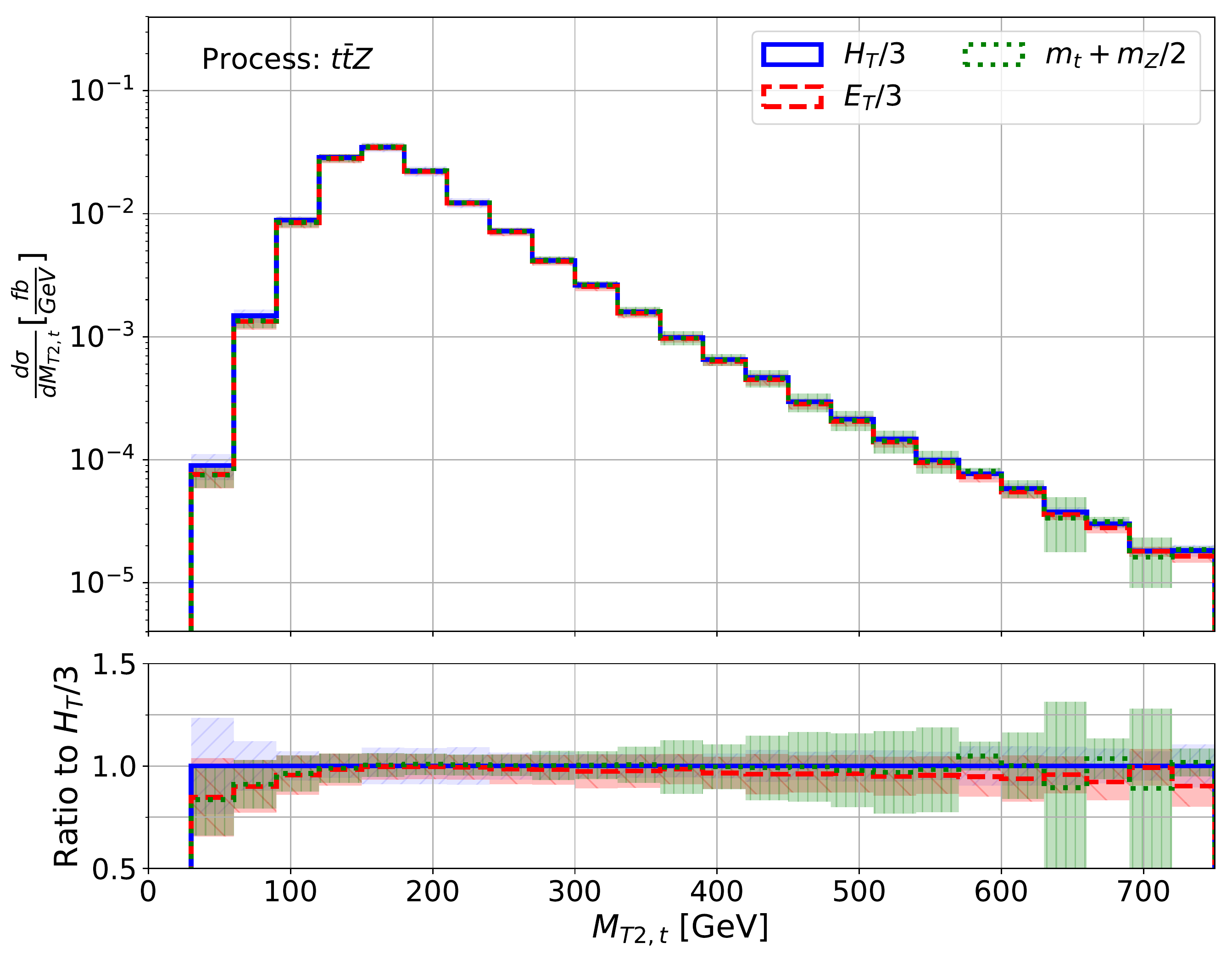}
	\includegraphics[width=.5\linewidth]{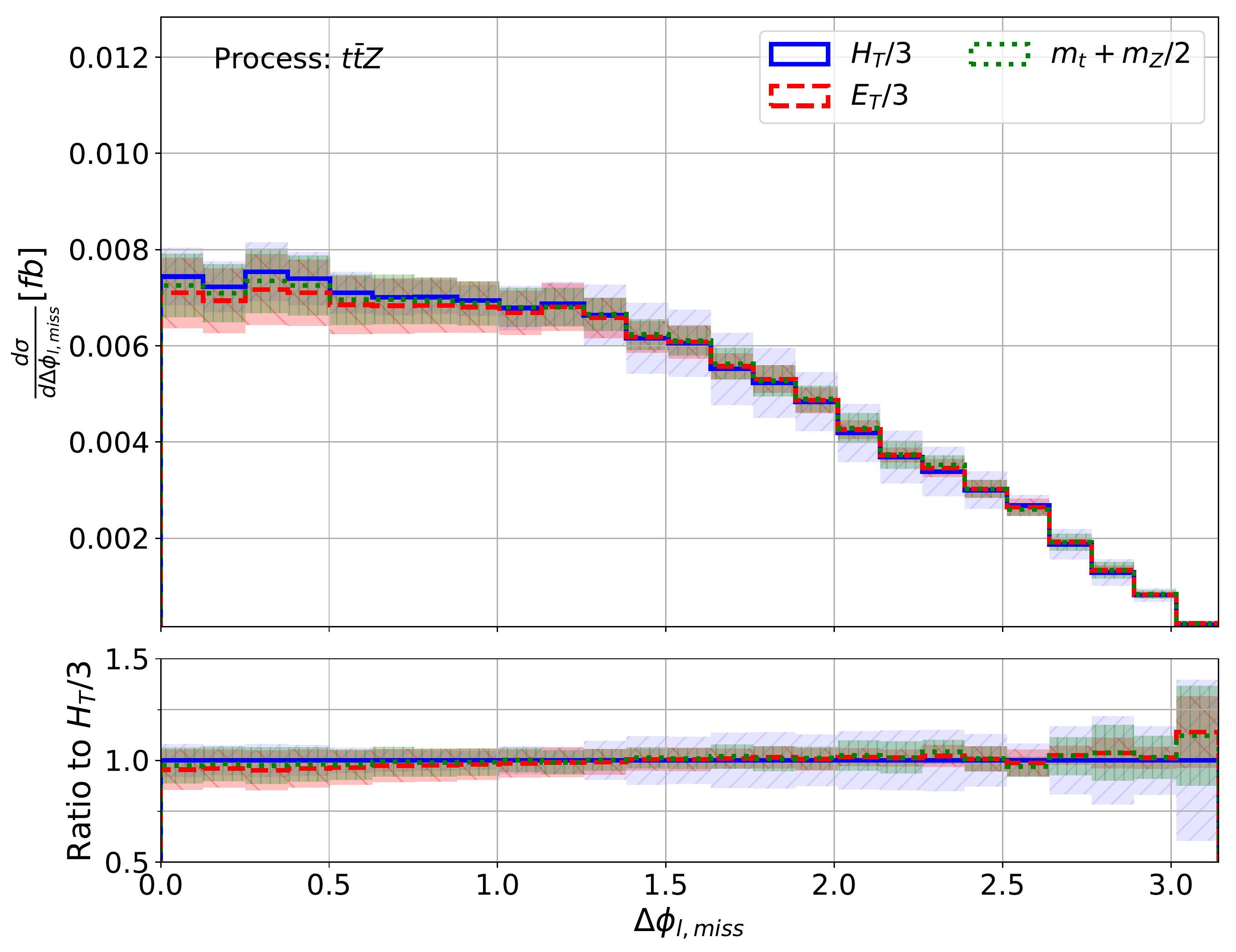}
	
	\caption{\textit{Comparison of differential NLO$_{\text{Off-shell}}$ distributions with different central scale choices for the $t\bar{t}Z$  background process. The samples have been generated with the NLO \textsc{CT14} PDF sets for the LHC with center of mass energy $\sqrt{s} = 13$ TeV. The error bands depict the respective scale uncertainties. In the lower panels we present the ratios to our default scale choice $H_T/3$.}}
	\label{fig:Bkg_Uncut_Scales_NLO}
\end{figure*}

In the final step of our assessment of the background processes, we want to briefly discuss the effects of choosing different central scales instead of $H_T/4$ and $H_T/3$. Most of the relevant distributions have  already been analysed in Ref. \citep{ttZ}. Thus,  we focus here on $M_{T2,t}$ and $\Delta \phi_{l,\text{miss}}$ which have not been previously discussed. The same is true for $M_{T2,W}$ and $C_{em,W}$ but we do not observe any significant differences for these observables. We should mention that the scale $E_T$ we use here corresponds to $E'_T$ in Ref. \citep{ttZ} with the minor change that we use the invariant  masses $M_i$ in the definition in Eq. (\ref{eq:scale_def}) instead of the on-shell masses $m_i$.

In Figure \ref{fig:Bkg_Uncut_Scales_NLO} we present the dependence of the above mentioned observables on the central scale choice for the NLO$_{\text{Off-shell}}$ $t\bar{t}Z$ background. In both cases, we only find minor changes in the distribution shapes which is mirrored by the corresponding $t\bar{t}$ distributions.  The main difference between the scales is the size of their respective scale uncertainties. In the high-$M_{T2,t}$ region, these are significantly larger for the fixed scale than the dynamical ones. For $\Delta \phi_{l,\text{miss}}$, we can observe the opposite behavior with $H_T$ yielding larger scale uncertainties for large angles. As exclusion limits are negatively impacted by large scale uncertainties, the right choice of the central scale is indeed relevant in their calculation.

\section{Modelling in the presence of exclusive cuts} \label{sec:Cut_effects}

Having established the general size and shape of the SM background in the previous section, we will now discuss the effects of applying a set of very exclusive selection cuts to the signal and background processes. Particular emphasis will be given to the impact of these additional cuts on the size of higher-order corrections and off-shell effects.

\subsection{Analysis strategy} \label{subsec:Cuts}
As one might expect, we make use of the kinematic edge in $M_{T2,W}$ as well as the shape differences of several other observables to significantly reduce the SM background, in particular $t\bar{t}$. For this, we follow the strategy outlined in Ref. \citep{Haisch_analysis} and employ the following additional cuts to both the signal and the SM background:

\begin{equation}
\begin{split}
	M_{T2,W}>90 \text{ GeV}, \;\;\;&\;\;\;p_{T,\text{miss}}>150 \text{ GeV}, \\
	\Delta \phi_{b,\text{miss}} > 0.2, \;\;\;&\;\;\;M_{ll} > 20 \text{ GeV}, \\
	\text{and} \;\;\; C_{em,W} &> 130 \text{ GeV}.
\end{split}
\end{equation}
Here, $\Delta \phi_{b,\text{miss}}$ is defined as the angle between the missing transverse momentum and the nearest $b$-jet, similar to the definition of $\Delta \phi_{l,\text{miss}}$. Note that only the cuts on $C_{em,W}$, $M_{T2,W}$ and $p_{T,\text{miss}}$ are actually used to suppress the background. The $\Delta \phi_{b,\text{miss}} > 0.2$ cut, on the other hand, is motivated experimentally and should limit the effect of $p_{T, {\rm miss}}$ resulting from $b$-jet-mismeasurement. The cut on $M_{ll}$ reduces effects from virtual $\gamma^* \to l^+ l^-$ splittings. Of course, the impact of this last cut is rather minor as we already imposed cuts on $p_{T,l}$ and $\Delta R_{ll}$. These are related to the $M_{ll}$ cut through

\begin{equation}
	\left(M_{ll}\right)_{\text{min}} = \left(p_{T,l}\right)_\text{min}
    \sqrt{2 \left(1 - \cos ((\Delta R_{ll})_{\rm min})\right)}\,.
\end{equation}
For the cuts specified in Eq. (\ref{eq:leptonic_cuts}) the latter already gives
$\left(M_{ll}\right)_{\text{min}} \approx 12$ GeV.

\subsection{Effects on cross sections} \label{subsec:xSec_Cut}

\begin{table*}
	\caption{\textit{Comparison of LO and NLO integrated cross sections for the two background processes in the NWA (top) and including full off-shell effects (bottom) before and after applying the additional cuts. All values are given for the LHC with a center of mass energy of $\sqrt{s} = 13\,\text{TeV}$. We employ the (N)LO \textsc{CT14} PDF set as our default PDF set. The numbers of events are given for an integrated luminosity of $L = 300 \,\text{ fb}^{-1}$ and include the lepton flavour factors ($4$ for the DM signal and $t\bar{t}$, and $12$ for $t\bar{t}Z$).}}
	\centering
	\begin{tabular}{lll@{\hskip 10mm}lll@{\hskip 10mm}l}
		\hline\noalign{\smallskip}
		Process  & Order & Scale & $\sigma_{\text{uncut}}$ [fb] & $\sigma_{\text{cut}}$ [fb] & $\sigma_{\text{cut}} / \sigma_{\text{uncut}}$ & Events for $L=300 \,\text{ fb}^{-1}$\\
		\noalign{\smallskip}\midrule[0.5mm]\noalign{\smallskip}
		\multirow{5}{*}{$t\bar{t}$ NWA}  & LO          & $H_T/4$    & $1061$    & $ 0 $    & $ 0.0 \%$  & $ 0 $\\
		& LO          & $E_T/4$    & $984$    & $ 0 $    & $ 0.0 \%$  & $ 0 $\\
		& LO          & $m_t$      & $854$    & $ 0 $    & $ 0.0 \%$  & $ 0 $\\
		& NLO         & $H_T/4$    & $1097 $   & $ 0 $    & $ 0.0 \%$    & $ 0 $\\
		& NLO, LO dec & $H_T/4$    & $1271 $   & $ 0 $    & $ 0.0 \%$    & $ 0 $\\
		\noalign{\smallskip}\hline\noalign{\smallskip}
		\multirow{5}{*}{$t\bar{t}Z$ NWA}  & LO         & $H_T/3$ & $0.1223 $ & $0.0130 $               & $ 11\%$     & $ 47$\\
		& LO         & $E_T/3$ & $0.1052 $ & $0.0116 $               & $ 11\%$     & $ 42$\\
		& LO         & $m_t+m_Z/2$ & $0.1094 $ & $0.0134 $               & $ 12\%$     & $ 48$\\
		& NLO        & $H_T/3$ & $ 0.1226 $  & $ 0.0130 $               & $ 11 \%$    & $ 47$\\
		& NLO, LO dec & $H_T/3$ & $ 0.1364 $  & $ 0.0140 $               & $ 10 \%$    & $ 50 $\\
		\noalign{\smallskip}\hline\noalign{\smallskip}
		\multirow{4}{*}{$t\bar{t}$ Off-shell}       & LO  & $H_T/4$ & $1067$ & $0.0144$ & $0.0013 \%$ & $17$\\
		& LO  & $E_T/4$ & $989 $ & $0.0131$ & $0.0013 \%$ & $16$\\
		& LO  & $m_t$   & $861$ & $0.0150$ & $0.0017 \%$ & $18$\\
		& NLO & $H_T/4$       & $1101 $ & $0.0156$ & $0.0014 \%$ & $19$\\
		\noalign{\smallskip}\hline\noalign{\smallskip}
		\multirow{4}{*}{$t\bar{t}Z$ Off-shell}  & LO  & $H_T/3$ & $0.1262 $ & $0.0135$ & $11 \%$ & $49$\\
		& LO  & $E_T/3$ & $0.1042 $ & $0.0115$ & $11 \%$ & $41$\\
		& LO  & $m_t+m_Z/2$ & $0.1135 $ & $0.0140$ & $12 \%$ & $50$\\
		& NLO  & $H_T/3$      & $0.1269$ & $0.0134$ & $11 \%$ & $48$\\
		\noalign{\smallskip}\hline
	\end{tabular}
	\label{table:total_xSec_NLO_cut}
\end{table*}

The impact that these more exclusive cuts have on the signal's and SM background's size is summarised in Table \ref{table:total_xSec_NLO_cut}. For the full off-shell predictions, we find that about $11 \%$ of $t\bar{t}Z$ events pass these cuts whilst only about one in $10^5$ $t\bar{t}$ events does so. As a result, the respective cross sections are very similar to each other after applying the extra cuts. If we now take into account the different lepton flavour factors for the two processes, $4$ for $t\bar{t}$ and $12$ for $t\bar{t}Z$, we actually end up with $t\bar{t}Z$ being the dominant SM background. This is in stark contrast to the naive expectation that $t\bar{t}$ should be the main background due to its significantly larger cross section before applying the analysis cuts (see Table \ref{table:total_xSec_FW_effects}). 

One could be led to conclude that the $t\bar{t}ZZ$ contribution we briefly mentioned earlier could be similarly enhanced and might thus also turn out to be an important background. However, even if every single $t\bar{t}ZZ$ event passed the additional cuts, we would still end up with $\mathcal{O}(10^{-2})$ fewer $t\bar{t}ZZ$ events compared to $t\bar{t}Z$ since the $\sigma_{t\bar{t}ZZ}$ contribution before the extra cuts is already three orders of magnitude smaller than $\sigma_{t\bar{t}Z}$. As this is well within the statistical uncertainties $\sqrt{N_{\text{Event}}}$ on the number of events $N_{\text{Event}}$, we do not need to consider $\sigma_{t\bar{t}ZZ}$ here. 

Assuming an integrated luminosity\footnote{If not stated otherwise, luminosity always refers to the integrated luminosity.} of $L = 300$ fb$^{-1}$, we get $66$ background events in total at LO and $67$ at NLO. The number of $t\bar{t}$ events is slightly larger at NLO than at LO due to amplified higher-order corrections for $t\bar{t}$. After the extra cuts are applied we have $K_{t\bar{t}} = 1.08$ compared to $K_{t\bar{t}} = 1.03$ before the cuts (see Section \ref{subsec:Bkg_xSec_and_gen}). This is a result of the large NLO corrections in the $p_{T,\text{miss}}$ tails meaning that for $p_{T,\text{miss}} > 150$ GeV the NLO corrections are much larger than for the full phase space. 
Though to a lesser extend, the same is true for the $t\bar{t}Z$ $p_{T,\text{miss}}$ distribution. However, the higher-order corrections in the high-$M_{T2,W}$ region are negative which seems to compensate the positive corrections in $p_{T,\text{miss}}$. As a result, we only find sub-percent NLO corrections to the integrated fiducial $t\bar{t}Z$ cross section.

The effects of the additional cuts on the top-quark background are even more severe in the NWA. Due to the missing tails in the $M_{T2,W}$ distribution we showed in Figure \ref{fig:Bkg_Uncut_Modeling_MTW}, not a single $t\bar{t}$ event passes the selection cuts, irrespective of the order at which we calculate $\sigma_{t\bar{t}}$. This would even be true if we used the NNLO predictions \citep{tt_NNLO_1,tt_NNLO_2,tt_NNLO_3} we mentioned in the introduction as the QCD corrections do not affect the $W$ decay and thus leave the kinematic edge in $M_{T2,W}$ unaltered. Let us mention that the same would have happened had we considered $tW$ production in the NWA as well, since $M_{T2,W} < m_W$ also holds for this process.
In contrast,  the off-shell effects for the integrated fiducial $t\bar{t}Z$ cross section are essentially the same as without the extra cuts, i.e. between $3\%-4 \%$. For the results with LO decays, they are slightly smaller than before at $8 \%$. So in the NWA, we have $47$ $t\bar{t}Z$ events at LO and for the full NLO. When considering NLO with LO decays this number is slightly higher at around $50$ events. These are also the total number of background events in all three cases.

At NLO, the  central scale choice has very little impact on the number of events. This is why
we do not even include different scale settings at NLO in Table \ref{table:total_xSec_NLO_cut}. At LO, however, we find that they are slightly lower if we use $E_T$ instead of $H_T$. This is mostly a result of the smaller overall cross section as the distribution shapes are very similar for the two dynamical scales. 
In contrast, using the fixed scale leads to larger contributions from the $p_{T,\text{miss}}$ tail which in turn yields an increased percentage of events passing the additional cuts. This compensates the smaller integrated cross sections before the cuts. Consequently, between the fixed and the $H_T$ scale setting the number of events only differs by two for $L = 300$ fb$^{-1}$ in the off-shell case.
\begin{figure*}
	\includegraphics[width=.5\linewidth]{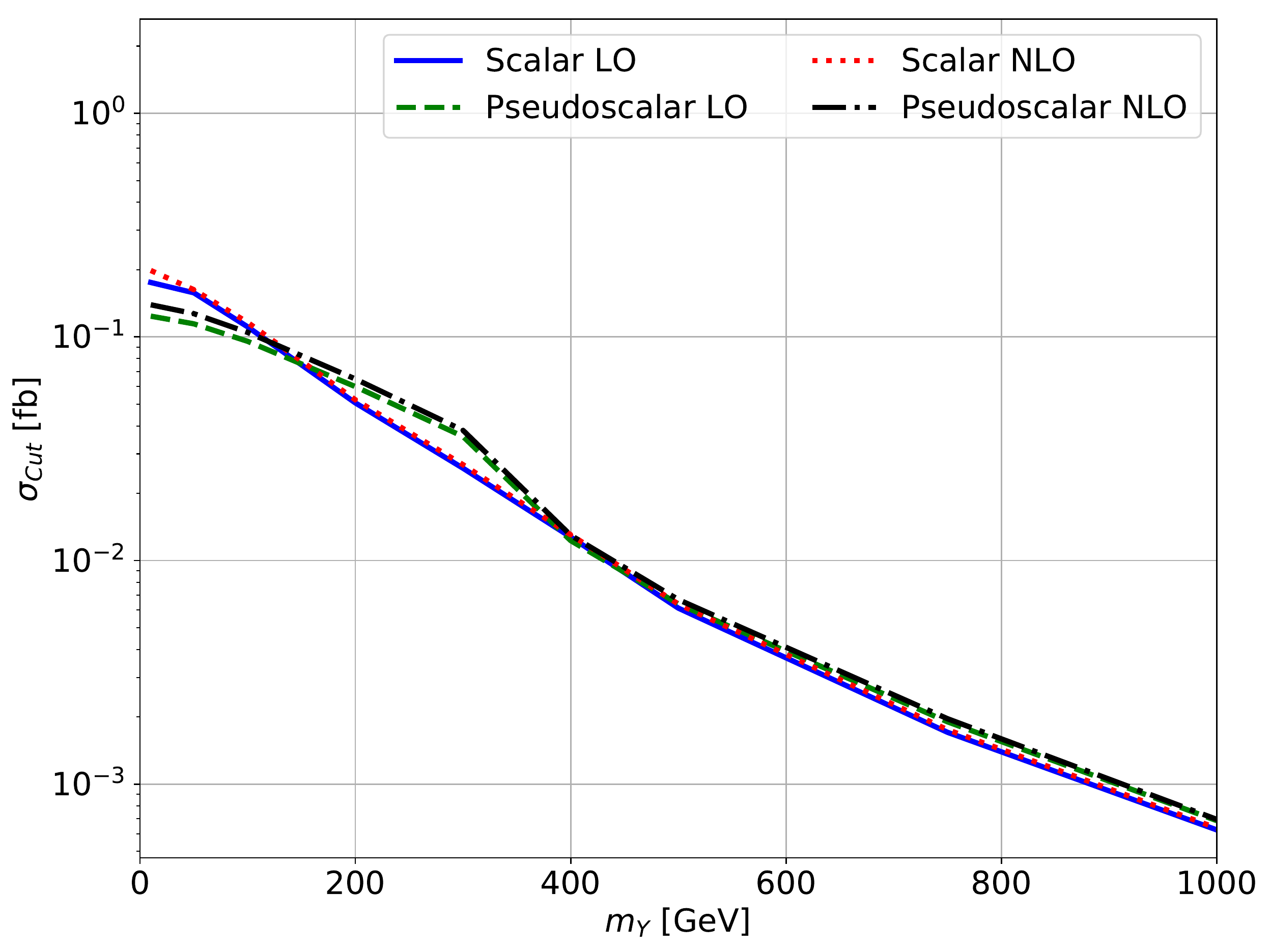}
	\includegraphics[width=.5\linewidth]{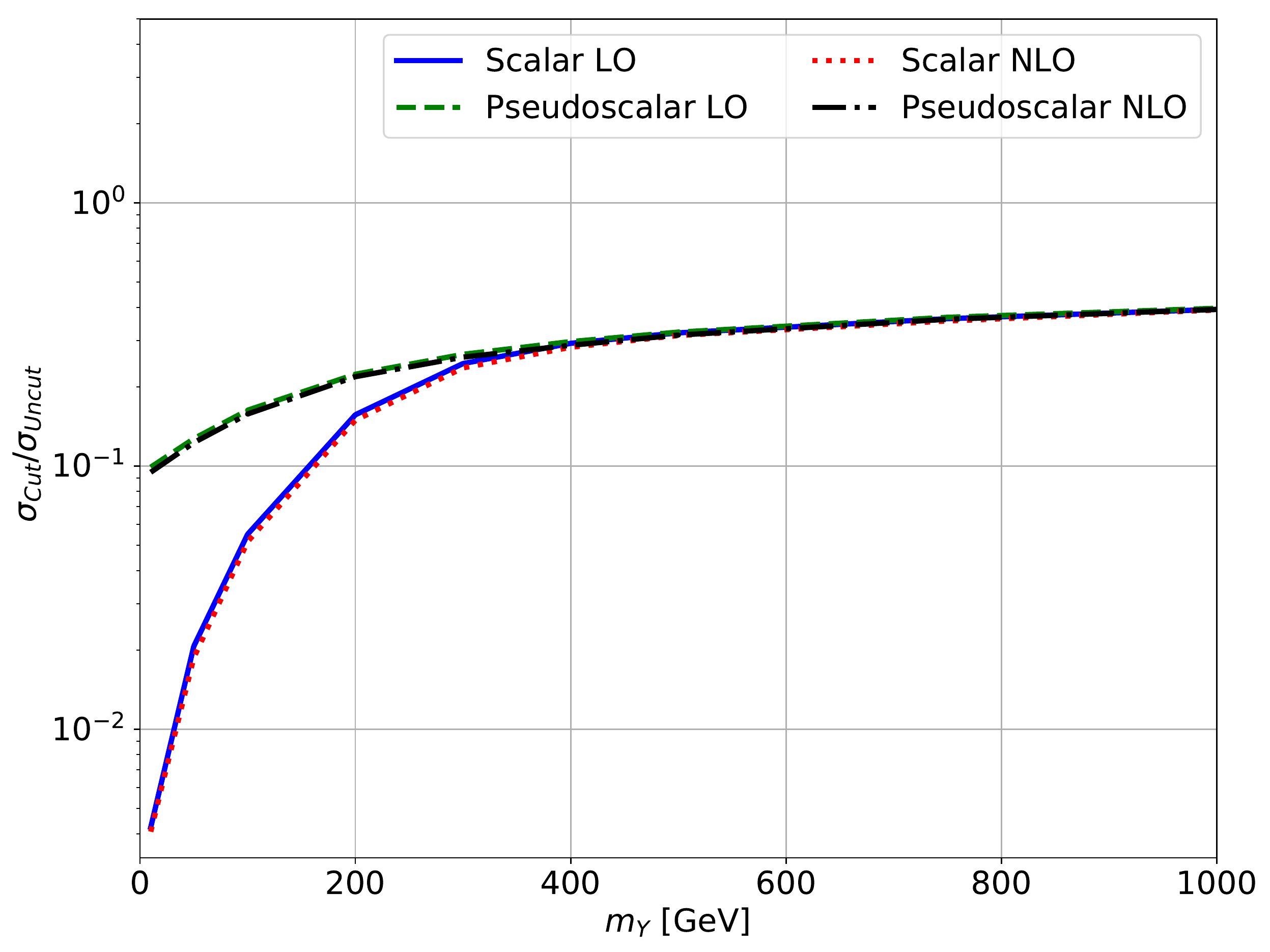}
	
	\caption{\textit{Production cross section for $p p \to b \bar{b} e^+ \mu^- \nu_e \bar{\nu}_\mu \chi \bar{\chi}$ for scalar and pseudoscalar mediators depending on the mass $m_Y$ of the mediator after applying the analysis cuts (left) and ratio between the cross sections before and after applying the additional cuts (right). The results have been generated using \textsc{MadGraph} for the LO / NLO production and \textsc{MadSpin} for the (LO) decays with the respective LO and NLO \textsc{CT14} PDF sets and a central scale $\mu_0^{\text{DM}} = E_T/3$.}}
	\label{fig:DM_total_xSec_Cut}
\end{figure*}

For the DM signal, the number of events depends heavily on the mediator mass. It ranges from about $700$ events for $m_Y = 10$ GeV to $2$ for $m_Y = 1$ TeV, with minor variations between scalar and pseudoscalar scenarios. The corresponding cross sections are plotted on the left hand side of Figure \ref{fig:DM_total_xSec_Cut}. If we compare this to our findings in Figure \ref{fig:DM_total_xSec}, we observe that the range of cross sections has been reduced significantly by about two orders of magnitude. This is a direct consequence of the different distribution shapes as these tend towards larger $p_{T,\text{miss}}$ and $M_{T2,W}$ values for heavier mediators which leads to more events passing the selection cuts. The percentage of events passing these cuts is shown on the right hand side of Figure \ref{fig:DM_total_xSec_Cut}. It spans from $0.4 \%$ for the lightest scalar mediator to $40\%$ for the heaviest one.

\subsection{Effects on distribution shapes}

\begin{figure*}
	\includegraphics[width=.5\linewidth]{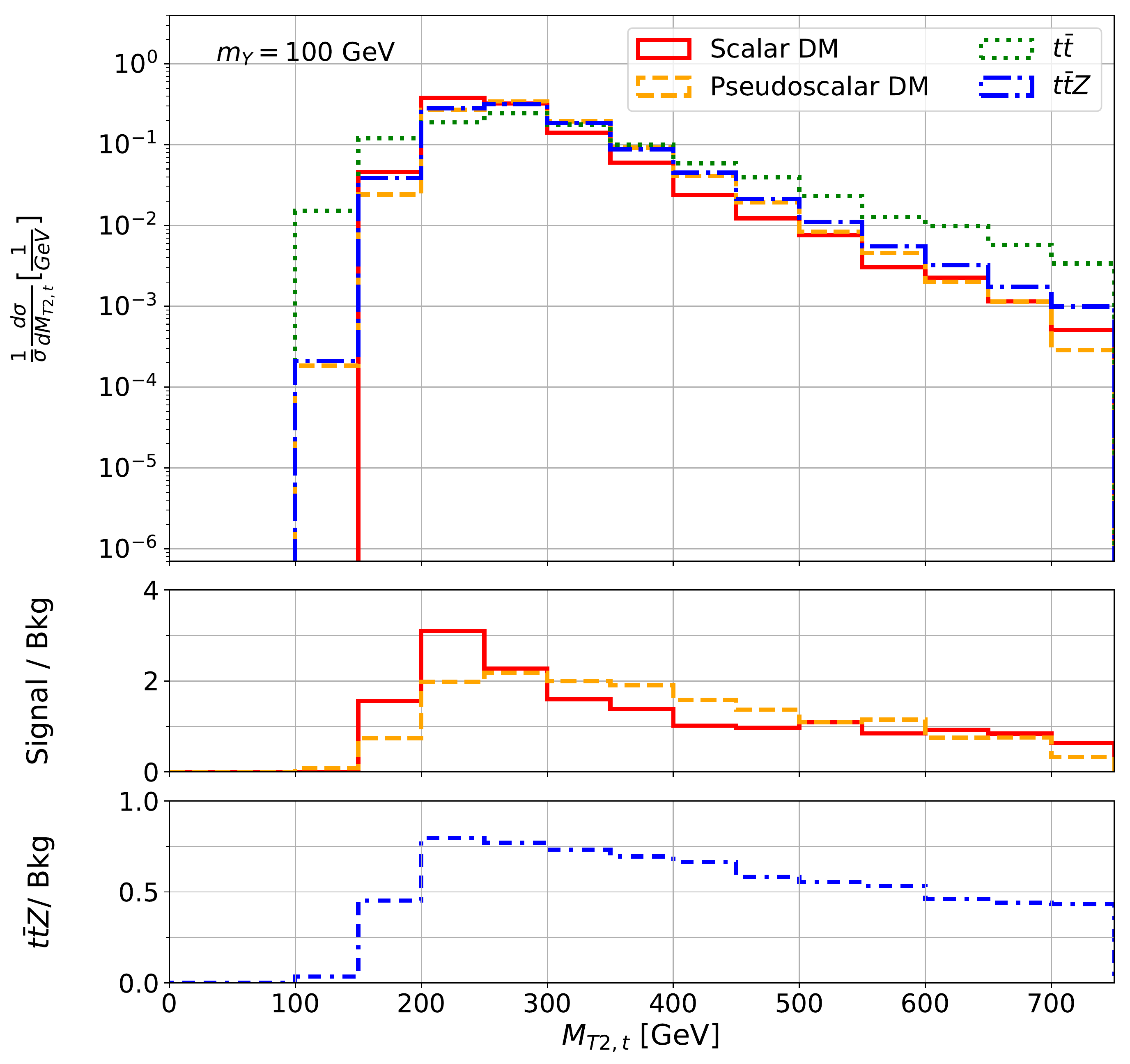}
	\includegraphics[width=.5\linewidth]{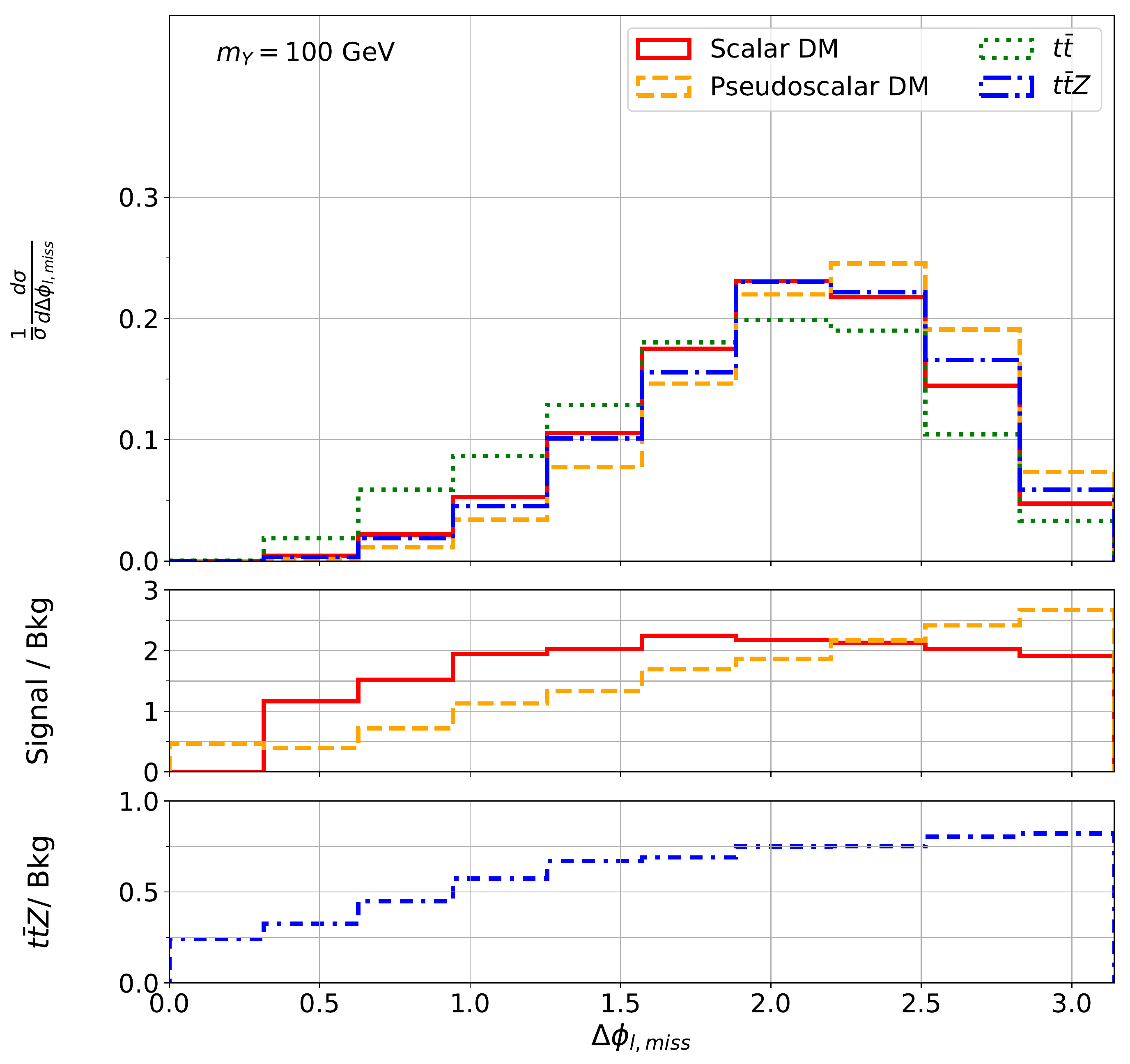}
	\caption{\textit{Comparison of normalised NLO differential distributions for the off-shell $t\bar{t}$ and $t\bar{t}Z$ background processes as well as scalar and pseudoscalar DM signals with $m_Y = 100$ GeV after applying the analysis cuts. The samples have been generated using the NLO \textsc{CT14} PDF set and our default scale choices for the LHC with center of mass energy $\sqrt{s} = 13$ TeV. In the respective central panels we show the signal-to-background ratio including the respective lepton flavour factors. The lower panels depict the fraction of the $t\bar{t}Z$ contribution to the total background.}}
	\label{fig:StBR_NLO_S_100_Cut}
\end{figure*}

In addition to the total number of events, we also want to discuss the effects that these additional cuts have on the shapes of various signal and background distributions. In Figure \ref{fig:StBR_NLO_S_100_Cut} we present normalised distributions for $M_{T2,t}$ and $\Delta \phi_{l,\text{miss}}$, just like in Figure \ref{fig:StBR_NLO_S_100_Uncut} but this time with the more exclusive cuts. One can clearly see that most of the shape differences present in Figure \ref{fig:StBR_NLO_S_100_Uncut} have disappeared, even for dimensionful observables such as $M_{T2,t}$. As a result, the signal-to-background ratio changes much less dramatically than without the additional cuts. This is a consequence of the drastic reduction in $t\bar{t}$ events as these previously dominated the signal-to-background ratio. The now dominant $t\bar{t}Z$ distributions were already much more similar to the signal before applying any additional cuts. The changes in $p_{T,\text{miss}}$ are very similar to those for $M_{T2,t}$.

Not only dimensionful observables are affected though.	
The change in $\Delta \phi_{l,\text{miss}}$ is also very notable with all distributions now peaking around $\sim 2.2 - 2.3$ instead of simply falling off towards larger angles and generally being much more akin to each other. 

These findings will make it harder to distinguish the signal from the SM background when calculating exclusion limits. On the other hand, $\cos ( \theta^*_{ll})$, the other angular observable that we are considering, has already been shown to keep its discriminating properties in distinguishing signal and background as well as the mediator parities \citep{Haisch_analysis}. Hence, this might be a more promising observable for calculating exclusion limits.

\begin{figure*}
	\includegraphics[width=.5\linewidth]{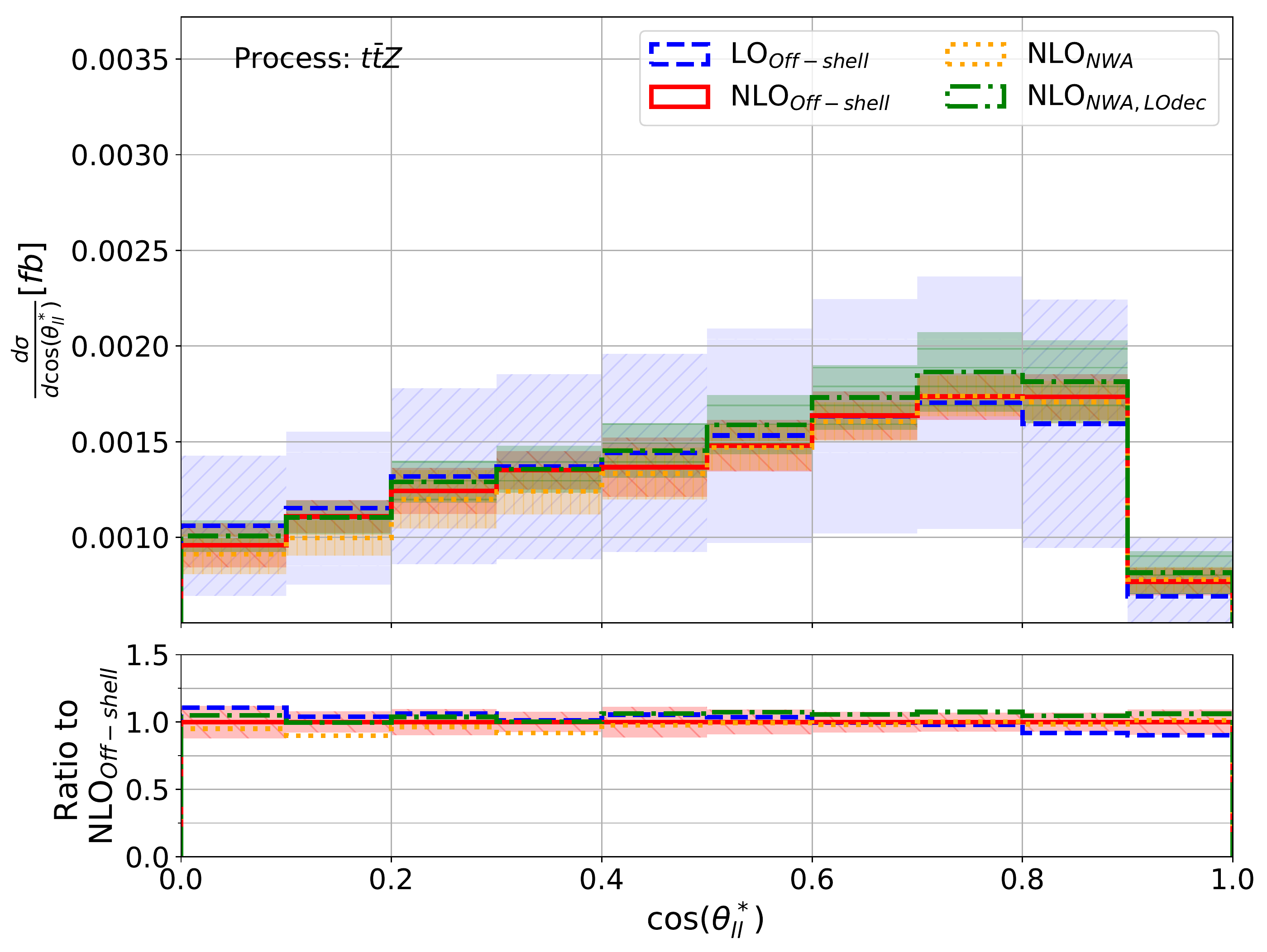}
	\includegraphics[width=.5\linewidth]{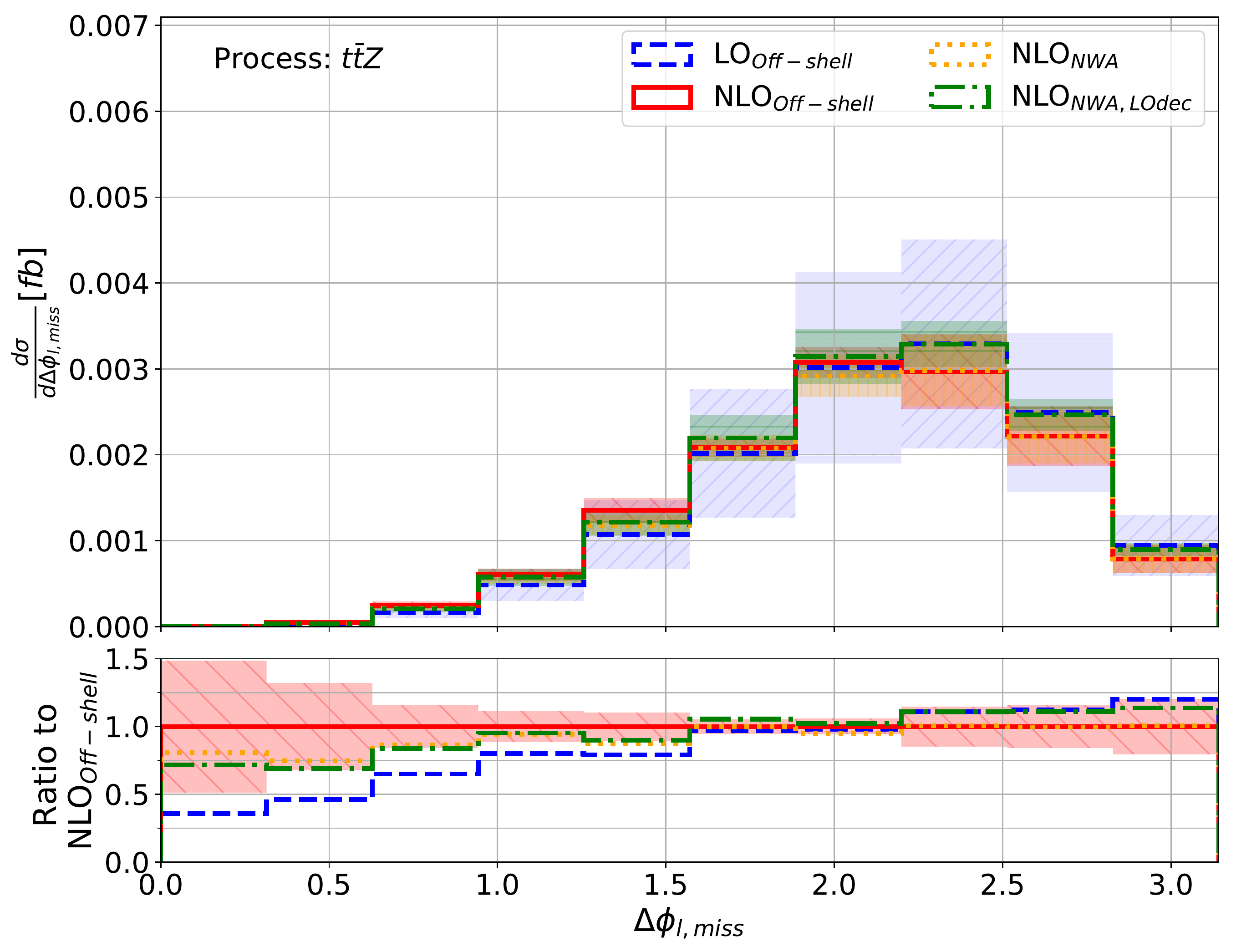}
	\includegraphics[width=.5\linewidth]{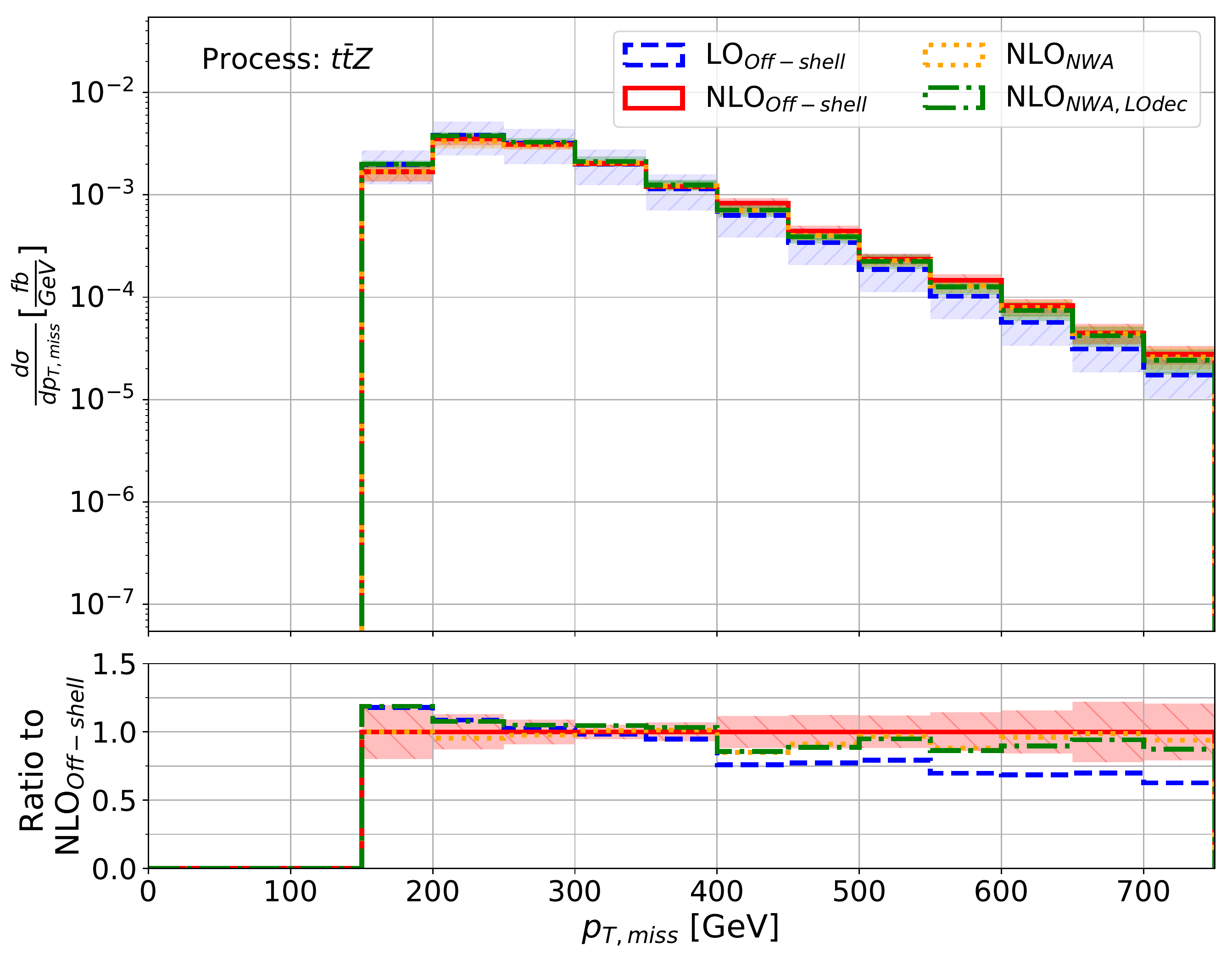}
	\includegraphics[width=.5\linewidth]{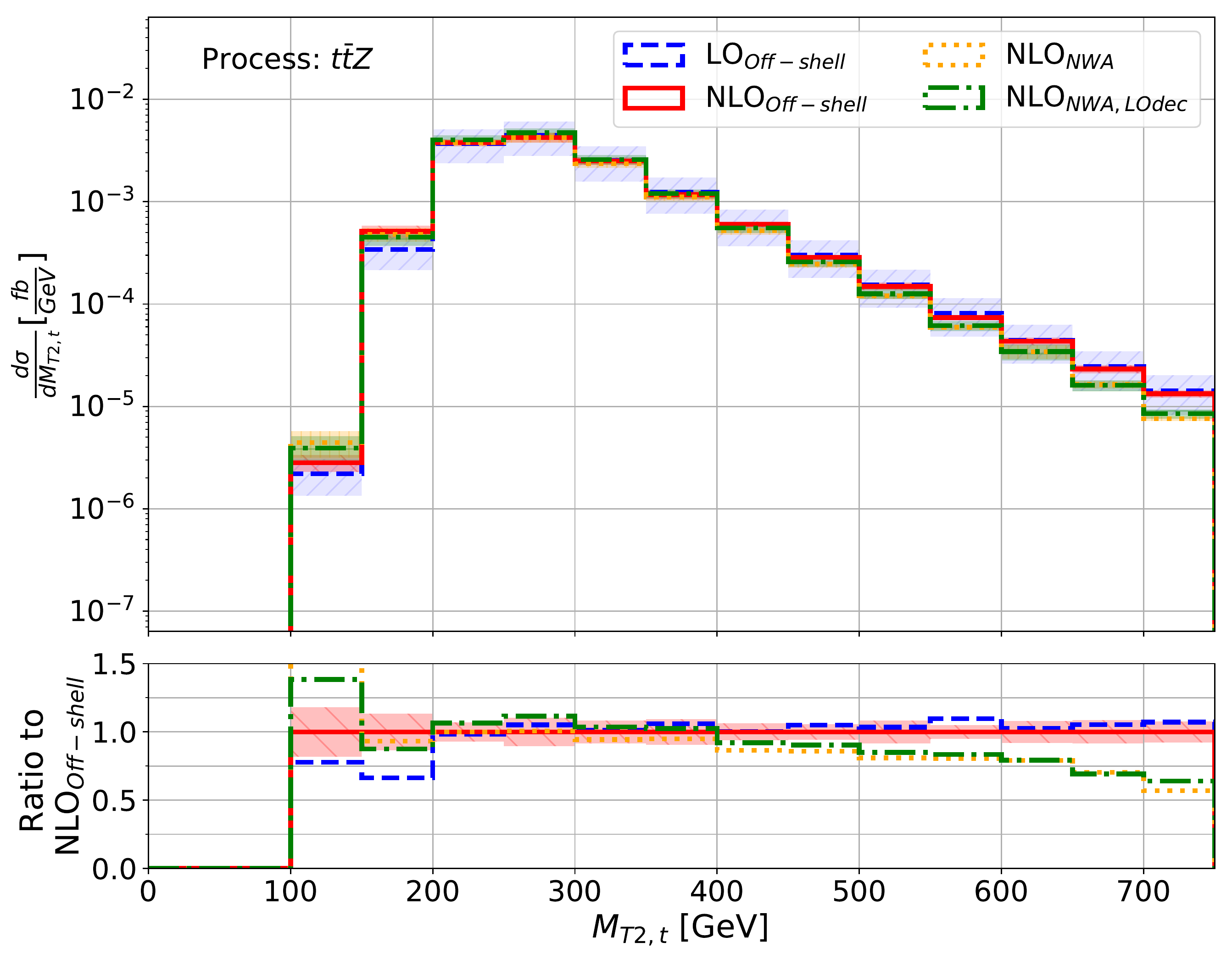}
	
	\caption{\textit{Comparison of differential distributions for the $t\bar{t}Z$ background process for different modelling approaches of the background after applying the analysis cuts. The samples have been generated using a central scale of $\mu^{t\bar{t}Z}_0 = H_T/3$ with the NLO \textsc{CT14} PDF sets for the LHC with center of mass energy $\sqrt{s} = 13$ TeV. The error bands depict the respective scale uncertainties. In the lower panels we present the ratios to the NLO$_{\text{Off-shell}}$ results.}}
	\label{fig:Bkg_Cut_Modeling_ttZ}
\end{figure*}

The way we model the background does not change this fact. Even with the more exclusive cuts, higher-order corrections and off-shell effects remain within a few percent for $\cos ( \theta^*_{ll})$, as can be seen from Figure \ref{fig:Bkg_Cut_Modeling_ttZ}. However, for $\Delta \phi_{l,\text{miss}}$ we find that both types of effects are significantly enhanced, even though this is an angular observable. For small angles, the $K$-factor can reach a value of up to $3$. We should note though that in this region the differential cross section is quite small for both LO and NLO. Off-shell corrections for this observable are also largest for small angles and reach up to $25 \%$.

In contrast, we find exactly the opposite phenomenon in $p_{T,\text{miss}}$ and $M_{T2,t}$. The off-shell effects remain below $20 \%$ for $p_{T,\text{miss}}$ and below $45 \%$ for $M_{T2,t}$. In both cases, this is much less significant than before, especially for $M_{T2,t}$. For the latter, the effects reach up to $75 \%$ without the additional cuts (compare Figure \ref{fig:Bkg_Uncut_Modeling_ttZ}). NLO QCD corrections are also reduced by the event selection and now the $K$-factor only reaches $1.6$ in $p_{T,\text{miss}}$ instead of $2.6$ without the additional cuts.

Let us mention that the bin sizes in Figure \ref{fig:Bkg_Cut_Modeling_ttZ} are larger than the ones in Figure \ref{fig:Bkg_Uncut_Modeling_ttZ} because our statistic is much smaller here due to the selection cuts.

As we have mentioned previously, the $t\bar{t}$ contribution vanishes in the presence of exclusive cuts when the NWA is employed. This makes off-shell effects indispensable for this process.
Concerning the higher-order corrections, we find a similar behavior as for the $t\bar{t}Z$ process with reduced corrections for $p_{T,\text{miss}}$ and $M_{T2,t}$ whilst they are slightly enhanced for $\Delta \phi_{l,\text{miss}}$.

\begin{figure*}
	\includegraphics[width=.5\linewidth]{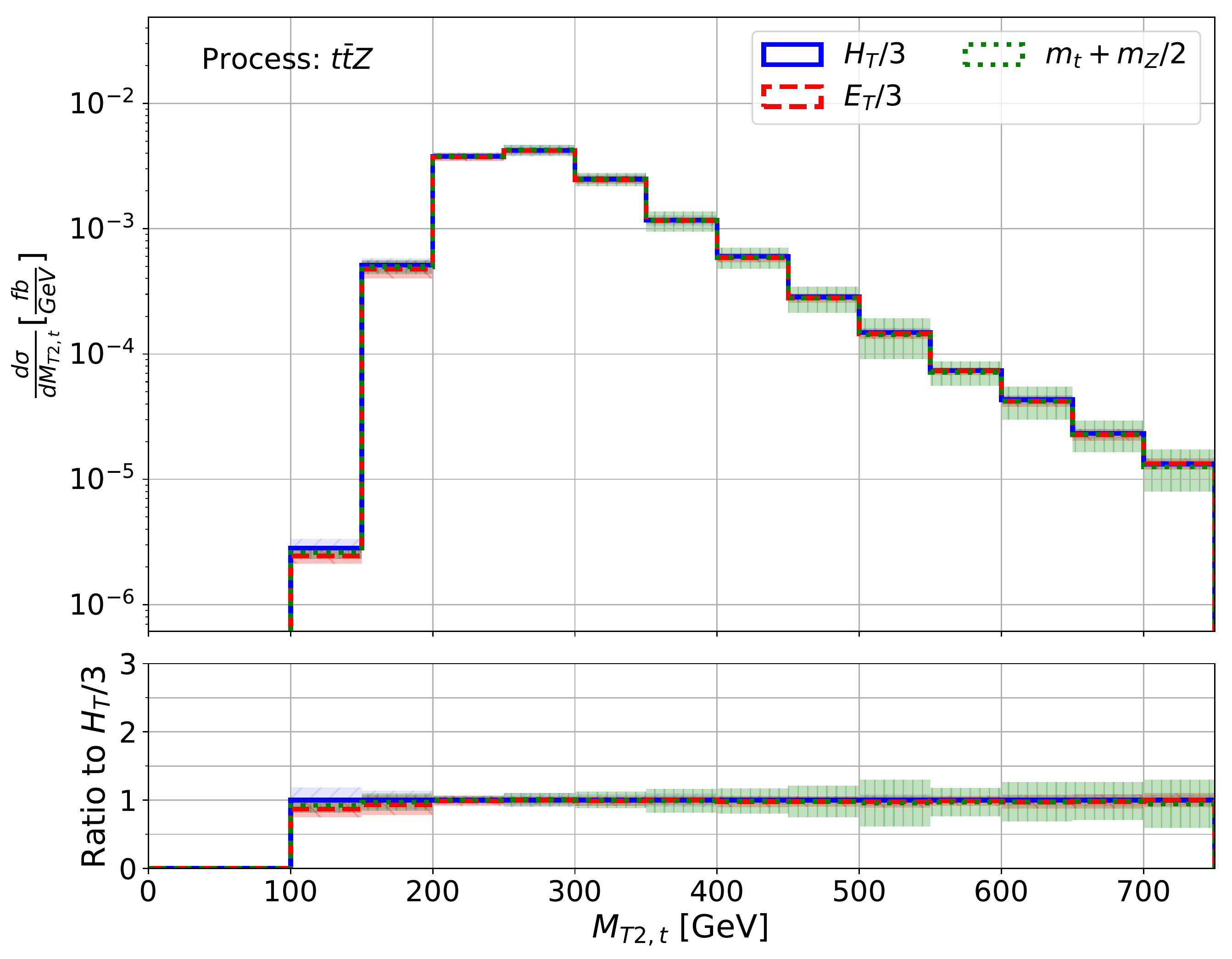}
	\includegraphics[width=.5\linewidth]{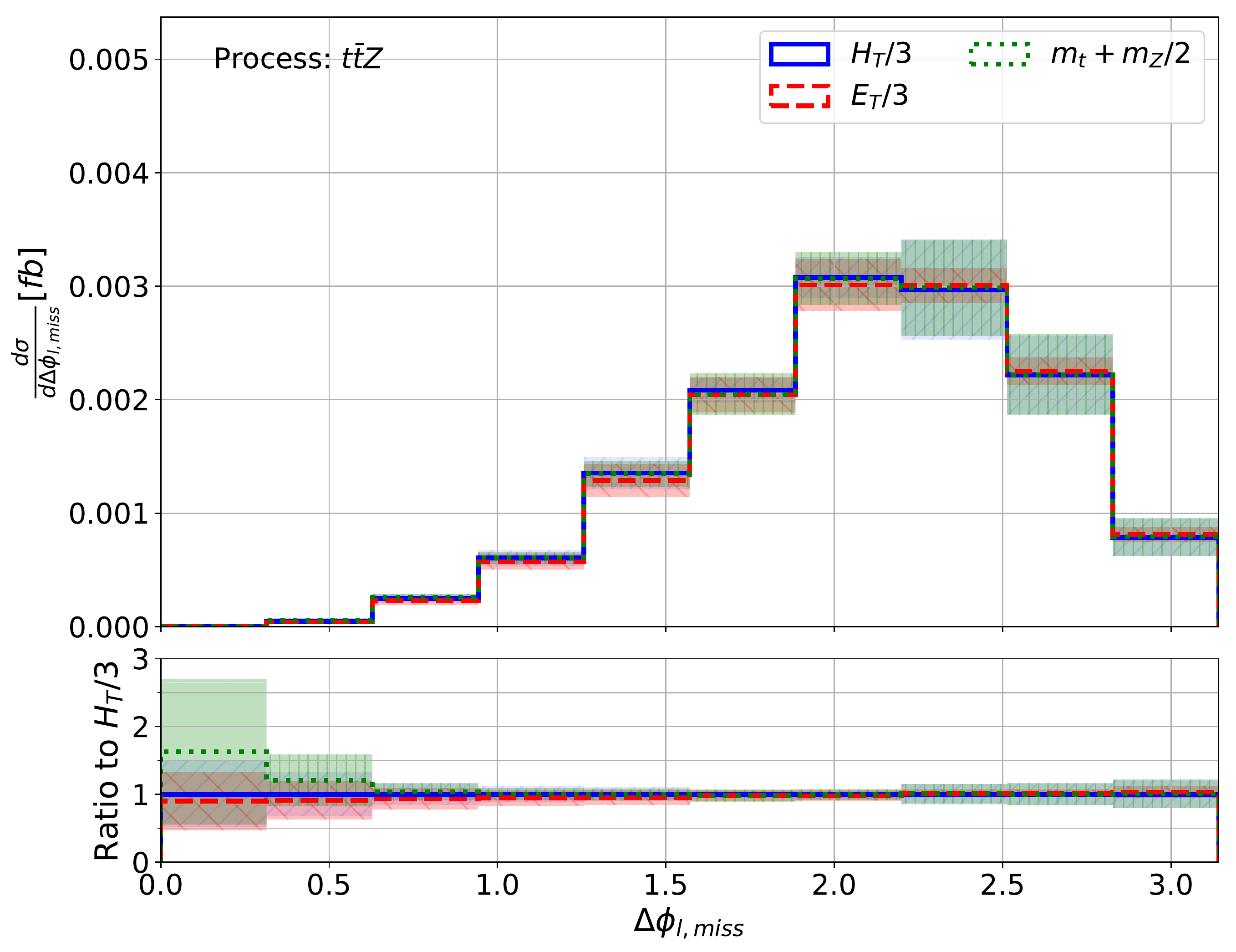}
	
	\caption{\textit{Comparison of differential NLO distributions with different central scale choices for the $t\bar{t}Z$  background process after applying the analysis cuts. The samples have been generated with the NLO \textsc{CT14} PDF sets for the LHC with center of mass energy $\sqrt{s} = 13$ TeV. The error bands depict the respective scale uncertainties. In the lower panels we present the ratios to out default scale choice $H_T/3$.}}
	\label{fig:Bkg_Cut_Scales_NLO}
\end{figure*}

For the dependence on the central scale choice which is shown in Figure \ref{fig:Bkg_Cut_Scales_NLO}, the changes are mostly limited to the normalisation, as listed in Table \ref{table:total_xSec_NLO_cut}. For LO and NLO predictions the shape differences are largely the same as without the selection cuts. However, the scale uncertainties for the fixed scale are significantly enhanced which is especially visible in $M_{T2,t}$. Even at NLO, they reach up to $37 \%$ for the fixed scale compared to $18 \%$ for $H_T$ and $15\%$ for $E_T$. 

Overall, we find that the more exclusive cuts have achieved what they are designed to do and the signal-to-background ratio has been significantly increased.
However, due to its similarity to the signal, the $t\bar{t}Z$ background is much less affected by the additional cuts than the a priori dominant top-quark background. This prevents us from further improving this ratio. Higher-order and off-shell effects are both reduced for $t\bar{t}Z$ when a suitable dynamical scale is chosen. In stark contrast, we actually have no contribution at all from $t\bar{t}$ in the NWA which results in a larger signal-to-background ratio. This should in principle result in more stringent limits compared to the off-shell case and would mean that using the NWA leads to underestimated limits on the signal strength or, conversely, overestimated limits on the mediator mass.

\section{Signal strength exclusion limits} \label{sec:Limits}
In this final part of our analysis, we evaluate whether our initial assumptions concerning off-shell effects and higher-order corrections and their role in calculating exclusion limits are indeed correct.
To this end, we compute signal strength exclusion limits $\mu^{95 \% \, CL}$ for our DM model using the \textsc{HistFitter} \citep{HistFitter} implementation of the $CL_s$-method \citep{CLs_ALRead}.
All values are computed for a $95 \%$ confidence level, i.e. $CL_s (\mu^{95 \% \,CL}) = 0.05$. This means that for a fixed DM model, all signal strengths $\mu > \mu^{95 \% \,CL}$ are said to be excluded at 95\% CL. Alternatively, one can turn this around and exclude all masses that yield a signal strength smaller than some reference value, usually $\mu^{95 \% \,CL} = 1$. In the following, we will primarily discuss the former interpretation and make comments on the mass limits where appropriate. Since this is primarily an analysis of the background, we always use the NLO  predictions for the $t\bar{t}+Y_{S/PS}\to t\bar{t}\, \chi\bar{\chi}$ signal, independently of the approach applied for the background modelling. For the computation we use five different observables: the integrated fiducial cross section $\sigma_{tot}$, $p_{T,\text{miss}}$, $M_{T2,t}$, $\cos (\theta^*_{ll})$, and $\Delta \phi_{l,\text{miss}}$.
The latter four have been chosen since they have exhibited significant shape differences between the DM signal and the SM background. 
On the other hand, $\sigma_{tot}$, which simply corresponds to the total number of events, is used as a reference value for the other observables. 
For each of these we take five equidistant bins which seems to be a good compromise between larger differences in the shape, the number of events in each bin and the runtime.  
The specific binnings used for each observable are summarised in Table \ref{table:Obs_binning}.
We also tried finer and coarser binnings but only found minor differences, if any. 
However, when going to too fine binnings one runs into the problem that Monte-Carlo uncertainties start to become relevant and misbinning\footnote{See e.g. Ref. \citep{Misbinning_Heymes_PhD} for an explanation of misbinning and Gaussian smearing. The latter is designed to combat the problem of misbinning.} can appear.

\begin{table}
	\renewcommand{\arraystretch}{2}
	\caption{\textit{Binning of the differential distributions used for the calculation of signal strength exclusion limits.}}
	\centering
	\begin{tabular}{l@{\hskip 10mm}l}
		\hline\noalign{\smallskip}
		Observable	&	Binning \\
		\noalign{\smallskip}\midrule[0.5mm]\noalign{\smallskip}
		$ \cos \left( \theta^*_{ll} \right) $ & $[0., 0.2, 0.4, 0.6, 0.8, 1.]$\\
		$ \Delta \phi_{l,\text{miss}} $		& $[0 , \pi/5, 2\pi/5, 3\pi/5, 4\pi/5, \pi]$\\
		$ p_{T,\text{miss}} $ 				& $[150, 250, 350, 450, 550, 650]$\\
		$ M_{T2,t} $ &  $[150, 250, 350, 450, 550, 650]$\\
		\noalign{\smallskip}\hline
		
	\end{tabular}
	\label{table:Obs_binning}
\end{table}

\subsection{Choice of observable} \label{subsec:Limits_Obs_choice}

\begin{figure*}
	\includegraphics[width=.5\linewidth]{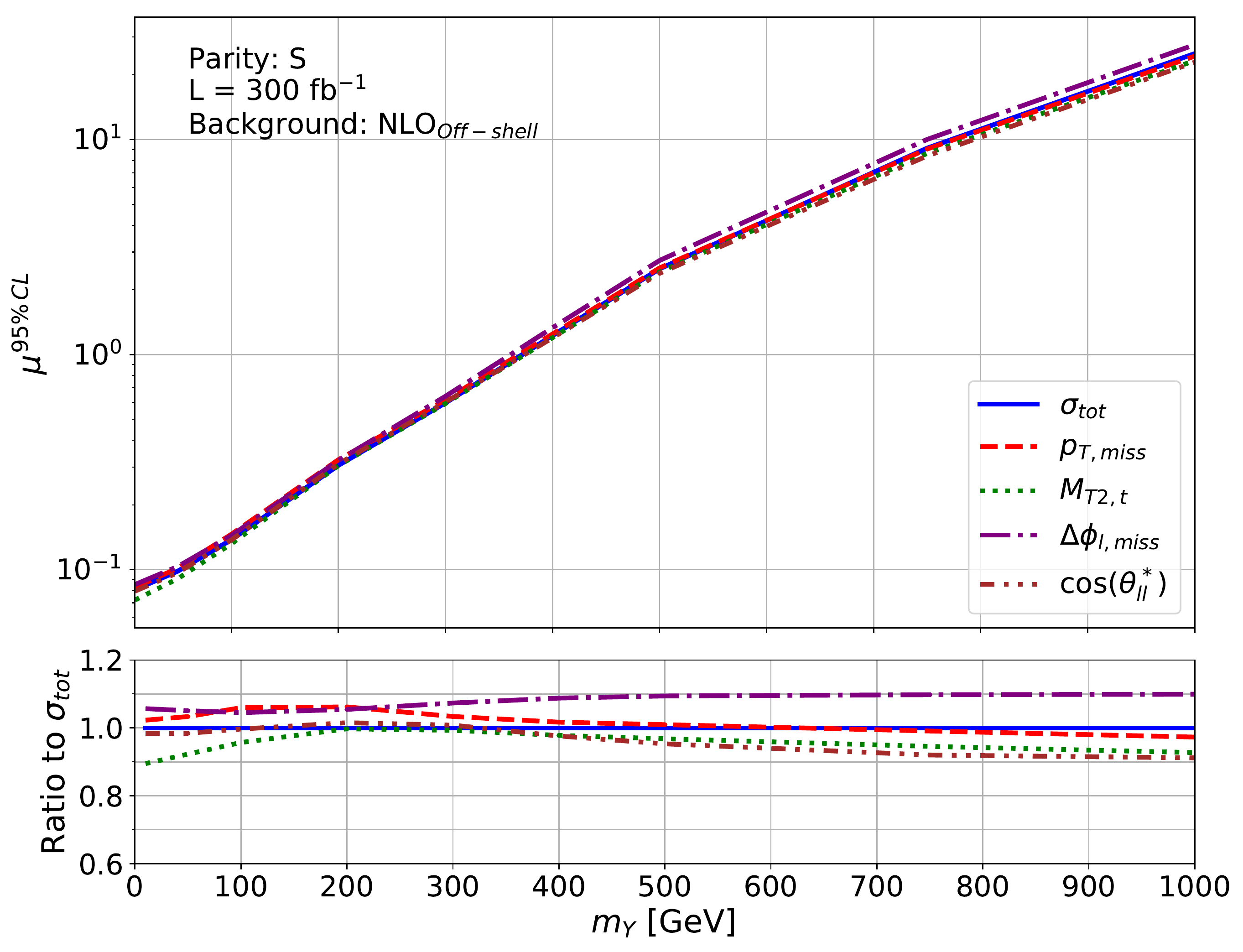}
	\includegraphics[width=.5\linewidth]{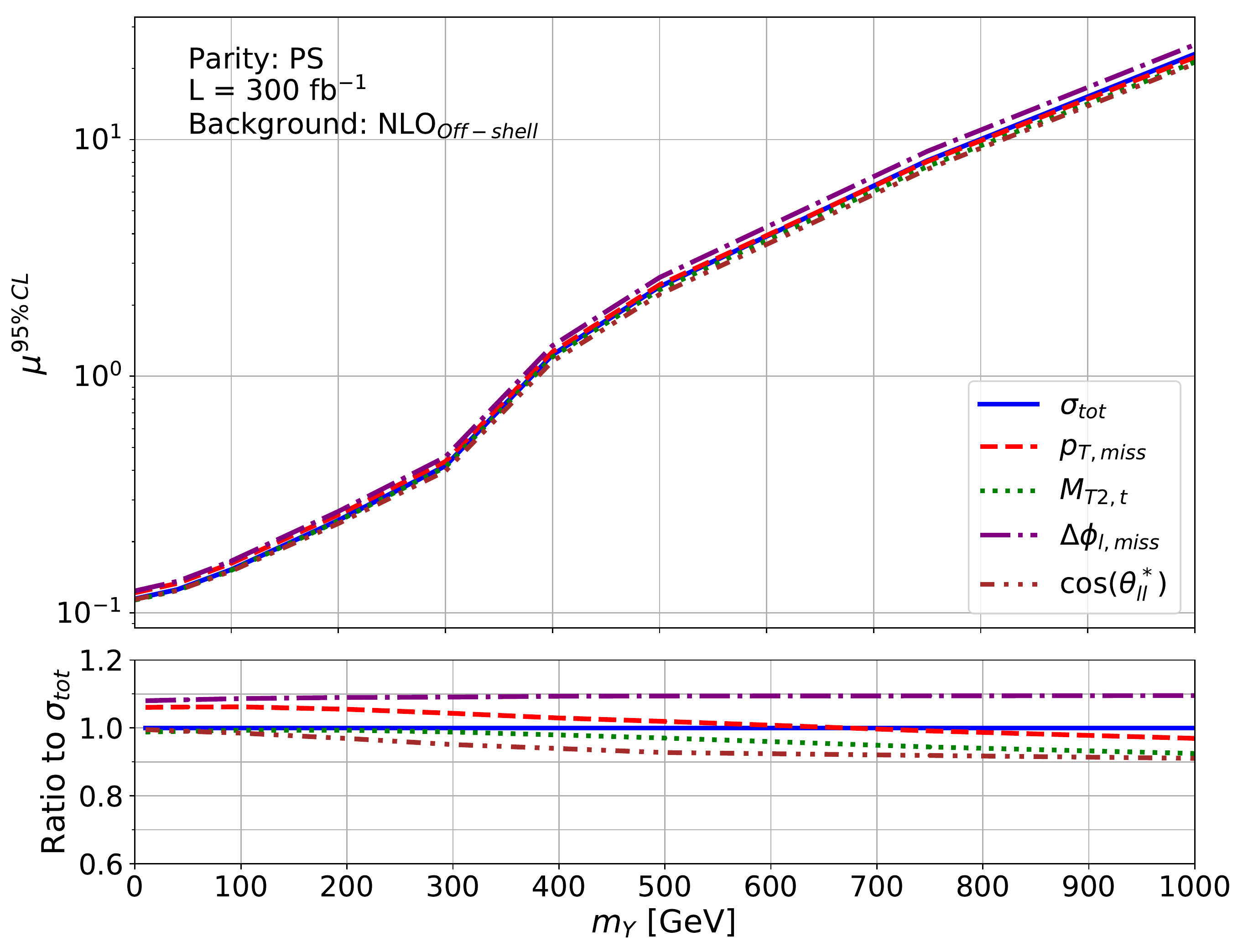}
	\includegraphics[width=.5\linewidth]{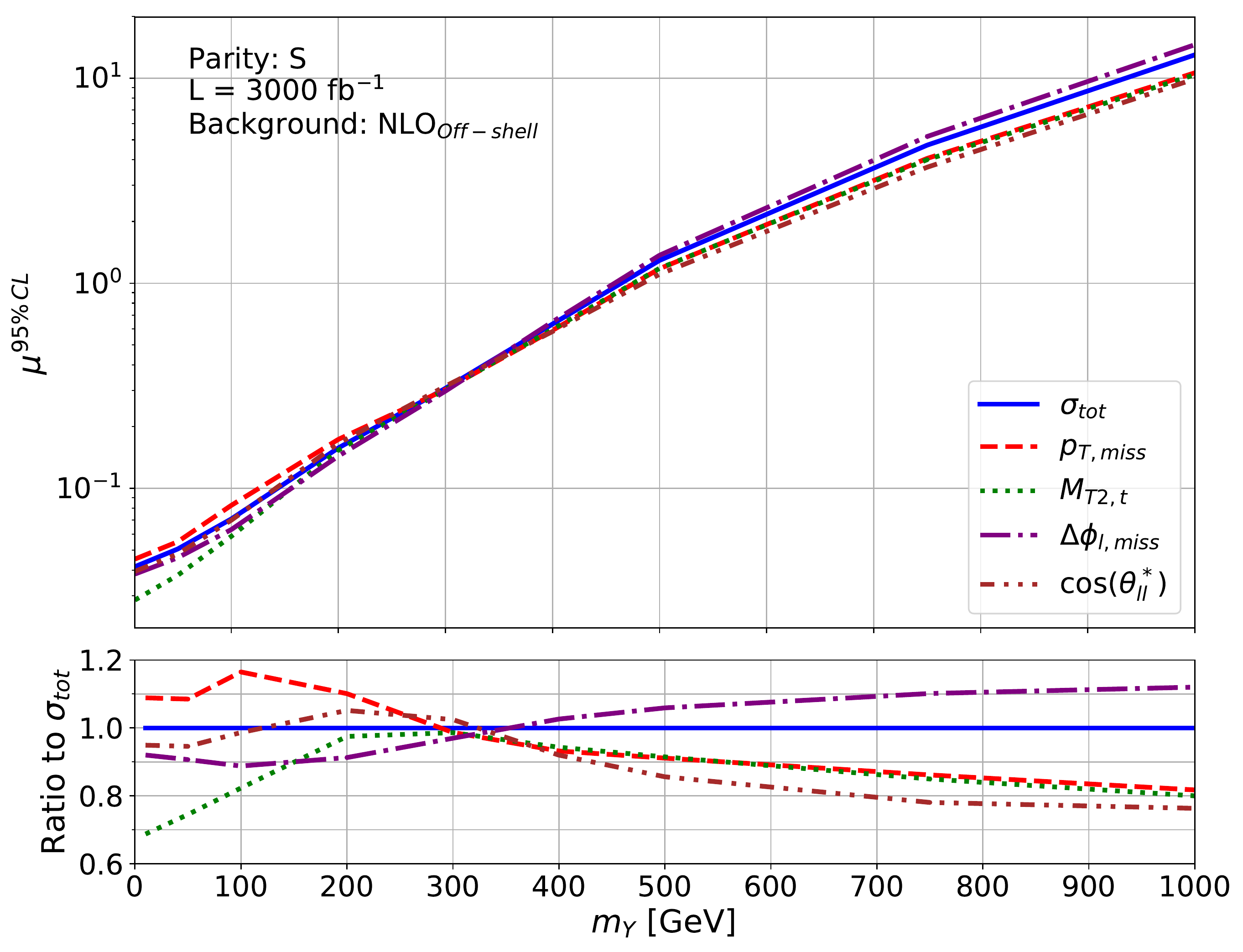}
	\includegraphics[width=.5\linewidth]{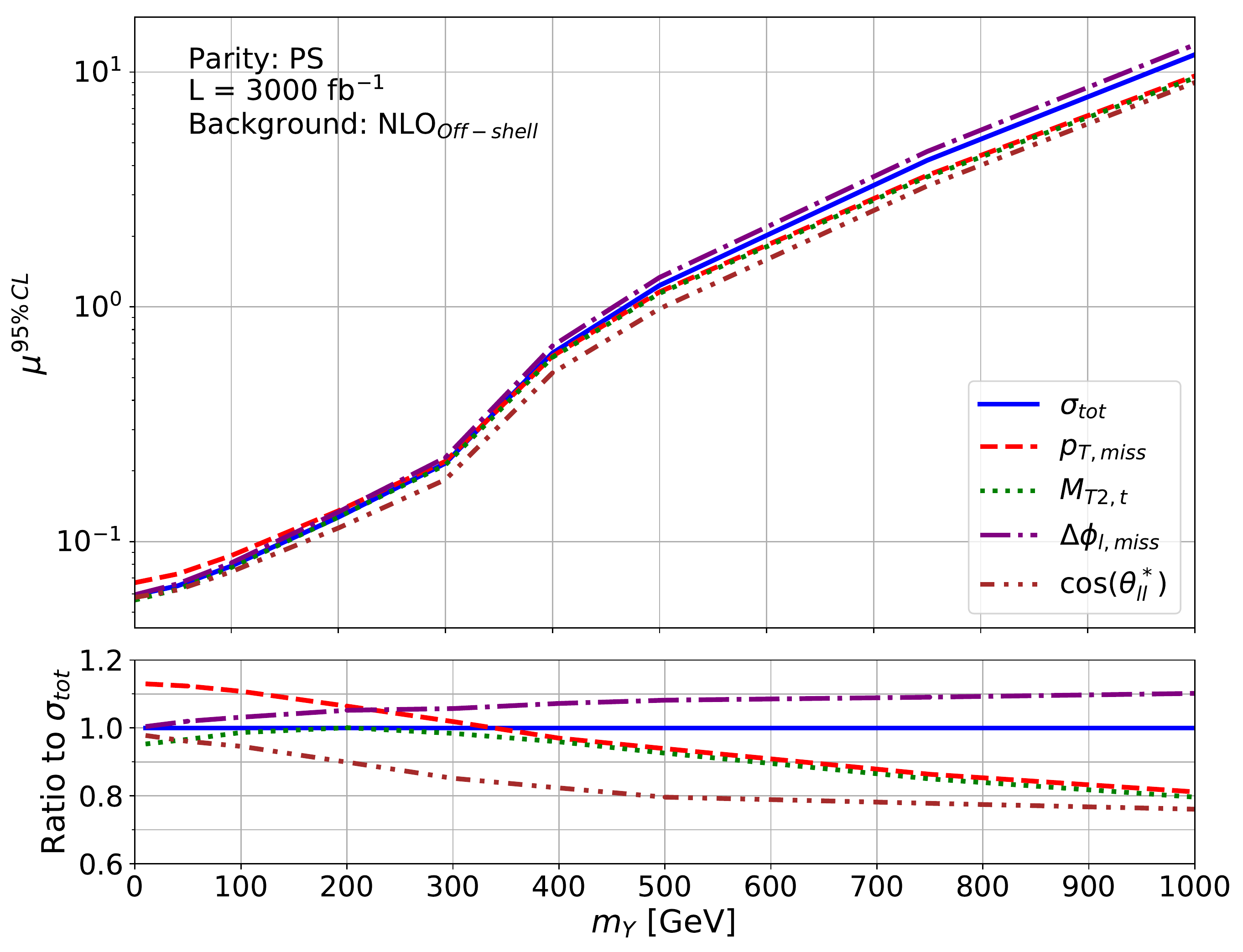}
	
	\caption{\textit{Comparison of signal strength exclusion limits computed with different observables for the scalar (left) and pseudoscalar (right) mediator scenario assuming luminosities of $L = 300 \text{ fb}^{-1}$ (first row) and $L = 3000 \text{ fb}^{-1}$ (second row).  In the lower panels we present the ratios to the limits obtained using just the integrated fiducial cross section.}}
	\label{fig:Limits_Obs_comp}
\end{figure*}
To begin our evaluation of exclusion limits, we first take a look at which observable provides the best, i.e. the most stringent, limits on the signal strengths for the full off-shell NLO background. The latter is the most precise background prediction that we have available so we will use it as our reference in the following.
 
The exclusion limits for all five observables are plotted in Figure \ref{fig:Limits_Obs_comp} depending on the mediator mass. We assume luminosities of $L = 300 \text{ fb}^{-1}$ (first row) and $L = 3000 \text{ fb}^{-1}$ (second row) for the calculation. In the lower panels of each plot we show the ratio to the limits obtained without any shape information, i.e. just using the total number of events.

In principle the binned observables should yield stronger limits than the total number of events as they contain additional information. One should, however, keep in mind that splitting the events into several bins results in fewer events in each bin and thus in larger statistical uncertainties. As this can compensate any advantage gained by the shape information, the comparison to the results for the total number of events might not always be favorable for the binned observables.

In the pseudoscalar mediator scenario (right column of Figure \ref{fig:Limits_Obs_comp}), we find that $\cos ( \theta^*_{ll})$ is the best observable throughout the considered mass spectrum, irrespective of the luminosity. However, the advantage that $\cos ( \theta^*_{ll})$ holds over the other observables narrows towards very light and very heavy mediators. Thus, one might have to choose a different observable, most likely $M_{T2,t}$, if one were to consider mediator masses outside of the presented range. 

For $L = 300 \text{ fb}^{-1}$, the difference between $\cos ( \theta^*_{ll})$ and $M_{T2,t}$ is fairly small and never exceeds $5\%$, so the latter would still give reasonable limits. However, the discrepancy becomes quite significant for the larger luminosity of $3000 \text{ fb}^{-1}$, in particular around the relevant mass region where $\mu^{95 \% \, CL}(m_Y) \sim 1$. In terms of the excluded mass range, i.e. the masses for which $\mu^{95 \% \, CL}(m_Y) \leq 1$, this difference translates into an improvement from about $m_Y = 475$ GeV when using $M_{T2,t}$ to $505$ GeV for $\cos ( \theta^*_{ll})$.

All other considered observables are consistently worse than $\cos ( \theta^*_{ll})$ and $M_{T2,t}$. 
Incidentally, $\Delta \phi_{l,\text{miss}}$ even provides limits that are worse than those computed using only the total number of events. This is a result of the above described effect of the analysis cuts in this observable. When these cuts are applied, the signal and background distributions behave very similarly to each other. The increased statistical uncertainties from using five bins instead of just one are thus more detrimental here than any advantage gained by the shape information.

For light mediators, we actually find the same phenomenon for $p_{T,\text{miss}}$. In contrast to $\Delta \phi_{l,\text{miss}}$, there are still significant differences to be observed in the normalised $p_{T,\text{miss}}$ distributions, even with the extra cuts, so this alone cannot be the reason for the poor performance. One should note, however, that these differences are mostly visible in the distribution tails where the number of events is very low. Fewer than one in a hundred events falls into the last bin for light mediators. This means that the shape differences are simply not significant enough in light of the substantial statistical uncertainties in that region. This changes if we go to heavier mediators since the tails are more pronounced for these. Thus, above $\sim 700 \,(350)$ GeV for $L = 300 \,(3000) \text{ fb}^{-1}$, the $p_{T,\text{miss}}$ limits are more stringent than those for $\sigma_{\text{tot}}$. The threshold above which $p_{T,\text{miss}}$ is better, is much lower for the larger luminosity because the statistical limitations are substantially smaller.

In the scalar mediator case, the observables behave very similarly to what is discussed above for the heavier mediators because in those cases, the signal distributions do not differ very much form the pseudoscalar case. This also means that $\cos ( \theta^*_{ll})$ provides us with the most stringent limits in that region. However, for lighter mediators, the best observable varies. Specifically, $M_{T2,t}$ outperforms the other observables for lighter mediators.
Here, the shape differences between the scalar DM signal and the SM background in $\cos ( \theta^*_{ll})$ are not as large as in the pseudoscalar case which results in the poorer performance of $\cos ( \theta^*_{ll})$ in this region. For $L = 3000 \text{ fb}^{-1}$ there is also a small mass window between $150$ and $300$ GeV in which $\Delta \phi_{l,\text{miss}}$ provides better limits on the signal strength.

Nevertheless, $\cos ( \theta^*_{ll})$ provides the most stringent limits on the mediator mass range for $\mu^{95 \% \, CL} = 1$, just like in the pseudoscalar case.
Using this observable, one should be able to exclude mediator masses up to $375$ ($385$) GeV for $L = 300 \text{ fb}^{-1}$ and around $485$ ($505$) GeV for $L = 3000 \text{ fb}^{-1}$ when considering the scalar (pseudoscalar) mediator model. Let us stress again that these results were computed assuming a perfect detector so they represent the most ideal case and are not necessarily fully realistic.

\subsection{Modelling of the background} \label{subsec:Limits_Modeling}

\begin{figure*}
	\includegraphics[width=.5\linewidth]{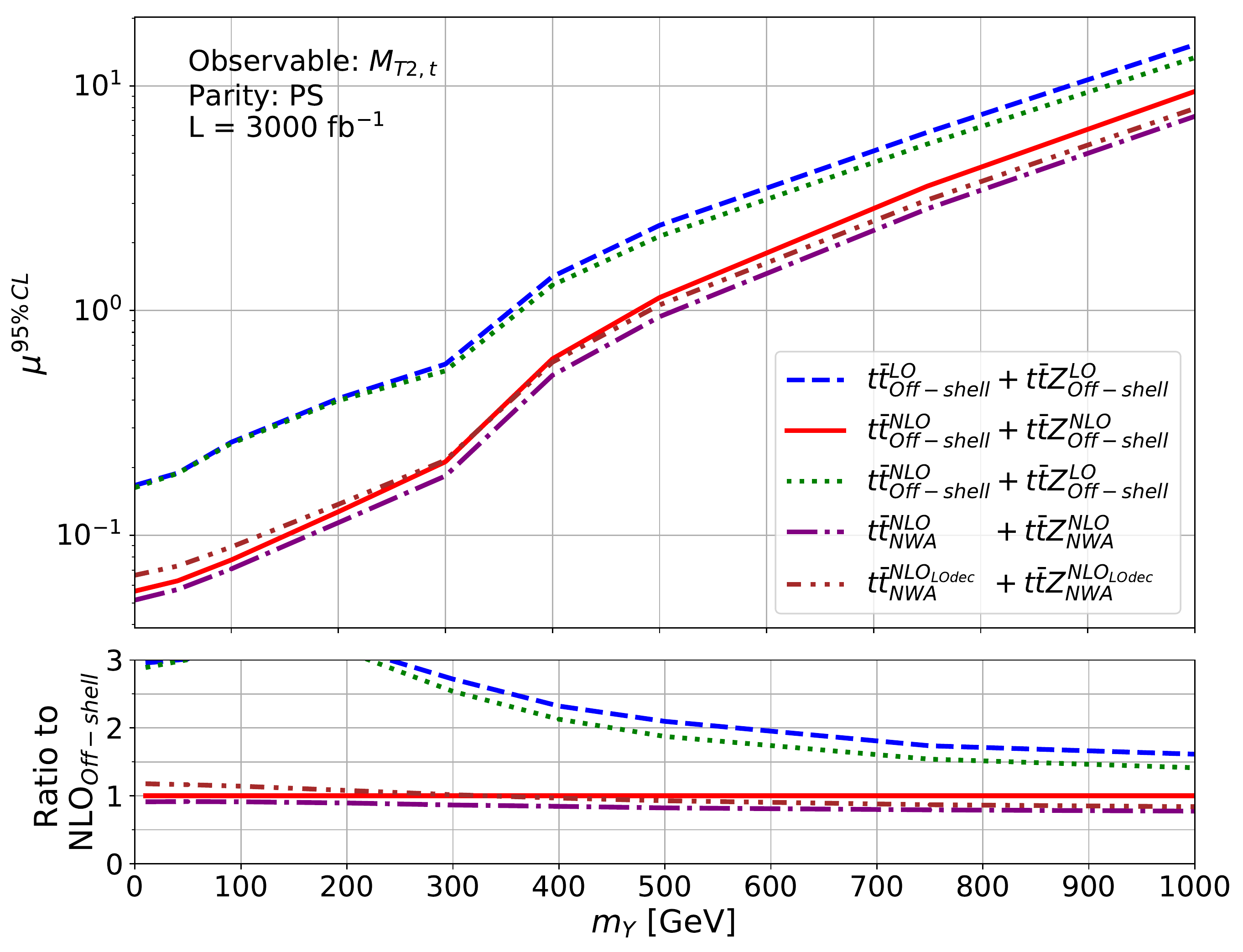}
	\includegraphics[width=.5\linewidth]{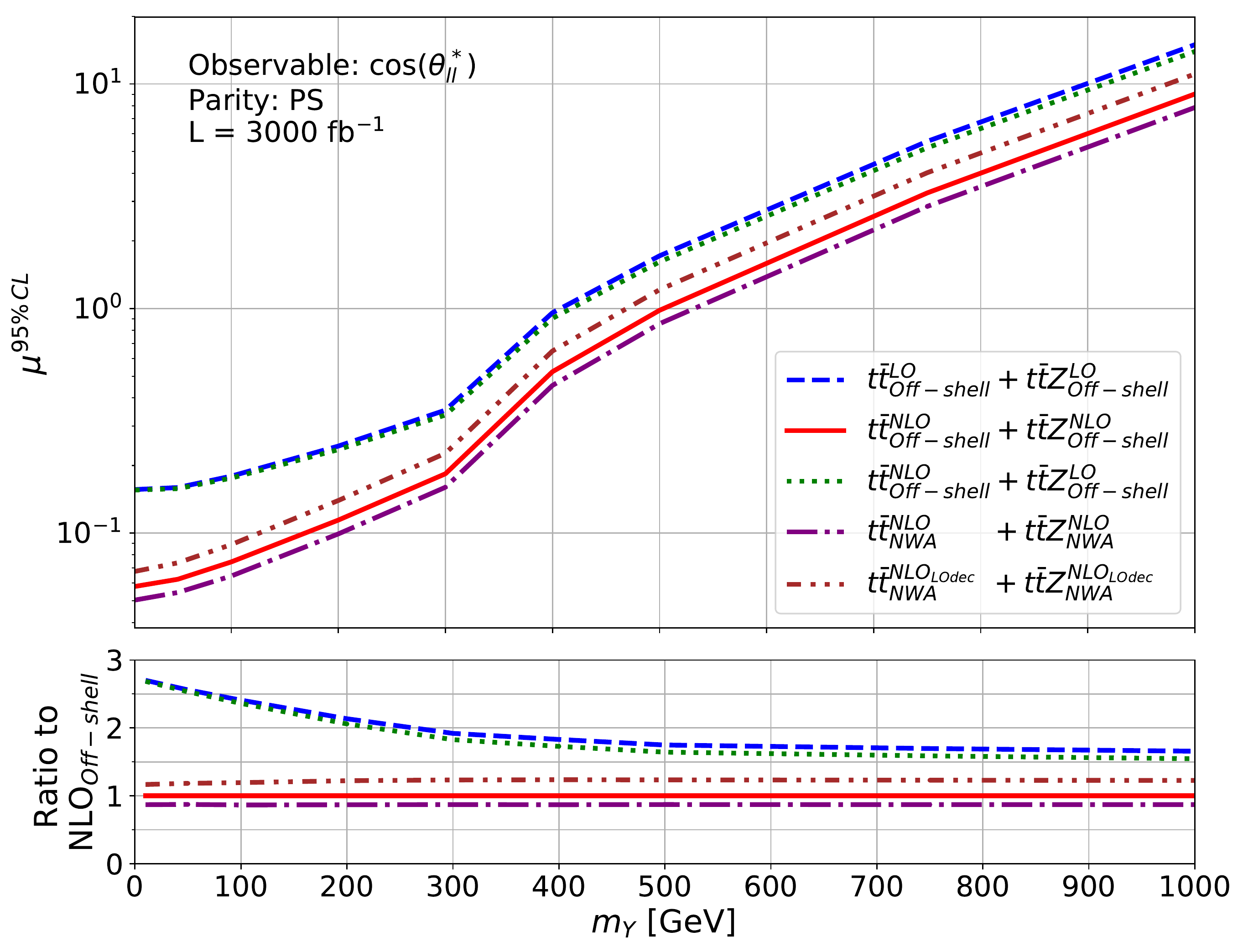}
	
	\caption{\textit{Comparison of signal strength exclusion limits computed with different background predictions for the pseudoscalar mediator scenario with a luminosity of $L = 3000 \text{ fb}^{-1}$ and using $M_{T2,t}$ (left) and $\cos ( \theta^*_{ll})$ (right) as observables. In the lower panels we present the ratios to the limits obtained using the NLO$_{\text{Off-shell}}$ background predictions.}}
	\label{fig:Limits_Bkg_comp}
\end{figure*}

\begin{figure*}
	\includegraphics[width=.5\linewidth]{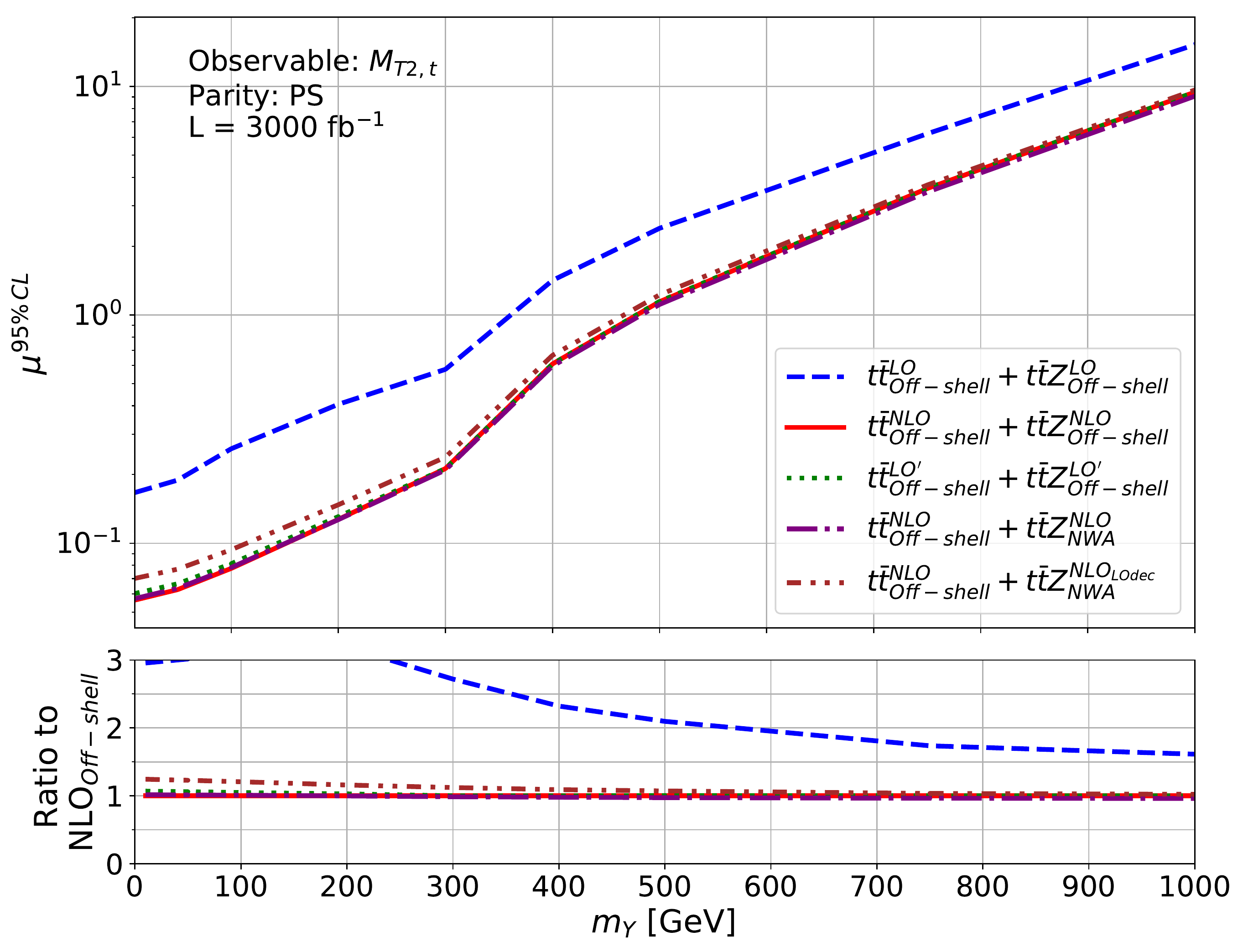}
	\includegraphics[width=.5\linewidth]{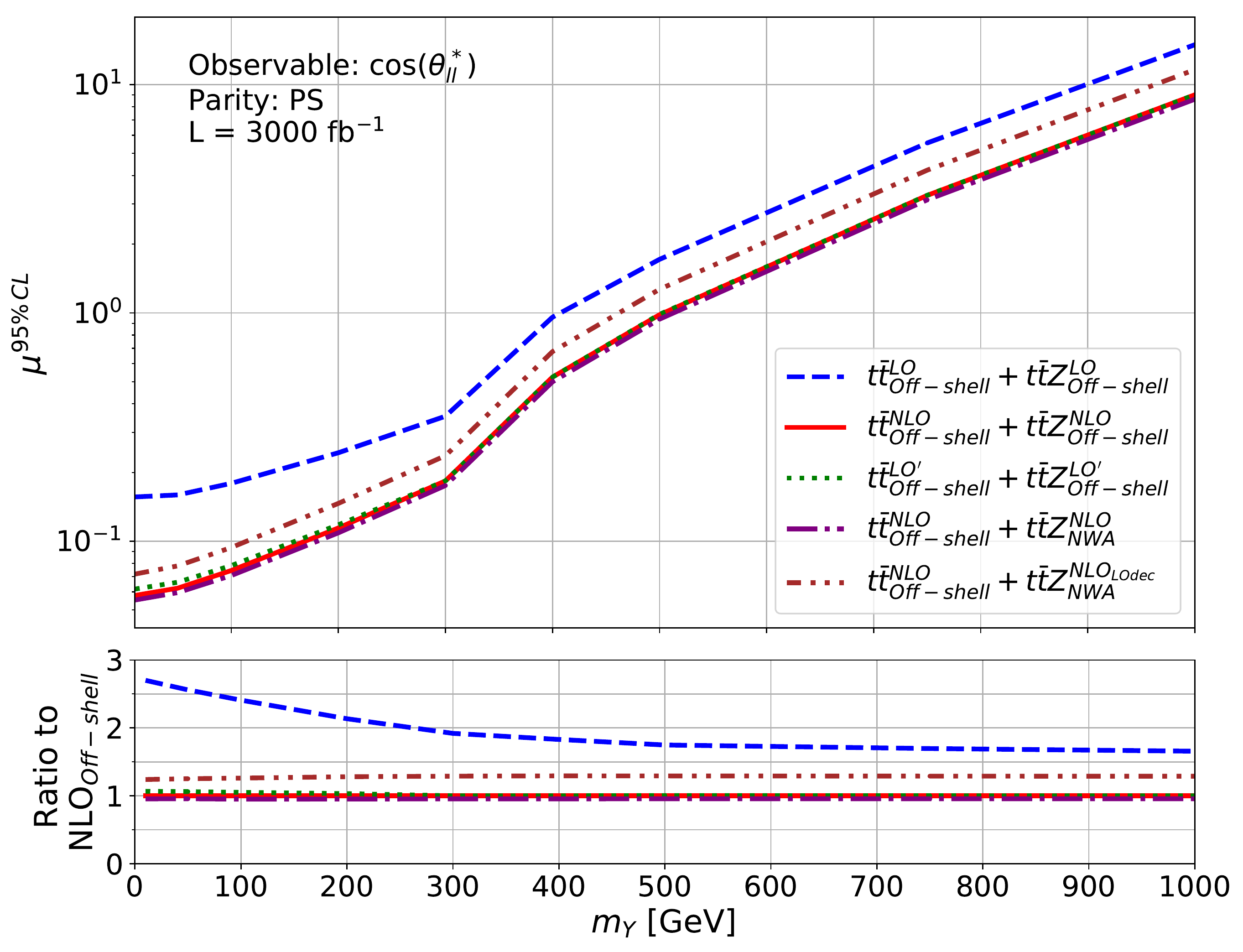}
	
	\caption{\textit{Comparison of signal strength exclusion limits computed with different background predictions for the pseudoscalar mediator scenario with a luminosity of $L = 3000 \text{ fb}^{-1}$ and using $M_{T2,t}$ (left) and $\cos ( \theta^*_{ll})$ (right) as observables. Here, we add the $t\bar{t}$ off-shell prediction to the $t\bar{t}Z$ results in the NWA to eliminate the effect from the missing $t\bar{t}$ contribution. In the lower panels we present the ratios to the limits obtained using the NLO$_{\text{Off-shell}}$ background predictions.}}
	\label{fig:Limits_Bkg_comp_mixed}
\end{figure*}

Since we have now established that $\cos (\theta^*_{ll})$ and $M_{T2,t}$ yield the most stringent limits on the signal strength, we use these to investigate the impact of higher-order and off-shell effects to the background when calculating these limits. 
To this end, we compare exclusion limits computed with the NLO$_{\text{Off-shell}}$ background to those for LO$_{\text{Off-shell}}$, NLO$_{\text{NWA}}$, and NLO$_{\text{NWA,LOdec}}$ in Figure \ref{fig:Limits_Bkg_comp}. As computing $t\bar{t}Z$ at NLO is much more involved than $t\bar{t}$, we also include a mixed case with $t\bar{t}$ at NLO and $t\bar{t}Z$ at LO. All of the limits are presented for the pseudoscalar mediator scenario for $L = 3000 \text{ fb}^{-1}$ but the effects are very similar for scalar mediators and different luminosities. In the lower panels, we show the ratios to the NLO$_{\text{Off-shell}}$ limits.

It is immediately apparent that using LO predictions is completely inadequate. Even combining NLO $t\bar{t}$ with LO $t\bar{t}Z$ predictions yields only minor improvements because $t\bar{t}Z$ is the dominant background process. NLO corrections to the latter are thus essential when computing exclusion limits. The large discrepancy between the LO and NLO results is mostly a consequence of the drastic reduction in scale uncertainties when higher-order corrections are included. In contrast, the shape distortions between the two orders in the perturbative expansion only play a minor role. Note that the latter are kept at a moderate level due to our scale choice and that different scale settings would significantly increase  the size of higher-order corrections.
The importance of scale uncertainties is emphasised by the observation that the gap between the LO$_{\text{Off-shell}}$ and NLO$_{\text{Off-shell}}$ curves decreases towards heavier mediators. Due to lower numbers of events compared to models with lighter mediators, the statistical uncertainties become more relevant which in turn diminishes the effect of reducing the scale uncertainties. Still, the LO limits on the signal strength are at least $65\%$ weaker in both observables for any considered mass point.

The impact of off-shell effects is significantly smaller but still relevant. At first glance, it seems like the `best' limits are obtained by using the full NWA at NLO for both background processes. However, this is not really an improvement but rather an underestimation of the signal strength exclusion limits when compared to the ones obtained with the full off-shell predictions. Thus, we put `best' in quotation marks here.
There are two sources for this difference. First and foremost is the fact that there is no $t\bar{t}$ contribution in the NWA which means that about a quarter of the background events is missing. Secondly, there are the off-shell effects in the remaining $t\bar{t}Z$ background which change the behavior of the observables. As these effects are much more significant for $M_{T2,t}$ than for $\cos (\theta^*_{ll})$, the corresponding limits are also affected more severely for the former. In addition to being small, the off-shell effects in the $\cos (\theta^*_{ll})$ distribution are also fairly uniform. As a consequence, the ratio between the limits obtained with the NLO$_{\text{NWA}}$ and NLO$_{\text{Off-shell}}$ modelling approaches is essentially flat as well. For $M_{T2,t}$, this ratio decreases with increasing mediator mass since the $M_{T2,t}$-tails are the region where off-shell effects are the most prominent. As these tails are more important for the limits on large-$m_Y$ models, off-shell effects are more relevant for these mass points.

In order to assess which of the two effects, the missing $t\bar{t}$ contribution or the shape difference in $t\bar{t}Z$, is the main source of the discrepancy, we re-perform the calculations. Specifically, we re-use  $t\bar{t}Z$ in the NWA but this time we include the $t\bar{t}^{NLO}_{\text{Off-shell}}$ prediction since it is clearly not suitable to use the NWA predictions for the $t\bar{t}$ process.
The results of this are shown in Figure \ref{fig:Limits_Bkg_comp_mixed}. One can see that the discrepancy between off-shell and narrow-width backgrounds is reduced significantly to only a few percent, even in $M_{T2,t}$. These differences are much smaller than one might have initially expected from the distributions shown in Figure \ref{fig:Bkg_Uncut_Modeling_ttZ}. However, we have already seen that the off-shell effects are reduced substantially when the additional cuts are applied to the event samples. In addition, the fits are most likely dominated by the low-$M_{T2}$ bins as the number of events in these bins is several orders of magnitude larger than in the tails and off-shell effects mostly manifest in those tails.

It therefore seems like it is sufficient to only consider the full off-shell background for $t\bar{t}$ and keep $t\bar{t}Z$ in the NWA. 
It is important though to use the full NLO NWA for the $t\bar{t}Z$ process and not the NWA with LO decays. The former gives us $25 \% - 30 \%$ better limits for $\cos (\theta^*_{ll})$. For $M_{T2,t}$, which becomes relevant for light mediators, the effects are a lot smaller in the large-$m_Y$ region, but we still find $15\% - 25 \%$ better limits for light mediators. These improvements are primarily a result of the larger scale uncertainties for NLO$_{\text{NWA,LOdec}}$ compared to the full NLO predictions.

Let us mention that the same general behavior can be observed in the other three observables as well. In all cases, the limits are essentially ordered according to the size of the scale uncertainties on the background. Moreover,  the limits computed for the NWA underestimate the full off-shell results. We also observe again that adding the $t\bar{t}_{\text{Off-shell}}$ predictions to those for $t\bar{t}Z_{\text{NWA}}$ eliminates most of the off-shell effects.

Figure \ref{fig:Limits_Bkg_comp_mixed} also includes an additional prediction which we call LO'. It combines the LO$_{\text{Off-shell}}$ distributions with the uncertainties of the NLO$_{\text{Off-shell}}$ results. This allows us to disentangle the two main differences between the LO and NLO predictions, i.e. the shape distortions and the reduced scale uncertainties. We can clearly see that the LO' curves are much closer to the NLO results than the LO ones. In fact, LO' and NLO agree almost perfectly for $\cos (\theta^*_{ll})$ and $M_{T2,t}$ for mediators heavier than about $300 \text{ GeV}$. For lighter mediators one can observe some small deviations but these remain within a few percent. For $\Delta \phi_{l,\text{miss}}$, however, these deviations reach up to $15 \%$ in the scalar mediator scenario which makes it necessary to also use the NLO distribution shape for this observable. 

These results lead us to conclude that the scale uncertainties are indeed the main reason for the differences between the limits calculated with LO and NLO background predictions. In contrast, shape distortions only play a minor role for all of the considered observables.

\subsection{Central scale choice} \label{subsec:Limits_Scale}

\begin{figure*}
	\includegraphics[width=.5\linewidth]{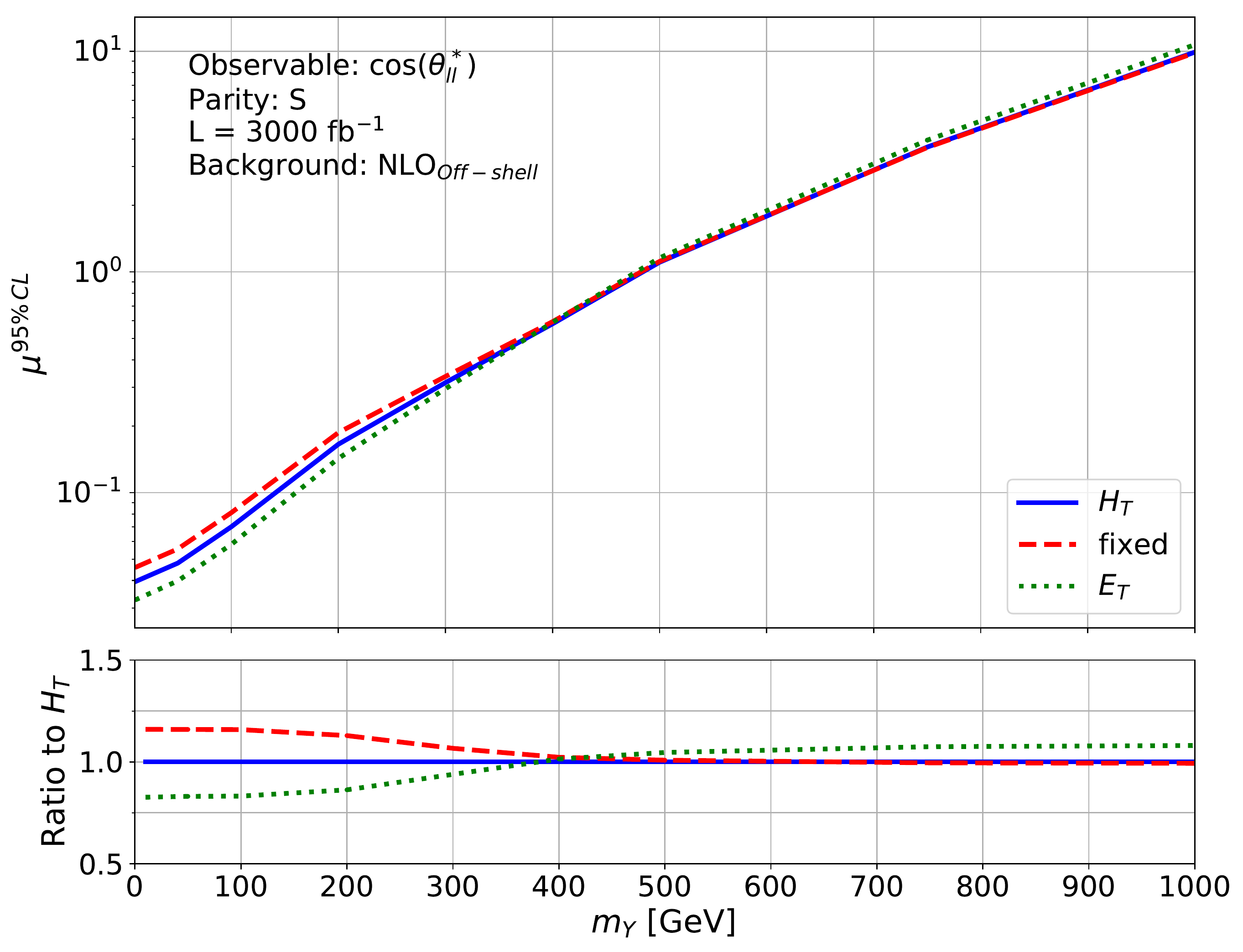}
	\includegraphics[width=.5\linewidth]{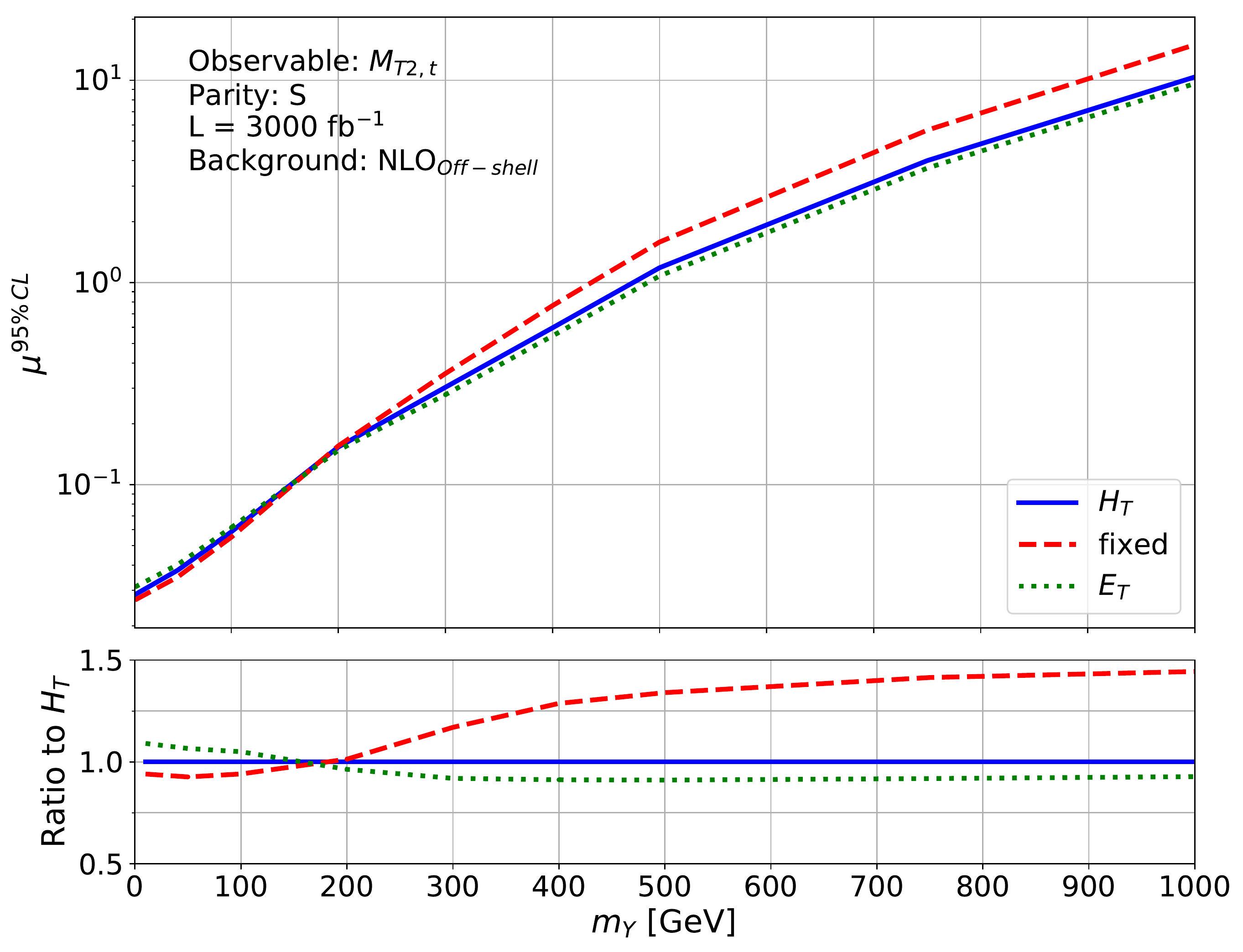}
	
	\caption{\textit{Comparison of signal strength exclusion limits computed with different central scale choices for the background using the scalar mediator DM model and a luminosity of $L = 3000 \text{ fb}^{-1}$. For this comparison we use the $NLO_{\text{Off-shell}}$ background. In the lower panels we present the ratios to the limits obtained using the default $H_T$ scale setting.}}
	\label{fig:Limits_Scale_comp}
\end{figure*}
Next to our default scale choice, we also perform the same calculations for the backgrounds with fixed and $E_T$ scale settings (see Table \ref{table:Central_scale}). As this is purely an evaluation of the background, we keep $E_T/3$ as the central scale choice for our signal. The resulting exclusion limits are compared to the default scale setting in Figure \ref{fig:Limits_Scale_comp} for $L = 3000$ fb$^{-1}$ and the NLO$_{\text{Off-shell}}$ background.

We find effects of a few percent when using a luminosity of $L = 300$ fb$^{-1}$ as the shape differences are only minor between the various scale choices. However, for $L = 3000$ fb$^{-1}$ the effects can become quite significant and even exceed $45 \%$ for the fits performed with $M_{T2,t}$. This is simply a result of the larger scale uncertainties in the tails of this particular observable when one uses the fixed scale which results in weaker limits. This effect does not manifest for the smaller luminosity since the statistical uncertainties are so large that the difference in scale uncertainties is inconsequential.
We also present the same comparison for $\cos ( \theta^*_{ll})$ which we have earlier deemed to be the most promising observable to compute the exclusion limits. Here, too, we find a significant dependence on the central scale choice, especially for lighter scalar mediators. Again, this is mainly due to the different size of the scale uncertainties. In the pseudoscalar case, the gaps are much smaller.


If we perform the same comparison at LO, the results mostly behave as expected, i.e. the gap between the scale settings increases. However, for $M_{T2,t}$, the observable where this gap is the most prominent at NLO, the difference between the scale settings is almost the same at LO and NLO. Just like at NLO, the main contribution to this difference comes from the size of the scale uncertainties. For the dominant $t\bar{t}Z$ process, they amount to $52 \%$ for the fixed scale and $42 \%$ for $H_T$ at LO. Nevertheless, this is less of a discrepancy than at NLO and cannot alone account for the $\sim 45 \%$ gap between the fixed and $H_T$ settings. The remaining part comes from an overestimation of the $M_{T2,t}$-tail in the fixed scale setting by up to $\sim 55 \%$. Together these two effects at LO result in a behavior that is very similar to the NLO results.



\subsection{Luminosity} \label{subsec:Limits_Lumi}

\begin{figure*}
	\includegraphics[width=.5\linewidth]{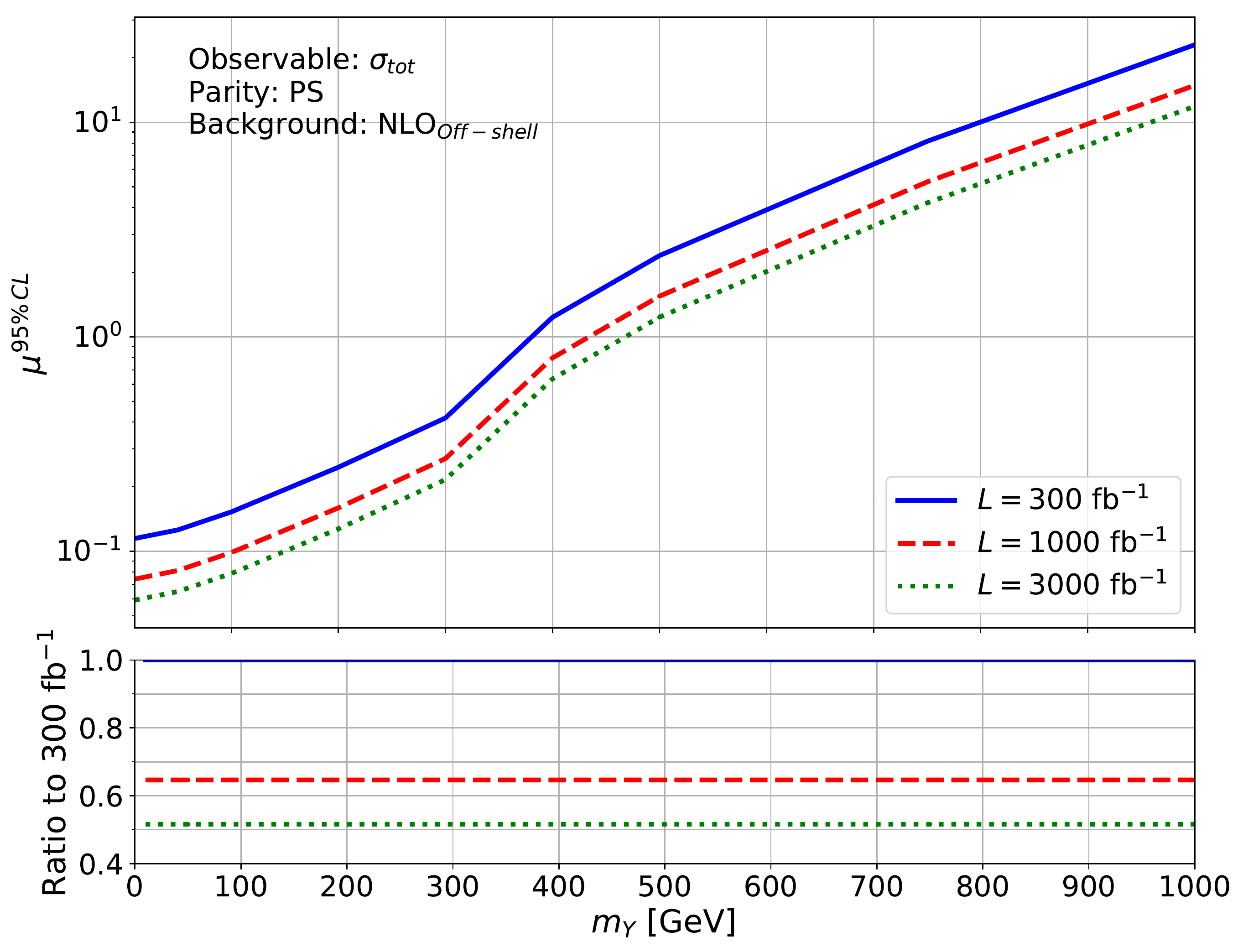}
	\includegraphics[width=.5\linewidth]{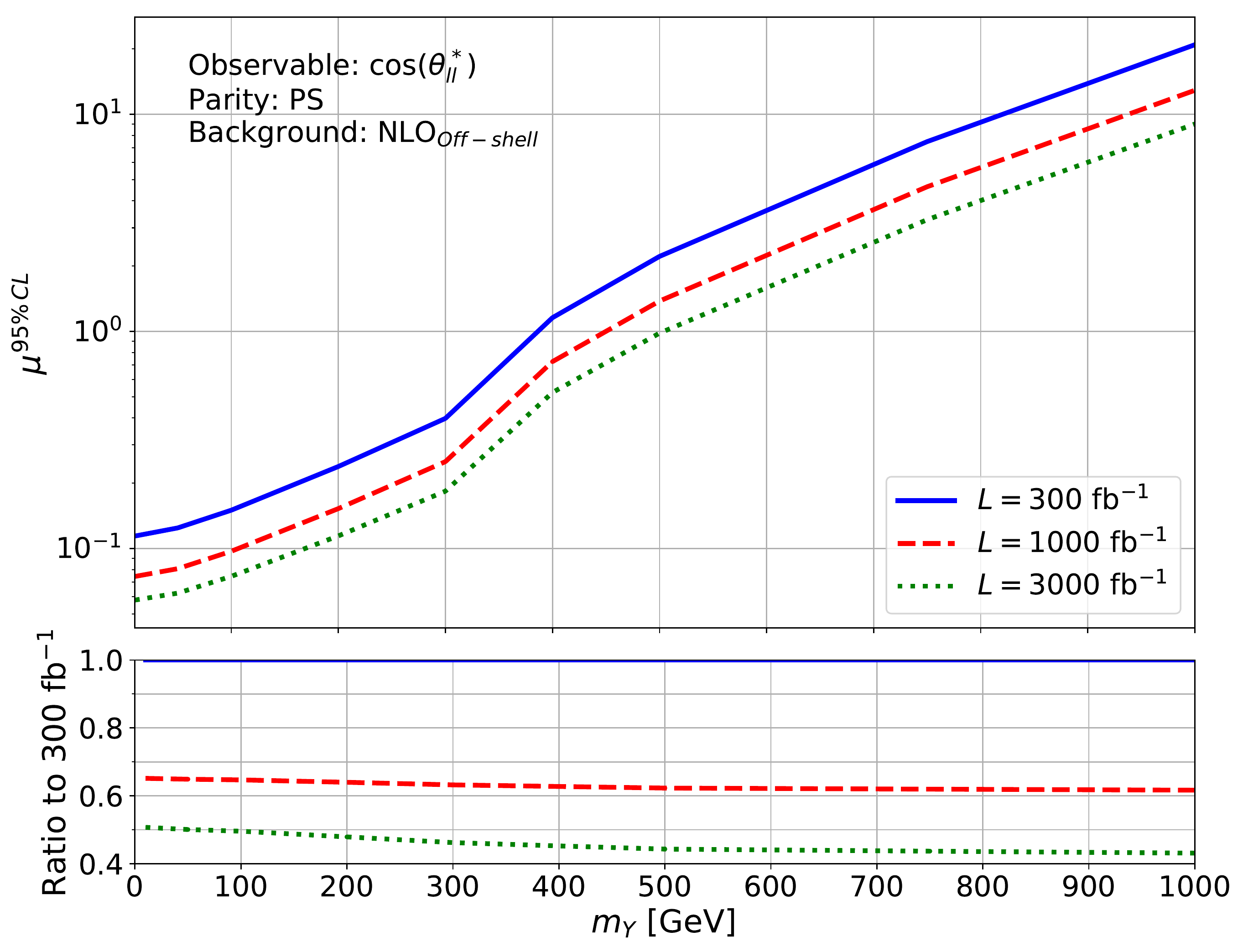}
	
	\caption{\textit{Comparison of signal strength exclusion limits for several different luminosities for the pseudoscalar mediator scenario computed with NLO$_{\text{Off-shell}}$ backgrounds. In the lower panels we present the ratios to the limits obtained using a luminosity of $L = 300 \,\text{ fb}^{-1}$.}}
	\label{fig:Limits_Lumi_comp_NLO}
\end{figure*}
So far, we mostly focused on an integrated luminosity of $L = 3000 \text{ fb}^{-1}$ whilst already touching upon some effects that come from changing the luminosity. In Figure \ref{fig:Limits_Lumi_comp_NLO} we present an explicit comparison of limits obtained with different luminosities for the full off-shell NLO background when using $\sigma_{\text{tot}}$ (left) or $\cos (\theta^*_{ll})$ (right). In both cases, the improvements resulting from reduced statistical uncertainties due to larger luminosities are immediately apparent. For $\sigma_{\text{tot}}$, we find $ 35 \%$ better limits for $L = 1000 \text{ fb}^{-1}$ and $ 48 \%$ for  $ 3000 \text{ fb}^{-1}$ when compared to $L = 300 \text{ fb}^{-1}$. This translates into an extension of the excludable mass range from up to $375$ GeV for $300 \text{ fb}^{-1}$ to $465$ GeV for $3000 \text{ fb}^{-1}$. These improvements for the signal strength limits are independent of the mediator mass and parity as the scale uncertainties are always the same and the statistical uncertainties are always reduced by the same percentage.

This changes when we consider differential distributions as not every bin has the same theoretical uncertainties. Thus, changing the statistical uncertainties can have a different impact on each bin. As a result, we find minor variations between the mediator masses. These changes also tend to be larger than for the integrated fiducial cross section since an individual bin always has fewer events than the total number of events. Consequently, reducing the statistical uncertainties has more of an impact. For $\cos^* (\theta_{ll})$, for example, the difference between the excludable masses is $120$ GeV instead of the $90$ GeV for $\sigma_{\text{tot}}$.

At LO, the effects of changing the luminosity are for the most part smaller than at NLO since LO scale uncertainties are much larger. Thus, reducing statistical uncertainties has less of an impact. For heavy mediators, however, one gets ratios comparable to those at NLO for the dimensionful observables since statistical uncertainties dominate the uncertainties in the tails, irrespective of the perturbative order.
We should also note that increasing the luminosity only improves the limits up to a certain point because the systematic uncertainties become the only limiting factor. This is more pronounced for LO predictions than for NLO ones since scale uncertainties are much larger for the former.
In contrast to changing the perturbative order, using the NWA instead of the full off-shell predictions has very little impact on the luminosity dependence as the uncertainties are largely independent of the modelling.

\section{Summary} \label{sec:Summary}
In this paper, we have presented a comprehensive study of higher-order corrections and off-shell effects for the dominant backgrounds in $t\bar{t}$ associated DM production. We have focused on the leptonic final state of the top quarks as this channel gives us access to several observables that are quite powerful in distinguishing signal and background processes. 

In the first step of our analysis, we have introduced the spin-0 $s$-channel mediator model which we have used to generate our DM signal. We have demonstrated that the shapes of key observables such as $p_{T,\text{miss}}$, $M_{T2,t}$, $M_{T2,W}$, $\cos (\theta^*_{ll})$, and $\Delta \phi_{l,\text{miss}}$ depend strongly on the mediator's mass and, to a lesser extend, on its parity. Specifically, we have found that tails in normalised distributions are much more pronounced for heavy mediators. Nevertheless, in absolute terms, light mediator models still yield larger cross sections, even in these regions.

We have then proceeded to show that the SM background is characterised by two very different processes, the top-quark background $t\bar{t}$ and the irreducible $t\bar{t}Z$ process. With inclusive cuts, the
top-quark background is very much the dominant process and its cross section is four orders of magnitude larger than the one for the $t\bar{t}Z$ process. However, we have also seen that the latter is much more akin to the signal than $t\bar{t}$ in all of the considered observables which makes it much harder to distinguish the two.
 
Higher-order corrections and off-shell effects have also proven to be of significance here as they substantially alter the shape of $t\bar{t}$ and $t\bar{t}Z$ distributions, particularly in their respective tails. For $t\bar{t}Z$, higher-order corrections exceed $150 \%$ in the high-$p_{T,\text{miss}}$ region while for $M_{T2,t}$, we have observed off-shell corrections of up to $75 \%$. 
Angular observables, on the other hand, remain largely unaffected by the modelling, as does the normalisation. For the latter, $K$-factors amount to $1.01$ and $1.03$ for $t\bar{t}$ and $t\bar{t}Z$, respectively. Furthermore, full off-shell effects at LO and NLO are at the per-mille level for $t\bar{t}$ and between $3\% - 4 \%$ for $t\bar{t}Z$. The differences are much larger for the NLO$_{\text{NWA,LOdec}}$ predictions which deviate by up to $20 \%$ from the LO results.

Additionally, we have also investigated the changes that appear when switching to a different central scale. As expected, the LO results change quite significantly and we found effects in excess of $20 \%$, even for the normalisation. These vanish at NLO but the scale uncertainties' size still heavily depends on the scale choice. When using the fixed scale, they can be more than twice as large in some bins as for our default scale setting $H_T$. 

The significant shape differences between signal and background distributions have been used further to disentangle the two by applying very exclusive cuts in $p_{T\text{miss}}$, $C_{em,W}$, and $M_{T2,W}$. The latter has proven to be especially useful as it completely eliminates the originally dominant top-quark  background if one works in the NWA. However, the kinematic edge in $M_{T2,W}$ that causes this phenomenon is attenuated when considering full off-shell effects so that a small fraction of $t\bar{t}$ events, around $0.0015 \%$, passes the additional cuts when  off-shell effects are taken into account. Due to the large $t\bar{t}$ cross section, this still constitutes around $1/4$ of the total number of events. Consequently,  the inclusion of off-shell effects for the $t\bar{t}$ process is indispensable.

We have also demonstrated that due to its similarity to the signal, the $t\bar{t}Z$ process is much less affected by the additional cuts and actually turns out to be the dominant background for our analysis.
As a result, the total SM background behaves much more akin to the signal and we have shown that signal-to-background ratios no longer change as dramatically in the considered observables. However, the extra cuts still enhance the signal-to-background ratio by several orders of magnitude for all considered mass points.

Off-shell effects have also turned out to be much less important for the $t\bar{t}Z$ process than for $t\bar{t}$. Actually, they are even reduced by the analysis cuts in all considered observables except $\Delta \phi_{l,\text{miss}}$. This stands in contrast to the observation that off-shell effects are most prominent in distribution tails. Hence, one might have expected their importance to increase when applying very exclusive cuts.

For both background processes, we have observed a similar phenomenon for the higher-order corrections. They, too, are significantly reduced for most observables. Again, this is contrary to the expectation that these corrections should be enhanced by exclusive cuts.

In the final part of our analysis, we have investigated how all of these effects impact the calculation of signal strength exclusion limits for our DM model. We have primarily used $\cos (\theta^*_{ll})$ and $M_{T2,t}$ for these comparisons as we have identified these to be the most promising observables. Assuming a luminosity of $L = 3000 \text{ fb}^{-1}$, we have compared exclusion limits in these observables computed with the state-of-the-art NLO$_{\text{Off-shell}}$ background to those using either LO or predictions in the NWA. The differences between LO and NLO predictions were found to be substantial even though the number of events is almost identical due to our scale choice. Instead, the gap is a result of the much larger uncertainties in the LO case. Thus, huge improvements can be made by taking into account NLO QCD corrections to $t\bar{t}$ and $t\bar{t}Z$.

The conclusion for off-shell effects is not quite as strong. We have observed significant changes between the full off-shell description and the NWA but these are mostly down to the missing $t\bar{t}$ contribution in the latter case. When off-shell effects are properly included for $t\bar{t}$, these differences are reduced to a few percent. Thus, we conclude that it is vital to include off-shell effects for the top-quark backgrounds but doing so for the $t\bar{t}Z$ process is less important. However, it is necessary to use the full NLO$_{\text{NWA}}$ description for the latter as modelling the top-quark decays at LO results in larger scale uncertainties which, in turn, leads to less stringent limits.

The central scale choice has also proven to be of importance, even at NLO. As a fixed scale choice results in larger scale uncertainties, the corresponding limits are worse than those computed with the dynamical scale.
In a similar fashion, the impact of changing the integrated luminosity has been investigated. As increasing the luminosity leads to smaller statistical uncertainties, the exclusion limits improve considerably. These changes are more substantial at NLO than at LO as the systematic uncertainties are smaller for the former.

To summarise, the most stringent exclusion limits can be obtained by using $\cos (\theta^*_{ll})$ for the computation. Including higher-order corrections for both $t\bar{t}$ and $t\bar{t}Z$ significantly improves these limits, as does the usage of an appropriate dynamical scale. The inclusion of off-shell effects for the $t\bar{t}$ process is indispensable. However, for the more complicated $t\bar{t}Z$ process it is sufficient to consider the NWA but with NLO QCD corrections to both the production and the top-quark decays. 

In principle, one should do the same for the signal.
	However, extending the 	state-of-the-art prediction with NLO production and LO decays to a full NLO calculation is beyond the scope of this paper.
	Even so, all of the above conclusions should be independent of the order at which the decays are modeled and whether all off-shell effects are taken into account. Doing so could only affect three things; the normalization, the distribution shape and the size of theoretical uncertainties.  
	Firstly, from the difference between the full off-shell NLO and NLO$_\text{LOdec}$ results shown in Tables \ref{table:total_xSec_FW_effects} and \ref{table:total_xSec_NLO_cut}, we would indeed expect the normalization, i.e. the integrated fiducial cross section, to change. 
	However, this would essentially just be a nearly flat adjustment to all signal strength exclusion limit curves so this would not change the above conclusions.
	Secondly, the shape distortions will most likely be similar to those we have observed for $t\bar{t}Z$ and we have already seen that their impact was rather small.
	And finally, theoretical uncertainties on the signal are not taken into account in this type of analysis, so reducing them does not have any impact on the results.
	
	Let us also stress ones more that the aim of this paper is not to provide realistic limits for a particular DM model but rather to highlight the importance of higher-order corrections and off-shell effects in this type of search. In this context, the model we have chosen is just one amongst many and most of the conclusions we have drawn here should be valid for any analysis that relies on high-$p_T$ tails or kinematic edges to distinguish the signal from the SM background, see e.g. Refs. \citep{Nambu_Goldstone,Stop_Reconstruction,Sleptons}.

\begin{acknowledgements}
 We would like to thank Lutz Feld and Danilo Meuser for suggesting the $\Delta \phi_{l,\text{miss}}$ observable. \\
  
  This work was supported by the Deutsche Forschungsgemeinschaft (DFG) under grant $396021762$ - TRR $257$: P$3$H - Particle Physics Phenomenology after the Higgs Discovery and under grant $400140256$ - GRK $2497$: The physics of the heaviest particles at the Large Hadron Collider. \\
  
  Support by a grant of the Bundesministerium f\"ur Bildung und Forschung (BMBF) is additionally acknowledged. \\
  
Simulations were performed with computing resources granted by RWTH Aachen University under projects ${\tt rwth0414}$.
\end{acknowledgements}


\begin{thebibliography}{10}
\providecommand{\url}[1]{{#1}}
\providecommand{\urlprefix}{URL }
\expandafter\ifx\csname urlstyle\endcsname\relax
  \providecommand{\doi}[1]{DOI \discretionary{}{}{}#1}\else
  \providecommand{\doi}{DOI \discretionary{}{}{}\begingroup
  \urlstyle{rm}\Url}\fi

\bibitem{LHC}
  L.~Evans, P.~Bryant, JINST \textbf{3} (2008) S08001.

\bibitem{CMS}
  S.~Chatrchyan, et~al., JINST \textbf{3} (2008) S08004.
 
\bibitem{ATLAS}
  G.~Aad, et~al., JINST \textbf{3} (2008) S08003.

\bibitem{CMS_EFT_1}
V.~Khachatryan, et~al., JHEP  \textbf{06} (2015) 121.

\bibitem{ATLAS_EFT_1}
G.~Aad, et~al., Eur. Phys. J. C \textbf{75}  (2015) 92.

\bibitem{DM_simp_1}
D.~Abercrombie, et~al., Phys. Dark Univ. \textbf{27} (2020) 100371.

\bibitem{DM_simp_2}
D.~Alves, J. Phys. G: Nucl. Part. Phys. \textbf{39} (2012) 105005.

\bibitem{DM_simp_3}
J.~Abdallah, et~al., Phys. Dark Univ. \textbf{9-10}  (2015) 8.

\bibitem{DM_simp_4}
S.A. Malik, et~al., Phys. Dark Univ. \textbf{9-10} (2015) 51.

\bibitem{CMS_simp_DM_1}
A.M. Sirunyan, et~al., Eur. Phys. J. C \textbf{77} (2017) 845.

\bibitem{CMS_simp_DM_2}
A.~Tumasyan, et~al.,  arXiv:2105.09178.

\bibitem{CMS_simp_DM_3}
A.M. Sirunyan, et~al., Eur. Phys. J. C \textbf{80}  (2020) 75.

\bibitem{CMS_simp_DM_4}
A.M. Sirunyan, et~al., JHEP \textbf{03}  (2020) 25.

\bibitem{ATLAS_simp_DM_1}
M.~Aaboud, et~al., Eur. Phys. J. C \textbf{78} (2018) 18.

\bibitem{ATLAS_simp_DM_2}
G.~Aad, et~al.,  arXiv:2104.13240.

\bibitem{ATLAS_simp_DM_3}
G.~Aad, et~al., JHEP \textbf{06} (2020) 151.

\bibitem{ATLAS_excl_limits}
D.~Bhattacharya, JHEP \textbf{04} (2021) 165.

\bibitem{CMS_excl_limits}
A.M. Sirunyan, et~al., JHEP \textbf{03} (2019) 141.

\bibitem{MFV}
G.~D'Ambrosio, G.F. Giudice, G.~Isidori, A.~Strumia, Nucl. Phys. B
  \textbf{645} (2002) 155.

\bibitem{cos_intro}
A.J. Barr, JHEP \textbf{02} (2006) 042.

\bibitem{Haisch_analysis}
U.~{Haisch}, P.~{Pani}, G.~{Polesello}, JHEP \textbf{02} (2017) 131.

\bibitem{MT2_1}
G.~Aad, et~al., JHEP \textbf{06} (2014) 124.

\bibitem{MT2_2}
  A.~Barr, C.~Lester, P.~Stephens, J. Phys. G: Nucl. Part. Phys. \textbf{29} (2003) 2343.

\bibitem{MT2_3}
  C.G. Lester, D.J. Summers, Phys. Lett. B \textbf{463} (1999) 99.

\bibitem{FeynGame}
R.~Harlander, S.~Klein, M.~Lipp, Comput. Phys. Commun. \textbf{256}  (2020) 107465.

\bibitem{tt_NLO_1}
A.~Denner, S.~Dittmaier, S.~Kallweit, S.~Pozzorini, Phys. Rev. Lett. \textbf{106}  (2011)  052001.

\bibitem{tt_NLO_2}
G.~Bevilacqua, M.~Czakon, A.~van Hameren, C.G. Papadopoulos, M.~Worek, JHEP
  \textbf{02} (2011) 083.

\bibitem{tt_NLO_3}
R.~Frederix, Phys. Rev. Lett.  \textbf{112} (2014) 082002.

\bibitem{tt_NLO_new}
A. Denner, S. Dittmaier, S. Kallweit, S. Pozzorini 
 	JHEP \textbf{10} (2012) 110.
  
\bibitem{tt_NLO_4}
G.~Heinrich, A.~Maier, R.~Nisius, J.~Schlenk, M.~Schulze, L.~Scyboz, J.~Winter,
  JHEP \textbf{07} (2018) 129.

\bibitem{tt_NLO_5}
T.~Jezo, J.M. Lindert, P.~Nason, C.~Oleari, S.~Pozzorini, Eur. Phys. J. C
  \textbf{76}  (2016) 691.

\bibitem{ttZ}
G.~{Bevilacqua}, H.B. {Hartanto}, M.~{Kraus}, T.~{Weber}, M.~{Worek}, JHEP
  \textbf{11} (2019) 001.

\bibitem{tt_NNLO_1}
  J.~Gao, A.S. Papanastasiou, Phys. Rev. D \textbf{96} (2017) 051501(R).

\bibitem{tt_NNLO_2}
A.~Behring, M.~Czakon, A.~Mitov, R.~Poncelet, A.S. Papanastasiou, Phys. Rev. Lett. 
  \textbf{123} (2019) 082001.

\bibitem{tt_NNLO_3}
M.~Czakon, A.~Mitov, R.~Poncelet,  JHEP \textbf{05} (2021) 212.

\bibitem{MadGraph}
J.~{Alwall}, et~al., JHEP \textbf{07} (2014) 079.

\bibitem{Denner_Dittmaier}
A.~{Denner}, S.~{Dittmaier}, S.~{Kallweit}, S.~{Pozzorini}, JHEP \textbf{10} (2012) 110.

\bibitem{LHAPDF}
  M.R. {Whalley}, D.~{Bourilkov}, {R.~C. Group},  hep-ph/0508110 [hep-ph].
  
\bibitem{CT14_PDF}
S.~Dulat, T.J. Hou, J.~Gao, M.~Guzzi, J.~Huston, P.~Nadolsky, J.~Pumplin,
C.~Schmidt, D.~Stump, C.P. Yuan, Phys. Rev. D \textbf{93} (2016) 033006.

\bibitem{HELAC_NLO}
G.~{Bevilacqua}, M.~{Czakon}, M.V. {Garzelli}, A.~{van Hameren}, A.~{Kardos},
  C.G. {Papadopoulos}, R.~{Pittau}, M.~{Worek}, Comput. Phys. Commun.
  \textbf{184} (2013) 986.
  
\bibitem{HELAC_NWA}
  G.~Bevilacqua, H.B. Hartanto, M.~Kraus, T.~Weber, M.~Worek, JHEP \textbf{03} (2020) 154.
  
\bibitem{anti-kT}
  M.~Cacciari, G.P. Salam, G.~Soyez, JHEP \textbf{04} (2008) 063.
  
\bibitem{DM_model_Backovic}
M.~Backovi\'c, M.~Kr\"amer, F.~Maltoni, A.~Martini, K.~Mawatari, M.~Pellen,
Eur. Phys. J. C \textbf{75} (2015) 482.

\bibitem{MadSpin}
P.~Artoisenet, R.~Frederix, O.~Mattelaer, R.~Rietkerk, JHEP \textbf{03} (2013) 015.

\bibitem{MadWidth}
J.~{Alwall}, C.~{Duhr}, B.~{Fuks}, O.~{Mattelaer}, D.G. {{\"O}zt{\"u}rk}, C.H.
  {Shen}, Comput. Phys. Commun. \textbf{197}  (2015) 312.

\bibitem{FastJet_manual}
  M.~Cacciari, G.P. Salam, G.~Soyez, Eur. Phys. J. C \textbf{72} (2012) 1896.
  
\bibitem{DM_MatrixElement}
T.~{Han}, J.~{Sayre}, S.~{Westhoff}, JHEP \textbf{04} (2015) 145.

\bibitem{MT2_calculation}
C.G. Lester, B.~Nachman, JHEP \textbf{03} (2015) 100.

\bibitem{HELAC_1loop}
  A.v. Hameren, C.~Papadopoulos, R.~Pittau, JHEP \textbf{09} (2009) 106.
  
\bibitem{HELAC_Dipoles}
  M.~{Czakon}, C.G. {Papadopoulos}, M.~{Worek}, JHEP \textbf{08} (2009) 085.
  
\bibitem{CutTools}
G.~Ossola, C.G. Papadopoulos, R.~Pittau, JHEP
\textbf{03} (2008) 042.

\bibitem{OneLoop}
  A.~van Hameren, Comput. Phys. Commun. \textbf{182} (2011) 2427.
  
\bibitem{NS_subtraction}
G.~Bevilacqua, M.~Czakon, M.~Kubocz, M.~Worek, JHEP 
  \textbf{10} (2013) 204.

\bibitem{CS_subtraction}
  S.~{Catani}, M.H. {Seymour},  Nucl. Phys. B \textbf{485} (1997) 291.
  
\bibitem{CS_subtraction_massive}
S.~Catani, S.~Dittmaier, M.H. Seymour, Z.~Trócsányi, Nucl. Phys. B
  \textbf{627} (2002) 189.

\bibitem{ttgamma}
G.~Bevilacqua, H.~Hartanto, M.~Kraus, T.~Weber, M.~Worek, JHEP \textbf{03} (2020) 154.

\bibitem{ttW}
G.~Bevilacqua, H.Y. Bi, H.B. Hartanto, M.~Kraus, M.~Worek, JHEP  \textbf{08} (2020) 043.

\bibitem{HistFitter}
M.~Baak, G.J. Besjes, D.~C\^ote, A.~Koutsman, J.~Lorenz, D.~Short, Eur. Phys.
J. C \textbf{75} (2015) 153.

\bibitem{CLs_ALRead}
  A.L. Read,  J. Phys. G: Nucl. Part. Phys. \textbf{28} (2002) 2693.


\bibitem{Misbinning_Heymes_PhD}
D.~Heymes, {\it A general subtraction scheme for next to next to leading order
  computations in perturbative quantum chromodynamics}, PhD thesis,
RWTH Aachen University, Germany (2015).

\bibitem{Nambu_Goldstone}
U.~Haisch, G.~Polesello, S.~Schulte, JHEP \textbf{09} (2021) 206.

\bibitem{Stop_Reconstruction}
T.~Plehn, M.~Spannowsky, M.~Takeuchi, D.~Zerwas, JHEP \textbf{10} (2010) 078.

\bibitem{Sleptons}
B.~Fuks,  K.~Nordström, R.~Ruiz, S.L.~Williamson, Phys. Rev. D \textbf{100} (2019) 074010.


\end{thebibliography}

%
%

\end{document}